\documentclass[aps,prb,twocolumn,groupedaddress,
showpacs,amsmath,amssymb]{revtex4}

\usepackage{graphicx}
\usepackage{dcolumn}
\usepackage{bm}
\begin{document}

\title{Transport and magnetic properties of GdBaCo$_{2}$O$_{5+x}$ single
crystals: A cobalt oxide with square-lattice CoO$_2$ planes over a wide
range of electron and hole doping}

\author{A. A. Taskin}
\email[]{kotaskin@criepi.denken.or.jp}
\author{A. N. Lavrov}
\altaffiliation{Present address: Institute of Inorganic Chemistry,
Novosibirsk 630090, Russia}
\author{Yoichi Ando}
\affiliation{Central Research Institute of Electric Power Industry,
Komae, Tokyo 201-8511, Japan}


\begin{abstract}

Single crystals of the layered perovskite GdBaCo$_{2}$O$_{5+x}$ (GBCO)
have been grown by the floating-zone method, and their transport,
magnetic, and structural properties have been studied in detail over a
wide range of oxygen contents, $0 \leq x \leq 0.77$. The obtained data
are used to establish a rich phase diagram centered at the ``parent''
compound GdBaCo$_{2}$O$_{5.5}$ -- an insulator with Co ions in the 3+
state. An attractive feature of GdBaCo$_{2}$O$_{5+x}$ is that it allows
a precise and {\it continuous} doping of CoO$_2$ planes with either
electrons or holes, spanning a wide range from the charge-ordered
insulator at 50\% electron doping ($x=0$) to the undoped band insulator
($x=0.5$), and further towards the heavily hole-doped metallic state.
This continuous doping is clearly manifested in the behavior of
thermoelectric power which exhibits a spectacular divergence with
approaching $x$=0.5, where it reaches large absolute values ($\pm800$
$\mu$V/K) and abruptly changes its sign. At low temperatures, the
homogeneous distribution of doped carriers in GBCO becomes unstable, as
is often the case with strongly correlated systems, and both the
magnetic and transport properties point to an intriguing nanoscopic
phase separation into two insulating phases (for electron-doped region)
or an insulating and a metallic phases (for hole-doped region). We also
find that throughout the composition range the magnetic behavior in GBCO
is governed by a delicate balance between ferromagnetic (FM) and
antiferromagnetic (AF) interactions, which can be easily affected by
temperature, doping, or magnetic field, bringing about FM-AF transitions
and a giant magnetoresistance (MR) phenomenon. What distinguishes GBCO
from the colossal-MR manganites is an exceptionally strong uniaxial
anisotropy of the Co spins, which dramatically simplifies the possible
spin arrangements. This spin anisotropy together with the possibility of
continuous ambipolar doping turn GdBaCo$_{2}$O$_{5+x}$ into a model
system for studying the competing magnetic interactions, nanoscopic
phase separation and accompanying magnetoresistance phenomena.

\end{abstract}


\maketitle

\section{INTRODUCTION}

Since the discovery of the high-$T_c$ superconductivity (HTSC) in
cuprates and shortly after of the colossal magnetoresistance (CMR) in
manganites, a great deal of experimental and theoretical efforts have
been made to clarify the nature of these phenomena. The research has
soon revealed that the unusual behavior of cuprates and manganites is
not limited to HTSC and CMR: These compounds based on seemingly simple
metal-oxygen planes turn out to possess very rich phase diagrams,
originating from strong electron correlations and involving spin,
charge, orbital, and lattice degrees of freedom.\cite{Imada, CMR_rev,
Kivelson} In particular, the strong electron correlations, which prevent
the electrons in partially filled bands from forming conventional
itinerant Bloch states, make these systems prone to nanoscopic phase
separation and self-organization of electrons into various
superstructures. The role of this electron self-organization still
remains controversial. It is often argued, for example, that the HTSC
and CMR would never be possible in a homogeneous system, and it is the
nanoscopic mixture of phases that stays behind these novel
phenomena.\cite{CMR_rev, Kivelson}

Ironically, the complexity of manganese and copper oxides that brings
about all the fascinating physics also makes these compounds very
difficult for understanding. The research thus naturally expanded
towards other transition-metal oxides, both because the comparison of
different systems could give a clue to the behavior of cuprates and
manganites, and because those less studied systems might be interesting
in their own right. Such exploration has indeed been proven to be very
fruitful, resulting, for instance, in the discovery of unconventional
superconductivity in a layered-perovskite ruthenium oxide\cite{Sr2RuO4}
and recently in a layered cobalt oxide.\cite{SC} A study of nickel and
cobalt oxides has also revealed the spin/charge ordering
phenomena\cite{LaNiO,CO_Co214, Vogt,Suard} and nanoscopic phase
separation\cite{cubic_Co, BiCo2201} closely resembling those in cuprates
and manganites; these observations confirm that the charge ordering is
indeed a generic feature of strongly correlated electrons. By now,
probably the most rich and intriguing behavior has been found in cobalt
oxides, which ranges from giant MR\cite{Martin, Troy1, Troy2, Maignan,
Respaud, GBC_PRL} and large thermoelectric power attributed to strong
electron correlations\cite{NaCoO, S_book, Koshibae, Ando_specific_heat}
to unconventional superconductivity.\cite{SC} Apparently, the cobalt
oxides, which are still much less studied than cuprates or manganites,
are meant to become the next primary field in investigations of the
strongly correlated electron systems.

Unlike cuprates and manganites, the layered cobalt oxides have two
substantially different crystallographic types: the layered perovskites
derived from {\it square}-lattice CoO$_2$ planes, similar to HTSC and
CMR compounds, and compounds like Na$_x$CoO$_2$ derived from {\it
triangular}-lattice CoO$_2$ planes. In both cases, the CoO$_2$ planes
can be doped with charge carriers over a remarkably wide range so that
the effective valence of Co ions varies from Co$^{2+}$ to
Co$^{4+}$.\cite{Imada} In other words, the doping level ranges from one
electron to one hole per Co ion, if the Co$^{3+}$ state with even number
of electrons is taken as the ``parent'' state. Empirically, the
square-lattice and triangular-lattice systems behave quite differently.
For example, the square-lattice cobalt oxides, such as
La$_{2-x}$Sr$_x$CoO$_4$, Bi$_2$Sr$_2$CoO$_{6+\delta}$, and
RBaCo$_2$O$_{5+x}$ (where R is a rare earth element), are usually
reported to be non-metallic,\cite{Co_214, BiCo2201, Martin, Troy1,
Troy2, Maignan, Respaud, GBC_PRL, Vogt, Moritomo, Moritomo2, OpticalSm}
with the exception of heavily hole-doped {\it cubic} perovskites
La$_{1-x}$Sr$_{x}$CoO$_3$.\cite{cubic_Co} In contrast, the
triangular-lattice cobaltites often appear to be fairly good
metals,\cite{NaCoO,misfitRc, misfit} with the hydrated Na$_x$CoO$_2$
being even a superconductor.\cite{SC} This difference in behavior may
come in part from a disparity in the doping level, given that the latter
compounds were usually studied in a more highly hole-doped region.
Nevertheless, the surprisingly robust non-metallic state in the
square-lattice cobalt oxides \cite{Co_214, BiCo2201} clearly points to a
more fundamental source, which is presumably a very strong tendency to
charge ordering. In the triangular-lattice compounds such tendency might
be considerably weaker. Indeed, the charge ordering usually gains
support from the antiferromagnetic (AF) exchange, while in a triangular
lattice the AF spin interactions should inevitably be frustrated. Thus
far, however, non of these systems has been systematically investigated
over the entire doping range to reveal a coherent picture of doped {\it
cobalt-oxygen} planes.

In this work, we undertake a systematic study of magnetic and transport
properties of a square-lattice cobalt oxide over a wide doping range.
For this purpose, we have selected the RBaCo$_2$O$_{5+x}$ compounds
(where R is a rare earth element), which have already attracted a lot of
attention owing to such fascinating features as the spin-state and
metal-insulator transitions, charge and orbital ordering phenomena, and
giant magnetoresistance (GMR).\cite{Martin, Maignan, Troy1, Troy2,
OpticalSm, Suard, Vogt, Moritomo, Akahoshi, Kusuya, Respaud, GBC_PRL,
Moritomo2, Frontera} The RBaCo$_2$O$_{5+x}$ compounds possess a layered
crystal structure which consists of square-lattice layers
[CoO$_2$]-[BaO]-[CoO$_2$]-[RO$_x$] stacked consecutively along the $c$
axis (Fig. 1) -- a so-called ``112''-type structure.\cite {Martin} This
structure is derived from a simple cubic perovskite
R$_{1-x}$Ba$_{x}$MO$_3$ (where M is a transition metal), but in contrast
to the latter, the rare-earth and alkali-earth ions are located in their
individual layers instead of being randomly mixed.

\begin{figure}
\includegraphics[width=8.6cm]{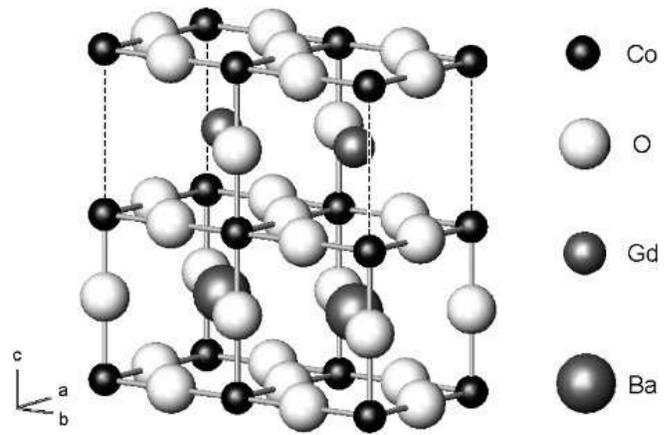}
\caption{Schematic picture of the RBaCo$_{2}$O$_{5+x}$ structure
for $x=0.5$. It is worth noting that the oxygen ions in [RO$_{x}$] layers
demonstrate a strong tendency to ordering; for $x=0.5$, for example, they
form alternating filled and empty rows running along the $a$ axis.\cite
{Martin, Moritomo, Akahoshi}}
\end{figure}

What makes RBaCo$_2$O$_{5+x}$ compounds particularly attractive for our
study is a large variability of the oxygen content: By changing the
annealing conditions, one can modify the oxygen concentration in the
rare-earth [RO$_x$] planes (Fig. 1) in a wide range, $0\leq x\leq 1$
($x$ depends also on the size of R$^{3+}$ ion).\cite {Martin, Akahoshi}
In turn, the oxygen content controls the nominal valence of Co ions,
which varies from 2.5+ to 3.5+, passing through $3+$ (``parent'' state)
at $x=0.5$. Ordinarily, experimental investigations of the $T-x$ phase
diagrams of solids are very time consuming, because they require to grow
many single crystals of different compositions. In contrast, the
composition of already grown RBaCo$_2$O$_{5+x}$ crystals can be modified
by annealing at various temperatures and oxygen partial pressures, which
may allow one to span the entire phase diagram using one and the same
single crystal. Furthermore, by varying the oxygen content one can tune
the doping level very smoothly, which gives a great advantage in
studying the critical regions of the phase diagram.

Upon choosing a compound from the RBaCo$_2$O$_{5+x}$ group, one would
prefer to have a non-magnetic R ion, such as Y, La, or Lu, to avoid
additional complication coming from the rare-earth magnetism.
Unfortunately, the growth of single crystals with these non-magnetic
elements turns out to be virtually impossible.\cite{YLaLu} Therefore, we
have selected the GdBaCo$_2$O$_{5+x}$ compound: Gd$^{3+}$, being a
4f$^7$ ion with zero orbital moment, is known to show rather simple
magnetic behavior in transition-metal compounds, making predominantly
paramagnetic contribution, which can be easily subtracted from the
overall magnetization. Also, owing to the intermediate size of Gd ion in
the series of rare-earth elements, GdBaCo$_{2}$O$_{5+x}$ allows a fairly
wide range of available oxygen concentration. Recently, we succeeded in
growing high-quality GdBaCo$_2$O$_{5+x}$ single crystals using the
floating-zone (FZ) technique, and studied the magnetic and transport
properties of the parent, $x=0.50$, composition.\cite{GBC_PRL}

Here, we present a systematic data on the evolution of transport,
magnetic, and thermoelectric properties of well-characterized
GdBaCo$_2$O$_{5+x}$ single crystals over the entire doping range
available for this compound, namely, $0 \leq x \leq 0.77$. This layered
cobalt oxide turns out to be a really fascinating filling-control system
which allows a {\it continuous ambipolar} doping: We have developed a
technique that provides an easy and precise tuning of the oxygen content
in GdBaCo$_2$O$_{5+x}$ single crystals and ceramics, and succeeded in
doping the parent semiconductor ($x=0.50$) with electrons ($x<0.50$) or
holes ($x>0.50$) with steps that could be as small as 0.001 per Co ion
($\Delta x \sim 0.001$). As a result, we could observe spectacular
singularities in the transport properties upon approaching the undoped
state, could compare the motion of doped holes and electrons, and study
the impact of a small amount of doped carriers on the competition
between the ferromagnetic and antiferromagnetic spin ordering in this
compound. Our study has revealed a rich phase diagram for this layered
cobalt oxide, which is found to include regions of an intriguing
nanoscopic phase separation over virtually the entire doping range.

This paper is organized as follows. In Section II, we describe the
growth of high-quality GdBaCo$_2$O$_{5+x}$ single crystals by the
floating-zone technique, the method of modifying their oxygen content,
and the detwinning technique used to obtain single-domain orthorhombic
crystals. The details of magnetic and transport measurements are also
presented in Section II. Section III starts with a brief summary of the
crystal structure of GdBaCo$_2$O$_{5+x}$ over the oxygen-concentration
range $0 \leq x \leq 0.77$, which is followed with the experimental
results on transport, thermoelectric, and magnetic properties of
GdBaCo$_2$O$_{5+x}$, studied in this wide range of oxygen
concentrations. The obtained data are used to establish an empirical
phase diagram presented at the end of this section. The implications of
our observations are discussed in Section IV. Based on the obtained
experimental results, first we propose a magnetic and electronic
structure of the parent GdBaCo$_2$O$_{5.50}$ compound, giving our
explanation of its transport behavior and the origin of GMR; then we
discuss the effects of doping in GdBaCo$_2$O$_{5+x}$; and, finally,
suggest an overall electronic phase diagram for this compound. Section V
summarizes our findings.

\section{EXPERIMENTAL DETAILS}

\subsection{Growth of GdBaCo$_{2}$O$_{5+x}$ crystals}

We have grown high-quality GdBaCo$_{2}$O$_{5+x}$ single crystals by the
floating-zone technique, using an infrared image furnace with two
halogen lamps and double ellipsoidal mirrors (NEC Machinery SC E-15HD).
A polycrystalline feed rod for the crystal growth was prepared by the
solid-state reaction of Gd$_2$O$_3$, BaCO$_3$, and CoO dried powders:
The mixture was successively calcined at $850^\circ$C, $900^\circ$C,
$950^\circ$C, and $1000^\circ$C, each time for 20 hours, with careful
regrinding after each sintering. Then the obtained homogeneous
single-phase GdBaCo$_2$O$_{5+x}$ powder was isostatically pressed (at
$\sim 70$ MPa) to form a rod with typical dimensions of 7
mm$\phi~\times$ 100 mm. Finally, the feed rod was annealed at
$1200^\circ$C in air to make it dense and hard.

The crystal growth was performed in a flow of dried air, at a constant
rate of 0.5 mm/h. We found that such rather slow growth rate was
essential for obtaining the ordered ``112'' crystal structure, and for
reducing the number of domains in the resulting crystalline rod; higher
rates inevitably caused multi-domain crystal growth. Using optical
microscopy and Laue x-ray back-reflection control, we selected
single-domain parts of the grown crystal rod and cut them into
parallelepiped samples suitable for structural, transport and
magnetization measurements. All the samples' faces were carefully
polished and adjusted to the crystallographic planes with a 1$^{\circ}$
accuracy.

\begin{figure}
\includegraphics[width=8.6cm]{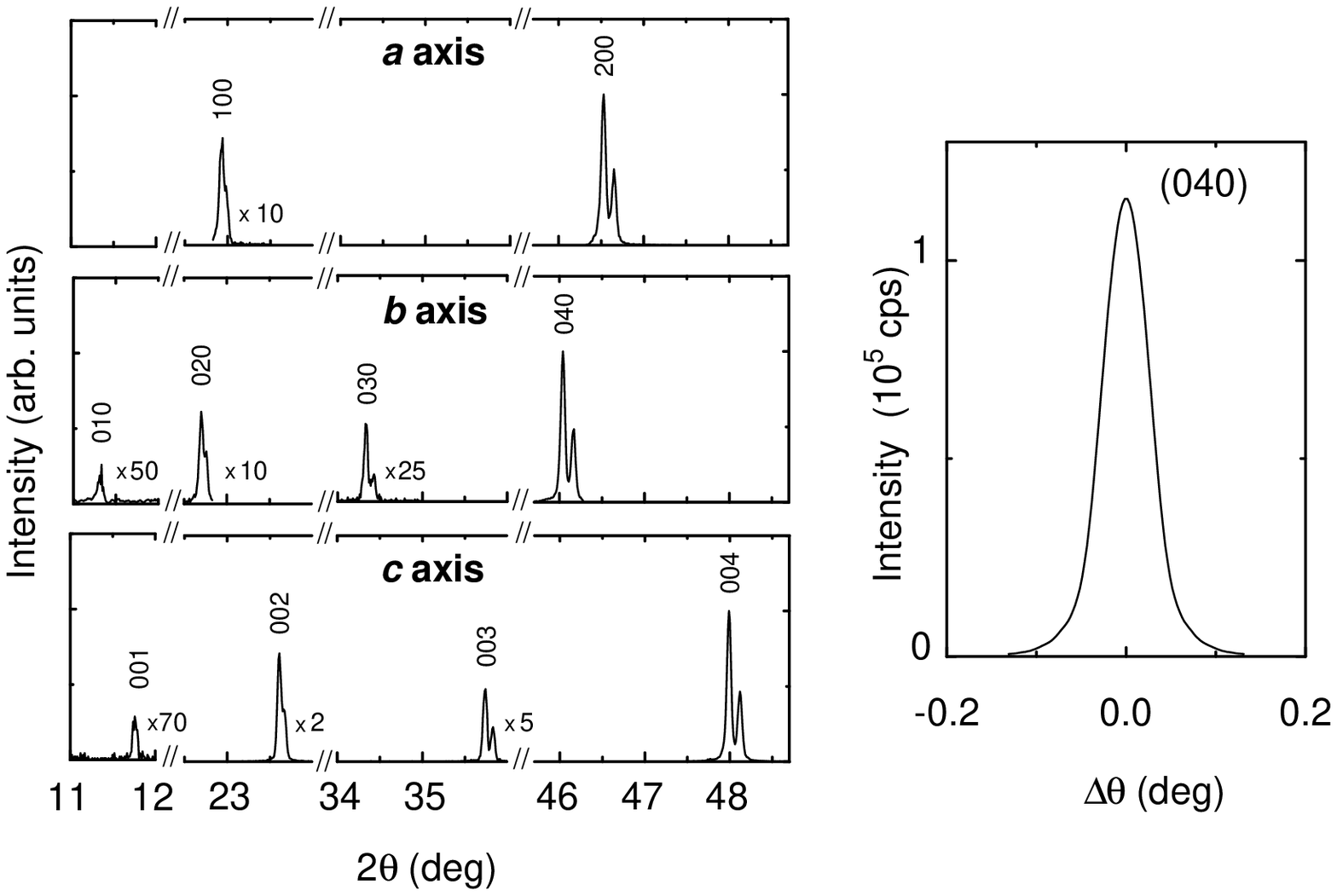}
\caption{(Left) X-ray Bragg's peaks for GdBaCo$_{2}$O$_{5+x}$
($x=0.5$) crystals, which demonstrate the unit-cell doubling along the
$b$ and the $c$ axis (each peak has CuK$\alpha_1$ and CuK$\alpha_2$
contributions to the diffraction pattern). For convenience, the peak
intensity is multiplied by a factor indicated near each peak. (Right)
X-ray (040) rocking curve.}
\end{figure}

The homogeneity and stoichiometry of the obtained GdBaCo$_{2}$O$_{5+x}$
crystals were analyzed by the electron-probe microanalysis (EPMA) and
inductively-coupled plasma (ICP) spectroscopy which confirmed that the
actual cation composition was uniform, corresponding to the nominal
1:1:2 ratio within the experimental accuracy. Another important issue is
whether Gd and Ba are well ordered in the lattice; in fact, we found
that large rare-earth ions, such as La and Pr, easily mixed with Ba,
resulting in a disordered cubic phase
R$_{0.5}$Ba$_{0.5}$CoO$_{3-\delta}$. In the case of
GdBaCo$_{2}$O$_{5+x}$ crystals, however, the x-ray diffraction data
demonstrate that Gd and Ba are indeed well ordered into consecutive
(001) layers, which results in the doubling of the unit cell along the
$c$ axis (Fig. 2). Moreover, for the oxygen concentration $x\approx
0.5$, the oxygen ions are found to form alternating filled and empty
chains running along the $a$ axis; this brings about the doubling of the
unit cell along the $b$ axis (Fig. 2). Apparently, such long-range
oxygen ordering would hardly be possible if considerable amount of Ba
were substituting Gd in GdO$_x$ layers. Figure 2 shows also a typical
x-ray rocking curve (040) for GdBaCo$_{2}$O$_{5.5}$, which has a
full-width-at-half-maximum (FWHM) of less than 0.1$^{\circ}$, indicating
that our crystals have few macroscopic defects. An additional evidence
for the macroscopic crystallographic perfection is the ease with which
the crystals could be cleaved, especially along the \{001\} planes,
exposing perfectly flat, shiny surfaces.

\subsection{Tuning the oxygen content in GdBaCo$_{2}$O$_{5+x}$}
\subsubsection{Equilibrium oxygen concentration as a function of
temperature and oxygen partial pressure}

Owing to the layered ``112'' crystal structure, the oxygen content in
RBaCo$_{2}$O$_{5+x}$ can be varied within a wide range $0<x<1$, where
the oxygen vacancies are located predominantly in the rare-earth RO$_x$
planes.\cite{Maignan} The most convenient way to modify the oxygen
stoichiometry is the high-temperature annealing, yet it requires a
detailed knowledge of how the equilibrium oxygen content depends on the
temperature and oxygen partial pressure. Besides, one needs to know the
kinetics of the oxygen exchange: The annealing temperature should be
carefully chosen so that the oxygen exchange is quick enough for the
equilibrium state to be reached in a reasonable time, but still slow
enough to avoid unwanted oxygen absorption during the subsequent cooling
of the sample.

In order to establish such dependences, we have performed special sets
of annealings, systematically varying temperature, oxygen partial
pressure $P_{\text{O2}}$, and annealing time, as exemplified by the data
shown in Fig. 3. By measuring the weight change of $\sim 1$-g
polycrystalline samples, as well as large single crystals, with a
0.1-$\mu$g resolution, we could evaluate the change in the oxygen
content $\Delta x$ with an accuracy better than 0.001. Although all the
relative oxygen variations are measured very accurately, there would
still be a large uncertainty in the absolute values of $x$, unless we
pin this relative scale to the absolute oxygen content at least at one
point. Fortunately, in GdBaCo$_2$O$_{5+x}$ there are two peculiar oxygen
concentrations that allow the absolute $x$ scale to be establish
unambiguously. First, in GdBaCo$_2$O$_{5+x}$ crystals, similar to other
RBaCo$_2$O$_{5+x}$ compounds with small rare-earth ions, \cite{Suard,
Akahoshi, Vogt} the oxygen content can be reduced down to $x=0$ by
annealing in vacuum, or inert atmosphere. One can naturally expect the
$x=0$ composition to be stable over rather broad range of parameters,
since in this phase all the weakly-bound oxygen is removed from GdO$_x$
layers, while the strongly-bound oxygen in BaO and CoO$_2$ layers is
still intact. Indeed, we observed that $x$ saturates, approaching some
lowest level [see Fig. 3(b)] as the oxygen partial pressure is reduced
and the annealing temperature is increased, and we attributed this
saturation value to $x=0$. Much more sensitive calibration can be done
at the $x=0.5$ point, which turns out to be critical for
GdBaCo$_2$O$_{5+x}$: Upon crossing this point, the mixed valence
composition of Co$^{2+}$/Co$^{3+}$ ions turns into the
Co$^{3+}$/Co$^{4+}$ one, and consequently the type of charge carriers
switches abruptly. As we will show below, several physical properties of
GdBaCo$_2$O$_{5+x}$ exhibit a very sharp singularity in their $x$
dependences near $x=0.5$, changing remarkably with a minute (by merely
0.001) modification of $x$. The calibrations that used the $x=0$ and
$x=0.5$ points gave the same results, which allows us to be confident in
absolute values of $x$ in our samples.

\begin{figure}
\includegraphics*[width=8.6cm]{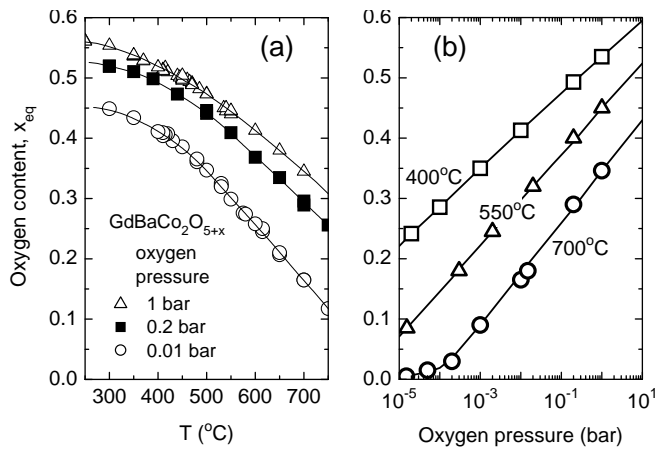}
\caption{ The equilibrium oxygen concentration $x_{\text{eq}}$
in GdBaCo$_{2}$O$_{5+x}$. (a) Temperature dependences of $x_{\text{eq}}$
for several values of the oxygen partial pressure. (b) Dependences of
$x_{\text{eq}}$ on the oxygen partial pressure at several temperatures.}
\end{figure}

Temperature dependences of the equilibrium oxygen content $x$ in
GdBaCo$_{2}$O$_{5+x}$ for different oxygen partial pressures are shown
in Fig. 3(a), which demonstrates that a large variation of $x$ can be
achieved by relatively simple means: by annealing in a flow of oxygen
and argon mixed in different proportions. Note that the shown data
correspond to the equilibrium state, and that all the changes in oxygen
concentration are therefore completely reversible. Our measurements have
shown that the equilibrium $x$ value at a given temperature is roughly
proportional to the logarithm of the oxygen partial pressure, as long as
$x$ is not too close to zero [Fig. 3(b)]. Therefore, in order to obtain
the lowest oxygen concentration $x=0$ we annealed samples in a flow of
argon or helium, and used strongly diluted (down to 10 ppm) oxygen-argon
mixtures to access the low-$x$ range, although precise tuning of $x$ in
this case becomes technically more difficult. In the opposite limit, the
samples with oxygen concentrations up to $x=0.77$ were prepared by
annealing in oxygen at high pressures (up to 70 MPa).

\subsubsection{Oxygen intercalation kinetics}

\begin{figure*}
\includegraphics*[width=13cm]{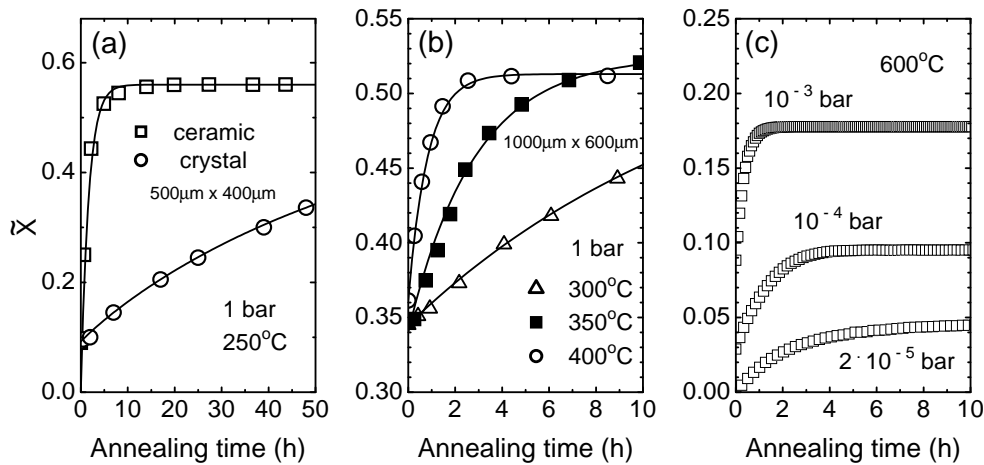}
\caption{Kinetics of the oxygen intercalation into GdBaCo$_{2}$O$_{5+x}$.
(a) Comparison of the oxygen uptake in a ceramic sample (with an average
grain size of $\sim 10$ $\mu$m) and a single crystal (with sizes in the
$ab$ plane of 500 $\mu$m$\,\times\,400$ $\mu$m) measured in the oxygen
flow at $250^\circ$C. Solid lines are the result of simulation, see
text. (b) The oxygen uptake in a single crystal with sizes in the $ab$
plane of 1000 $\mu$m$\,\times\,600$ $\mu$m measured in the oxygen flow
at different temperatures (solid lines are the result of simulation).
(c) Dependence of the oxygen-uptake kinetics in a ceramic samples at
$600^\circ$C on the oxygen partial pressure.}
\end{figure*}

As we have mentioned above, there is one more point to be concerned
about, upon choosing the proper annealing conditions, besides the
equilibrium $x$ value: it is the oxygen-intercalation kinetics. If the
oxygen diffusion is too slow, one will inevitably end up with a crystal
having a large composition gradient; if, in contrast, the oxygen uptake
is too fast, it will be difficult to preserve the achieved state even by
very fast cooling. By measuring how the sample's mass evolves with time
$t$, we have determined the time dependences of the average oxygen
content $\tilde{x}(t)$ for different annealing conditions, as shown in
Fig. 4. Apparently, the proper duration of the heat treatment necessary
to obtain a homogeneous oxygen distribution should be several times
longer than the characteristic time $\tau$ of the oxygen exchange
process at a given temperature and an oxygen partial pressure. We have found
that for ceramic samples the kinetics of oxygen intercalation in the
entire studied ranges of temperature and $P_{\text{O2}}$ follows a
simple exponential law,
\begin{equation}
\tilde{x}(t)=x_{\infty}-[x_{\infty}-x_{0}]e^{-t/\tau},
\end{equation}
where $x_{0}$ and $x_{\infty}$ are the initial and the equilibrium
oxygen contents, $\tau$ is a time constant that depends on temperature
and $P_{\text{O2}}$. Such simple exponential dependence can be naturally
attributed to the oxygen exchange limited by a surface energy barrier.
The time constant $\tau$ turns out to be less than 3 hours at
temperatures down to $250^\circ$C (at 1 bar oxygen pressure), so that
the equilibrium state in ceramic samples can be easily achieved by
one-day annealing [Fig. 4(a)].

In the case of single crystals, however, the oxygen intercalation
deviates from the simple exponential behavior, and goes at a noticeably
slower rate, as illustrated in Fig. 4(a). We have found that for $\sim
500$ $\mu$m sized crystals, two days of annealing at $250^\circ$C is far
from being enough to reach the equilibrium oxygen concentration [Fig.
4(a)]; such annealing turns out to be insufficient even for $\sim 100$
$\mu$m samples. This difference in the kinetics of oxygen absorption is
apparently caused by an additional limitation imposed by the bulk oxygen
diffusion: While the crystal surface layer is readily filled with
oxygen, it takes much more time for oxygen to diffuse towards the inner
part of the crystal. For rectangular-shape crystals the oxygen exchange
kinetics has an analytical solution,\cite{Lane, Yasuda} which depends on
the surface exchange coefficient $K$, characterizing the oxygen exchange
at the interface between the gas and the solid, and the chemical
diffusion coefficient $D$, besides the sample size. These parameters can
be extracted by fitting theoretical curves to experimental data, as
exemplified in Figs. 4(a) and 4(b). For example, the data shown in Fig.
4(a) provide the diffusion-coefficient value of $D=3 \times 10^{-8}$
cm$^2/$s at $250^\circ$C.\cite{cond-mat} With increasing temperature,
the diffusion coefficient grows rapidly and so does the rate of the
oxygen uptake, making it possible to achieve a homogeneous oxygen
distribution in $\sim 1$ mm sized crystal already after several hours
annealing at $400^\circ$C [Fig. 4(b)]. Note that the oxygen kinetics
does not show any detectable difference upon varying the sample size
along the $c$ axis, implying an essentially 2D character of the oxygen
diffusion typical for layered oxides.\cite{YBCO}

We should note that the diffusion coefficient found in
GdBaCo$_{2}$O$_{5+x}$ is remarkable in its own right, being unusually
large for such low temperatures; the oxygen diffusivity appears to be
comparable with that in best superionic conductors.\cite{cond-mat} One
additional implication of the high oxygen mobility, which is important
to the present study, is that the oxygen ions may be capable of
rearranging even at room temperature or below, and one should expect the
oxygen to form ordered superstructures or to participate in mesoscopic
phase separation.

After the homogeneous oxygen distribution in a crystal is reached, an
important issue to be concern about is the cooling procedure, which must
be as fast as possible to avoid any further oxygen uptake. To obtain the
most homogeneous samples we used the following rules upon selecting the
annealing conditions. The annealing temperature and the partial oxygen
pressure were selected so that: i) they provided the required oxygen
content in the crystal, and ii) the annealing time necessary to reach
the equilibrium oxygen distribution was in the range from several hours
to several days. After the annealing was completed, the samples were
quickly quenched to room temperature without changing the atmosphere.
Upon quenching, the crystal temperature dropped by $100-200^\circ$C in
several seconds, guaranteeing the blocking of any further oxygen
exchange. The huge difference in the time scales -- hours and days for
modifying the oxygen concentration and seconds for quenching
-- ensured us that the unwanted oxygen uptake could affect no more than a
fraction of percent of the crystal's volume. An important point to be
specially emphasized is that the oxygen exchange at the interface
between the gas and the solid is strongly suppressed at low oxygen
partial pressures. As illustrated in Fig. 4(c), by reducing the oxygen
partial pressure one can arrange a slow oxygen uptake in a ceramic
sample even at $600^\circ$C. This feature turns out to be very useful in
obtaining homogeneous samples within the low oxygen concentration range.

\subsection{Detwinning of crystals}

The oxygen ordering is indeed observed in GdBaCo$_{2}$O$_{5+x}$ at
$x\approx 0.5$, where the oxygen ions in GdO$_x$ planes order into
alternating filled and empty chains, causing a
tetragonal-to-orthorhombic (T-O) transition and doubling of the unit
cell along the $b$ axis (Fig. 2). Usually, this T-O transition is
accompanied by heavy twinning of crystals that mixes the $a$ and $b$
orthorhombic axes; one therefore needs to perform a detwinning procedure
to get a single-domain orthorhombic crystal for studying the in-plane
anisotropy. To detwin crystals, we slowly cooled them under a uniaxial
pressure of $\sim 0.15$ GPa from $260^{\circ}$C, using a polarized-light
optical microscope to control the twin removing. \cite{twins} We should
note, however, that the GdBaCo$_{2}$O$_{5+x}$ crystals are very fragile,
and it was virtually impossible to complete the detwinning procedure
without having them cleaved into two or more pieces. These pieces were
fairly well detwinned (according to x-ray measurements, the remaining
fraction of misoriented domains usually did not exceed 4-5\%) and
suitable for magnetization measurements. Thus far, however, we have not
succeeded in detwinning the crystals with already prepared electrical
contacts (see below), which are necessary for transport measurements.

\subsection{Details of measurements}

To characterize the physical properties of GdBaCo$_{2}$O$_{5+x}$, we
have carried out resistivity, magnetoresistance (MR), Hall, thermopower,
and magnetization measurements within the 2-400 K temperature range.

The in-plane ($\rho_{ab}$) and out-of-plane ($\rho_c$) resistivity was
measured using a standard ac four-probe method. Good electric contacts
were obtained by drawing with a gold paint on polished crystal surfaces,
and subsequent heat treatment. For current contacts, the whole area of
two opposing side faces was painted with gold to ensure a uniform
current flow through the sample. In turn, the voltage contacts were made
narrow ($\sim 50$ $\mu$m) in order to minimize the uncertainty in
absolute values of the resistivity. It is important to note that the
gold contacts were prepared {\it before} all the heat treatments that
were used to vary the oxygen content. After the required oxygen
concentration was set by annealing, thin gold wires were attached to the
contact pads using a room-temperature-drying silver paste (DuPont 4922),
which electrically and mechanically bound the wire to the sample. As the
measurements were completed, the wires were removed and the crystal was
reannealed to get the next $x$ value.

The MR measurements were done either by sweeping the magnetic field
between $\pm$14 T at fixed temperatures stabilized by a capacitance
sensor with an accuracy of $\sim$1 mK,\cite{MR_accur} or by sweeping the
temperature at a fixed magnetic field. Both $\Delta\rho_c/\rho_c$ and
$\Delta\rho_{ab}/\rho_{ab}$ have been measured for ${\bf
H}$$\,\parallel\,$$ab$ and ${\bf H}$$\,\parallel\,$$c$.

The Hall resistivity was measured using a standard six-probe technique
by sweeping the magnetic field ${\bf H}$$\,\parallel\,$$c$ to both plus
and minus polarities at fixed temperatures; the electric current was
always along the $ab$-plane. After the symmetric (MR) contribution
coming from a slight misplacement of the contacts was subtracted from
the raw data, the Hall resistivity appeared to be perfectly linear in
magnetic field for all measured temperatures, implying that the
anomalous Hall effect was negligible.

In order to determine the thermoelectric power (Seebeck coefficient)
$S$, we generated a slowly oscillating thermal gradient along the sample
(within $\sim 1$ K), and measured an induced periodic voltage. This
allowed us to get rid of thermoelectric contributions generated in the
remaining circuit. A chromel-constantan thermocouple employed for
measuring the thermal gradient was attached to the heat source and to
the heat sink, which were fixed to the sample by a silver paste. To
obtain the absolute value of the thermoelectric power, a contribution
from the gold wires ($\sim 2$ $\mu$V/K) used as output leads was
subtracted. Thermopower measurements were mostly performed on long ($>2$
mm) samples by applying a temperature gradient along the $a$ or the $b$
axis.

Magnetization measurements were carried out using a SQUID magnetometer
at fields up to 7 T applied along one of the crystallographic axis.
Measurement modes included taking data upon heating the sample after it
was cooled down to 2 K in zero field (ZFC), upon cooling the sample in
magnetic field from 400 K (FC), and upon sweeping the field at fixed
temperatures. Throughout this paper, the magnetization coming from Co
ions is determined by subtracting the contribution of Gd ions, assuming
their ideal paramagnetic (PM) behavior with total spin $S=7/2$; the
latter is a good approximation since no ordering of Gd$^{3+}$ moments in
any of the samples is detected down to 1.7 K.

\section{RESULTS}

\subsection{Crystal structure}
\label{sec:Str}

\begin{figure}[b]
\includegraphics[width=8.6cm]{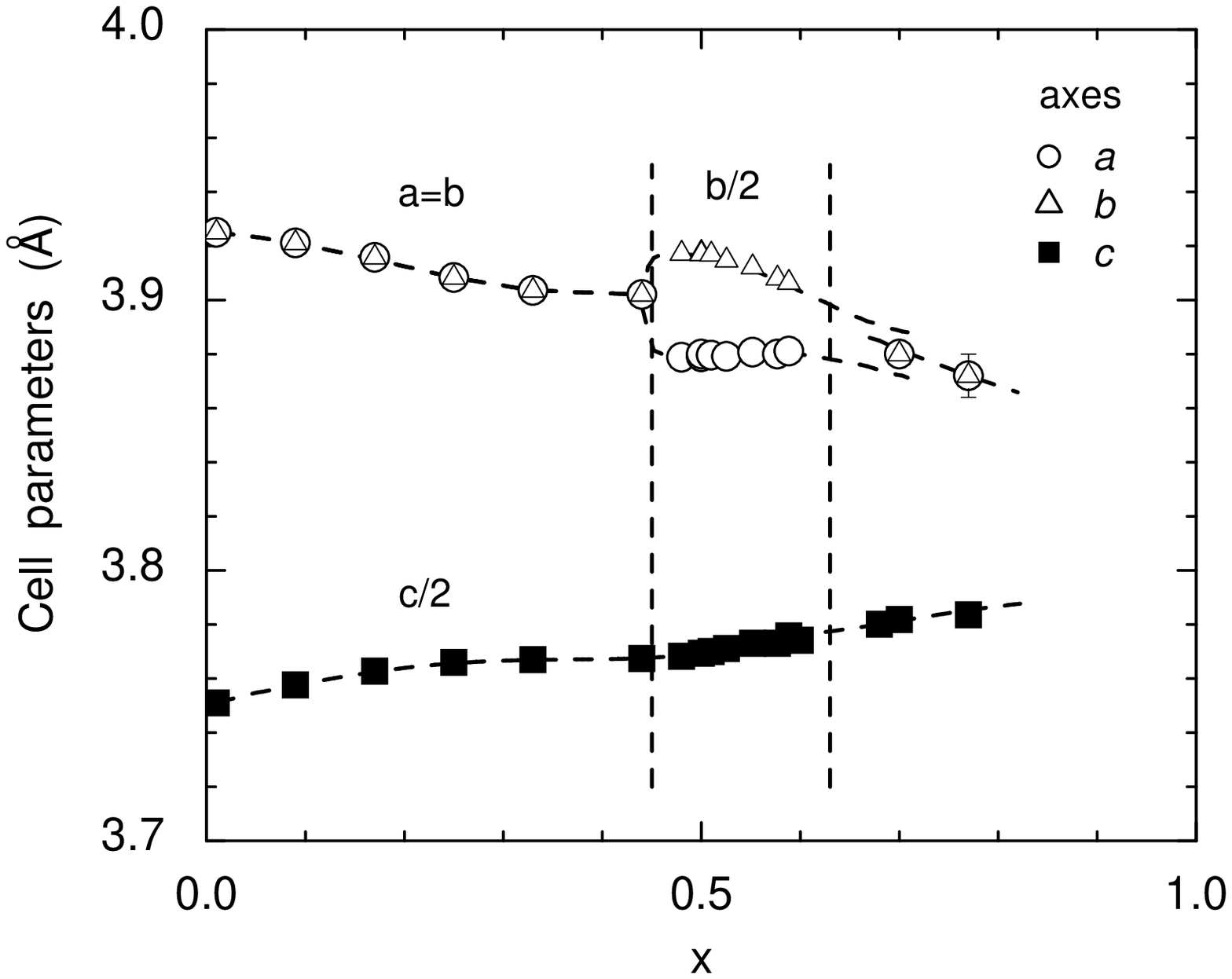}
\caption{Room-temperature parameters of the GdBaCo$_{2}$O$_{5+x}$ unit
cell as a function of the oxygen content.}
\end{figure}

Figure 5 shows the evolution of the GdBaCo$_{2}$O$_{5+x}$ crystal
structure as a function of the oxygen concentration $x$. We can clearly
distinguish three different composition regions:
\begin{itemize}
\item $0\leq x < 0.45$, the system keeps a macroscopically tetragonal
structure, where the unit cell smoothly expands in the $c$ direction and
shrinks in the in-plane directions with increasing $x$;

\item $0.45 < x < 0.60$, the oxygen ions order into alternating filled
and empty chains running along the $a$ axis, which results in the
orthorhombic structure and in the doubling of the unit cell along the $b$
axis (see Fig. 2);

\item $x > 0.60$, the system evolves towards a macroscopically
tetragonal symmetry, though in quite a complicated way. For crystals
located at the lower edge of this composition range, we observed at the
same time the x-ray diffraction peaks corresponding to the orthorhombic
structure with the unit cell doubling, and those related to the
tetragonal structure. This two-phase state is clearly intrinsic and
unrelated to a macroscopic oxygen-concentration gradient that would
emerge if crystals were improperly annealed. Indeed, the signs of the
two-phase state were reproducibly observed only in a quite narrow range
of $x$ and disappeared as the oxygen content was further increased. At
the highest achieved oxygen concentration, we could distinguish only the
tetragonal, albeit somewhat broadened, diffraction peaks.
\end{itemize}
We should keep in mind, however, that this dividing is based only on the
macroscopic symmetry, and thus it should not be taken too literally; the
behavior of the local structure may in fact be much more tricky. For
example, it is quite likely that oxygen ordering or mesoscopic phase
separation takes place at small $x$ values as well, but the ordered
domains are too small to be discernible by conventional x-ray
diffraction. The same is also true for high oxygen contents where one
may expect fairly ordered phases at $x\approx 2/3; 3/4$, as well as
various mesoscopic phase mixtures.

\subsection{Resistivity}
\label{sec:Res}

In perovskite compounds, the electron-band filling can be modified by a
partial substitution of cations with elements having a different
valence, as in the case of the La$_{1-x}$Sr$_{x}$$M$O$_3$ compounds ($M$
is a transition metal), and by introducing cation or anion vacancies and
interstitials as in V$_{2-y}$O$_3$, LaTiO$_{3+\delta}$, and
YBa$_{2}$Cu$_{3}$O$_{6+x}$. \cite{Imada} At integer filling (integer
number of electrons per unit cell) these compounds are usually band or
Mott insulators, yet a metallic state often emerges upon changing the
filling level, that is, when electron or holes are doped into the parent
insulator.

In GdBaCo$_{2}$O$_{5+x}$, it is obviously the $x=0.5$ composition that
is the parent compound, where all the Co ions are nominally in the 3+
state. The limiting phases $x=0$ and $x=1$ should correspond to the 1:1
mixtures Co$^{2+}$/Co$^{3+}$ and Co$^{3+}$/Co$^{4+}$, respectively, or
in other words to the doping levels of 0.5 electrons and 0.5 holes per
Co ion. It would not be surprising, therefore, if the $x=0.5$
composition were insulating, and a metallic behavior were emerging when
the oxygen content deviates from $x=0.5$ towards lower or higher values.
The actual behavior of GdBaCo$_{2}$O$_{5+x}$, however, turns out to be
more complicated: In contrast to the naive expectations, it never
becomes a true metal (Fig. 6), even though we change the doping level in
a very broad range from 0.5 electrons per Co ($x=0$) up to 0.27 holes
per Co ($x=0.77$).\cite{oxygen} At low temperatures, the in-plane
resistivity $\rho_{ab}$ exhibits an insulating behavior throughout the
whole accessible range of $x$. Moreover, by looking at the
low-temperature region ($T<150$ K) in Fig. 6, one may easily find out
that the resistivity even {\it increases} as the oxygen content is
reduced below 0.5 (``electron doping''); clearly, the observed evolution
differs remarkably from what one usually expects to see upon doping an
insulator.

\begin{figure}[!t]
\includegraphics*[width=8.6cm]{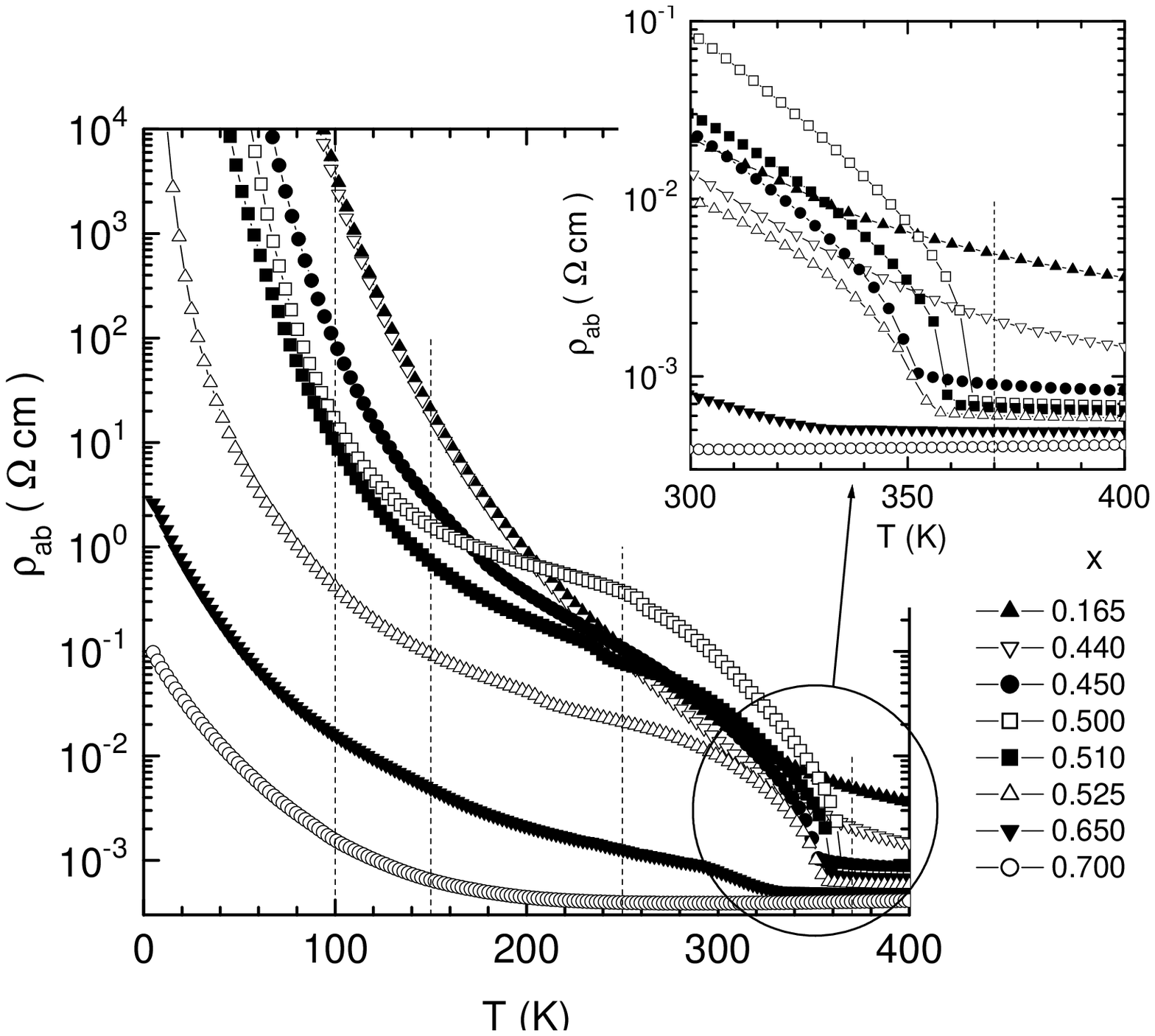}
\caption{In-plane resistivity $\rho_{ab}(T)$ of GdBaCo$_{2}$O$_{5+x}$
crystals with different oxygen concentrations. Inset: expanded view of
$\rho_{ab}(T)$ in the vicinity of the ``metal-insulator'' transition.}
\end{figure}

\begin{figure}[!t]
\includegraphics*[width=8.6cm]{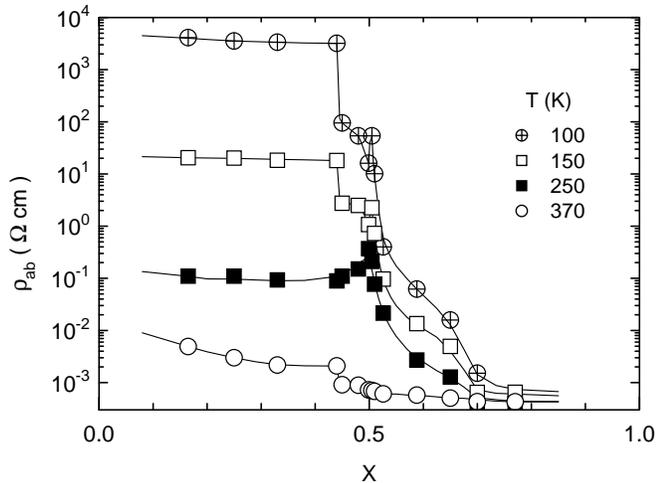}
\caption{Doping dependences of $\rho_{ab}$ in GdBaCo$_{2}$O$_{5+x}$
at several temperatures.}
\end{figure}

Although GdBaCo$_{2}$O$_{5+x}$ never behaves as a normal metal, the
temperature dependence of its resistivity does not follow that of a
simple insulator or semiconductor either. As the temperature increases
above $200-250$ K, the in-plane resistivity $\rho_{ab}(T)$ shows a
gradual crossover (for $x\leq 0.44$) or a sharp transition at $T\approx
360$ K (for $x$ close to 0.5) into a metal-like state, see inset of Fig.
6. This sharp ``metal-insulator'' transition (MIT) at $x\approx 0.5$ is
similar to what has been reported for air-sintered RBaCo$_{2}$O$_{5+x}$
ceramics. \cite{Martin, Maignan, Troy1, Troy2} Although the resistivity
$\rho_{ab}\sim 400-600$ $\mu\Omega\,$cm on the metal side of the MIT is
still relatively large, in transition-metal oxides such resistivity
values are actually more consistent with a metallic, rather than a
hopping transport.\cite{Imada} As can be seen in the inset of Fig. 6,
the metal-insulator transition is the sharpest and takes place at the
highest temperature $T_{\text{MIT}}\approx 364$ K for $x=0.50$, while
any deviation from this stoichiometry blurs the MIT and shifts it to
lower temperatures. This is particularly clear for high oxygen contents:
$T_{\text{MIT}}$ is reduced to $\sim 325$ K for $x=0.65$, and no
distinct transition can be found for $x \geq 0.70$. In addition to the
MIT, a clear, albeit small, kink is seen on the $\rho(T)$ curves in the
temperature range 100-260 K, which is related to a magnetic transition
as will be discussed below.

\begin{figure*}[!t]
\includegraphics*[width=0.95\linewidth]{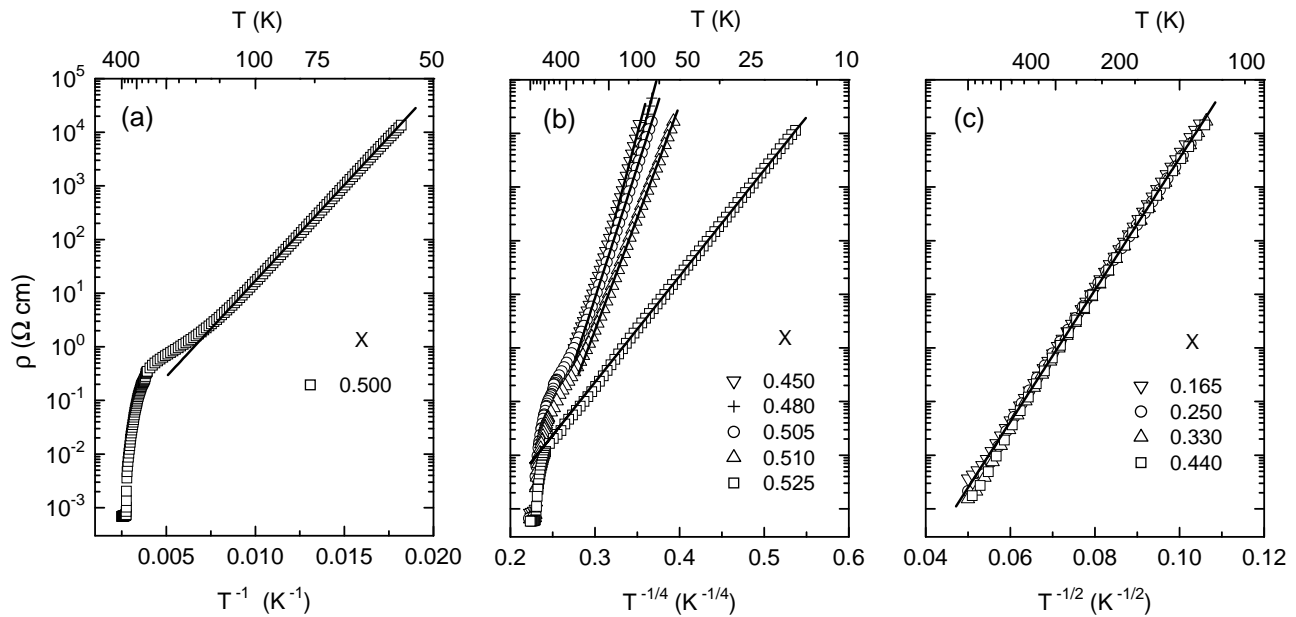}
\caption{Low-temperature resistivity in GdBaCo$_{2}$O$_{5+x}$:
(a) activation behavior for $x = 0.50$; (b) 3D variable-range hopping
for $x = 0.45 - 0.525$; and (c) Efros-Shklovskii behavior for $x = 0.165
- 0.440$.}
\end{figure*}

To make it more transparent how the resistivity in GdBaCo$_{2}$O$_{5+x}$
evolves with oxygen content, in Fig. 7 we plot $\rho_{ab}$ at several
temperatures as a function of $x$. Regardless of the temperature, the
$x$-dependence of resistivity splits into roughly the same three
composition regions as were observed in the behavior of the crystal
structure (Fig. 5). For both low ($0\leq x < 0.45$) and high ($x > 0.7$)
oxygen concentrations, the resistivity appears to be almost
$x$-independent, while in the region $0.45 < x \leq 0.7$ it changes
rather steeply, by orders of magnitude. This step-like resistivity drop,
taking place upon going from ``electron-doped'' to ``hole-doped''
compositions, tends to hide the singularity expected for the parent
compound. Only very detailed data collected with a step of $\delta x
\leq 0.01$ make this singularity discernible on the low-temperature
$\rho_{ab}-x$ curves, where it is manifested as a narrow peak in the
vicinity of $x=0.50$.

In fact, the most unusual and intriguing feature in Fig. 7 is the
asymmetry with respect to the oxygen concentration $x=0.50$. The
resistivity evolution at $x \geq 0.50$ looks rather conventional:
Indeed, doping of $\sim 2.5$\% of holes ($x=0.525$), which turns Co-ions
into a mixed 3+/4+ valence state, dramatically improves the
conductivity. As can be seen in Fig. 7, the in-plane resistivity at 100
K drops by more than two orders of magnitude as $x$ increases from
$\approx 0.50$ to 0.525, and by three orders of magnitude more upon
further increasing $x$. In contrast, upon electron doping (reducing $x$
below 0.50), an initial decrease of the resistivity almost immediately
turns into a resistivity growth. Moreover, in a wide region $x \leq
0.44$ both the absolute value of the resistivity and its temperature
dependence are virtually independent of the oxygen content (Figs. 6, 7).
Such insensitivity of the conductivity to doping is very unusual, and
may only be possible if the electrons released upon removing the oxygen
do not participate in the charge transport. This clearly speaks against
the ``rigid-band'' picture, where electrons are expected to fill the
lowest-energy unoccupied states. Apparently, the electron doping is
accompanied by developing of microscopic insulating states or mesoscopic
regions, where the electrons are immediately trapped. At this point it
is useful to remind that the insulating nature of the limiting $x=0$
composition has been already understood as coming from the charge
ordering of electrons in CoO$_2$ planes into a unidirectional
charge-density wave. \cite{Vogt, Suard} It is therefore reasonable to
suggest that the robust insulating behavior observed in the wide range
$0\leq x \leq 0.44$ is also associated with some kind of charge ordering
among the doped electrons.

It is interesting to analyze the low-temperature behavior of the
resistivity, which can give information on the transport mechanisms
operating in the system. Data fitting has shown that a crystal with the
precisely tuned $x=0.50$ composition exhibits a simple activation
behavior $\rho_{ab} \propto \exp(\Delta/T)$ with $\Delta \approx 70$ meV
[Fig. 8(a)], which is well consistent with the understanding of this
composition as a parent insulator (narrow-gap semiconductor). However,
when the oxygen content deviates somewhat from $x=0.50$, the resistivity
behavior almost immediately switches into the 3D variable-range hopping
(VRH) mode,\cite{Mott} $\rho_{ab} \propto \exp [(T_0/T)^{1/4}]$, as
shown in Fig. 8(b). This type of conduction is typical for disordered
systems where the charge carriers move by hopping between localized
electronic states. The formation of such localized states may be rather
simply conceived of in the following way: Initially, at the oxygen
composition of $x=0.50$, GdBaCo$_{2}$O$_{5+x}$ possesses a well-ordered
crystal structure, where the oxygen ions form perfect filled and empty
chains alternating along the $b$ axis (Fig. 1). When the oxygen
concentration deviates from $x=0.50$, this results either in vacancies
emerging in the filled chains, or in oxygen ions that go into the empty
chains. While these oxygen defects inevitably generate electrons or
holes in the CoO$_2$ planes, they also produce a poorly screened Coulomb
potential that may well localize the generated carriers, so that some of
the adjacent Co ions acquire the Co$^{2+}$ or Co$^{4+}$ state. Then the
conductivity occurs through hopping (tunneling) motion of such localized
Co$^{2+}$ or Co$^{4+}$ states. As can be seen in Fig. 8(b), the slope of
the resistivity curves monotonically decreases with increasing $x$,
implying that the localization length of holes (Co$^{4+}$ states) is
noticeably larger than that of electrons.

\begin{figure*}[!]
\leftskip80pt
\includegraphics*[width=0.66\linewidth]{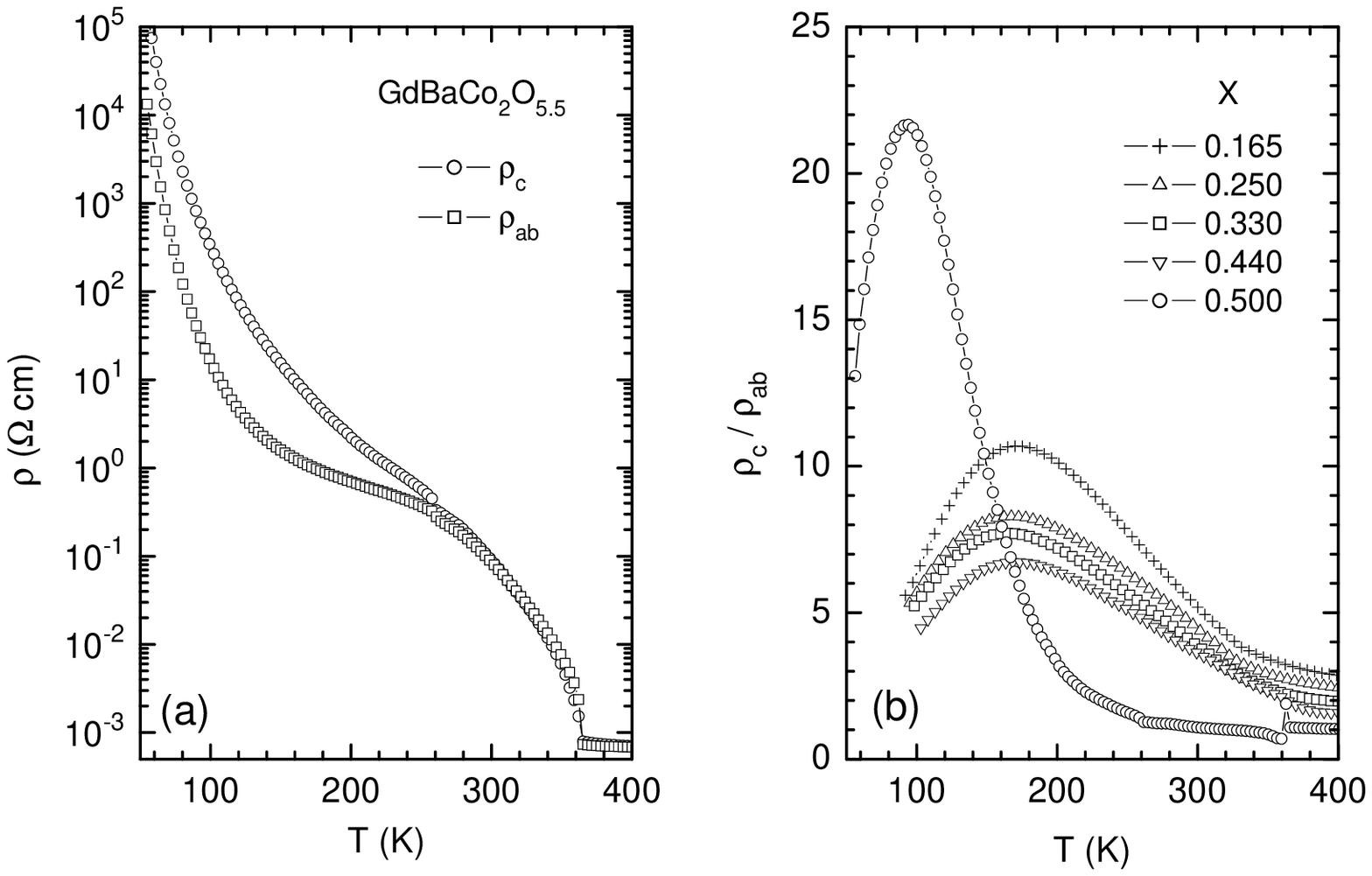}
\caption{The resistivity anisotropy in GdBaCo$_{2}$O$_{5+x}$.
(a) Resistivity of GdBaCo$_{2}$O$_{5.50}$ measured along the $c$ axis
and along the $ab$ plane. (b) The resistivity anisotropy ratio
$\rho_{c}$/$\rho_{ab}$ for $x = 0.165 - 0.440$ and 0.500. }
\end{figure*}

Upon further decreasing the oxygen content below $x\approx 0.45$, the
temperature dependence of the resistivity becomes steeper than expected
for the Mott's VRH regime, $\rho_{ab} \propto \exp [(T_0/T)^{1/4}]$, and
we find that it is better fitted by the Efros-Shklovskii expression for
the hopping conductivity,\cite{Efros} $\rho_{ab}
\propto \exp[(T^\prime_0/T)^{1/2}]$, see Fig. 8(c). The latter behavior
is usually observed when the Coulomb interaction starts to play a key
role in carriers hopping, bringing about a strong depletion in the
density of states (Coulomb gap) near the Fermi energy. The data in Fig.
8 imply that the Coulomb-repulsion effects in GdBaCo$_{2}$O$_{5+x}$ gain
strength as the oxygen content is reduced below $\approx 0.45$.

The unusual resistivity behavior at high oxygen concentrations (Fig. 6)
is also worth mentioning. For both the $x=0.65$ and $x=0.70$ crystals,
$\rho_{ab}$ smoothly increases by $2-3$ orders of magnitude upon
decreasing temperature, and reaches apparently non-metallic values of
the order of 1 $\Omega \,$cm. Nevertheless, it becomes also clear from
Fig. 6 that, at least in the temperature range studied, the resistivity
tends to saturate at a finite value, instead of diverging at $T=0$; such
behavior would indicate a metallic ground state, if the resistivity
values were not that large. A possible solution of this puzzle is an
intrinsic mesoscopic phase separation, that makes the carriers to move
along filamentary conducting paths. \cite{Mesosc} In this case, the
zero-temperature conductivity, being determined by the topology of the
metallic phase, may be arbitrary small and can easily violate the Mott
limit\cite{Mott} that sets the minimum metallic conductivity for
homogeneous systems. This picture is quite plausible, since, as we have
mentioned above, the crystal structure at these high oxygen
concentrations actually bears signs of phase separation.

Thus far we discussed the resistivity behavior along the CoO$_2$ planes,
which is however not the full story, since one might anticipate a strong
electron-transport anisotropy to be brought about by the layered crystal
structure of GdBaCo$_{2}$O$_{5+x}$. Indeed, many layered
transition-metal oxides, such as high-$T_c$ cuprates,\cite{cupr_Rc}
manganites,\cite{mang_Rc} or even cobaltites built from
triangular-lattice CoO$_2$ planes,\cite{NaCoO, misfitRc} exhibit huge
anisotropy values and contrasting temperature dependences of the
in-plane and out-of-plane resistivity. In the case of
GdBaCo$_{2}$O$_{5+x}$, where the GdO$_x$ layers with variable oxygen
content are located in between the CoO$_2$ planes (Fig. 1), it would be
natural to expect a kind of 3D-to-2D transition to occur upon reducing
$x$ from 1 to 0, that is, as the oxygen ions binding the CoO$_2$ planes
are removed. However, a comparison of the in-plane and out-of-plane
resistivity in Fig. 9 shows that this is not really the case. The
anisotropy $\rho_{c}/\rho_{ab}$ indeed increases slightly as the oxygen
concentration is reduced from $x \approx 0.44$ towards zero [Fig. 9(b)],
but remains rather moderate. In fact, a considerable anisotropy is
observed only at intermediate temperatures, while at both high and low
temperatures GdBaCo$_{2}$O$_{5+x}$ tends to become virtually isotropic.
It turns out therefore that the oxygen-depleted GdO$_x$ layers do not
constitute a serious obstacle for the $c$-axis electron motion.

Somewhat different behavior of the parent GdBaCo$_{2}$O$_{5.50}$
compound provides a clue to understand the mechanism responsible for
resistivity anisotropy. As can be seen in Figs. 9(a), 9(b), the in-plane
and out-of-plane resistivities stay virtually indistinguishable from
each other down to $T\approx 260 K$, where the FM-AF transition takes
place.\cite{GBC_PRL} Upon further decreasing the temperature, the curves
sharply diverge and $\rho_{c}/\rho_{ab}$ grows from $\approx 1$ at $T
\geq 260 K$ up to $\approx 22$ at 100 K. This indicates that the charge
motion in GdBaCo$_{2}$O$_{5+x}$ is very sensitive to the arrangement of
Co spins, and it is the peculiar spin ordering in this system that is
responsible for the appearance of conductivity anisotropy.

\subsection{Thermoelectric power}
\label{sec:Term}

Among the features that currently attract a lot of attention to
transition-metal oxides are their unusual and potentially useful
thermoelectric properties.\cite{NaCoO,S_book,LaSrTiO,Proceed} The
peculiar thermoelectric behavior in TM oxides is often attributed to
strong electron correlations,\cite{Koshibae, Ando_specific_heat} though
the picture still remains far from being clear. In this respect, the
RBaCo$_{2}$O$_{5+x}$ compounds, being capable of smoothly changing the
doping level in a very broad range, appear to provide a good testing
ground for studying the problem.

\begin{figure}[!tb]
\includegraphics*[width=8.6cm]{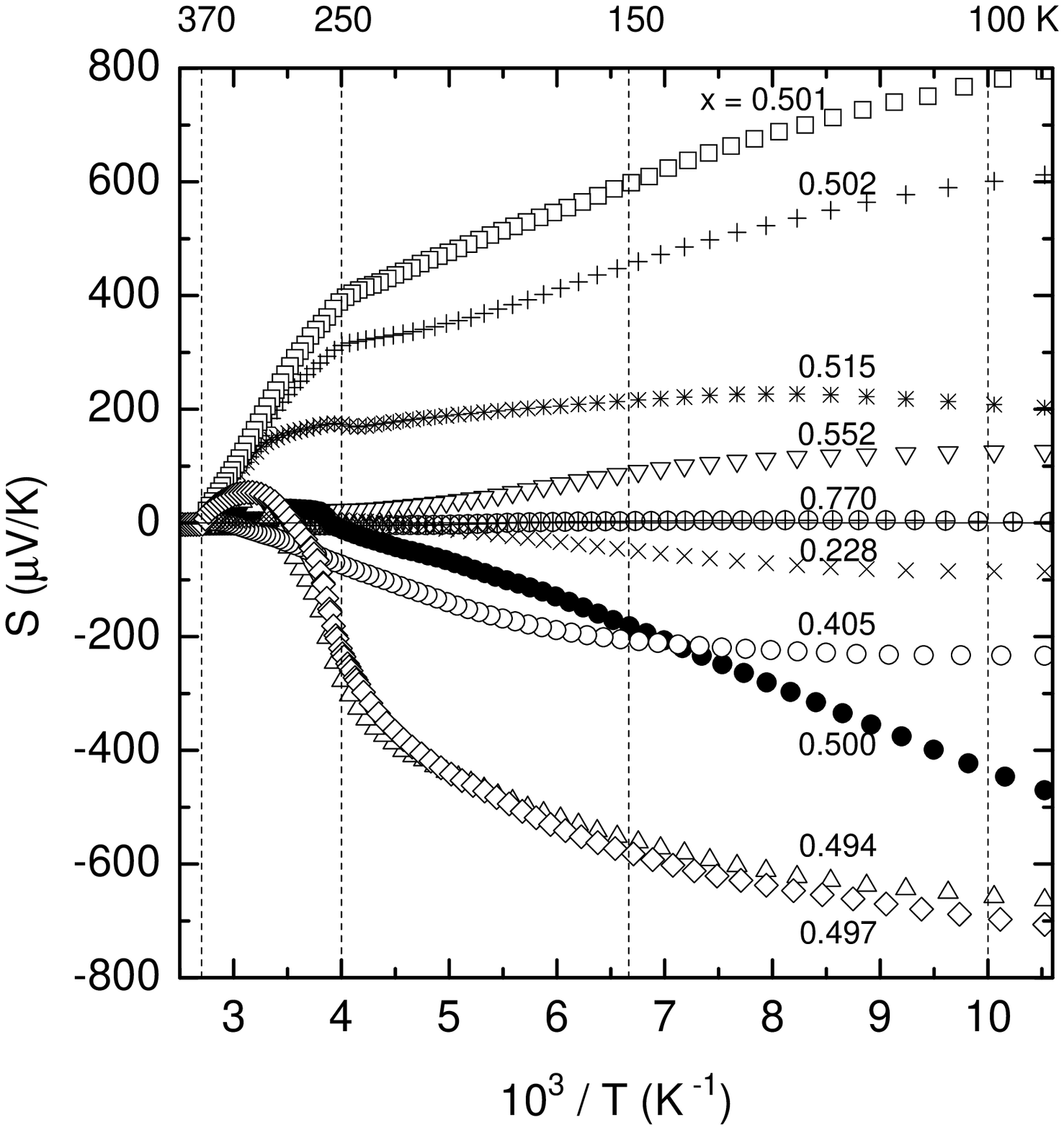}
\caption{Thermoelectric power of GdBaCo$_{2}$O$_{5+x}$ for different
oxygen concentrations plotted as a function of inverse temperature.}
\end{figure}

The thermoelectric power $S(T)$ measured on both single-crystal and
ceramic GdBaCo$_{2}$O$_{5+x}$ samples\cite{S_cer} throughout the
available oxygen-concentration range is presented in Fig. 10. Following
the usual approach in analyzing the behavior of semiconductors, we have
plotted the data in the inverse-temperature scale, since in
semiconductors the thermoelectric power is expected to be linear in
$1/T$ with a slope reflecting the activation energy for charge carriers
$\Delta_s$, $S(T) \approx A \pm
(k_{\text{B}}/e)(\Delta_s/k_{\text{B}}T)$. It becomes immediately clear
from Fig. 10 that the only composition that exhibits a conventional
semiconducting behavior is $x=0.500$, with the activation energy
$\Delta_s \approx 70$ meV being in good agreement with that deduced from
the resistivity data in Sec.\ \ref{sec:Res}. Just a subtle deviation of
the oxygen concentration from $x=0.500$ (by merely 0.001) qualitatively
changes the thermoelectric behavior: $S$ becomes almost temperature
independent at $\approx 100$ K, indicating that the electron transport
is no longer governed by the band gap. It is worth noting that the
resistivity of GdBaCo$_{2}$O$_{5+x}$ also switches its behavior from a
simple activation one into the variable-range hopping mode upon
deviation from the parent composition (Fig. 8).

The salient feature in Fig. 10 is a very abrupt change in sign of the
low-temperature thermoelectric power upon crossing $x=0.500$; $S$(100 K)
jumps from $\approx - 700$ $\mu$V/K at $x=0.497$ up to $\approx + 800$
$\mu$V/K at $x=0.501$. In fact, this sign change unambiguously indicates
that the type of change carriers sharply switches from electrons ($x <
0.5$) to holes ($x > 0.5$) without any messy intermediate state. It
turns out to be possible, therefore, to drive the doping level in
GdBaCo$_{2}$O$_{5+x}$ continuously across the parent insulating state
(so that the Fermi level jumps across the gap), which is really unusual
among transition-metal oxides. The feature that makes possible such
continuous and smooth doping of GdBaCo$_{2}$O$_{5+x}$ is the
metal-insulator transition at $T_{\text{MIT}}$: At temperatures where
the oxygen concentration is modified, the gap in the electronic band
structure is closed and, therefore, the chemical potential changes
gradually with the changing of the carriers concentration.

\begin{figure}[!t]
\includegraphics*[width=7.3cm]{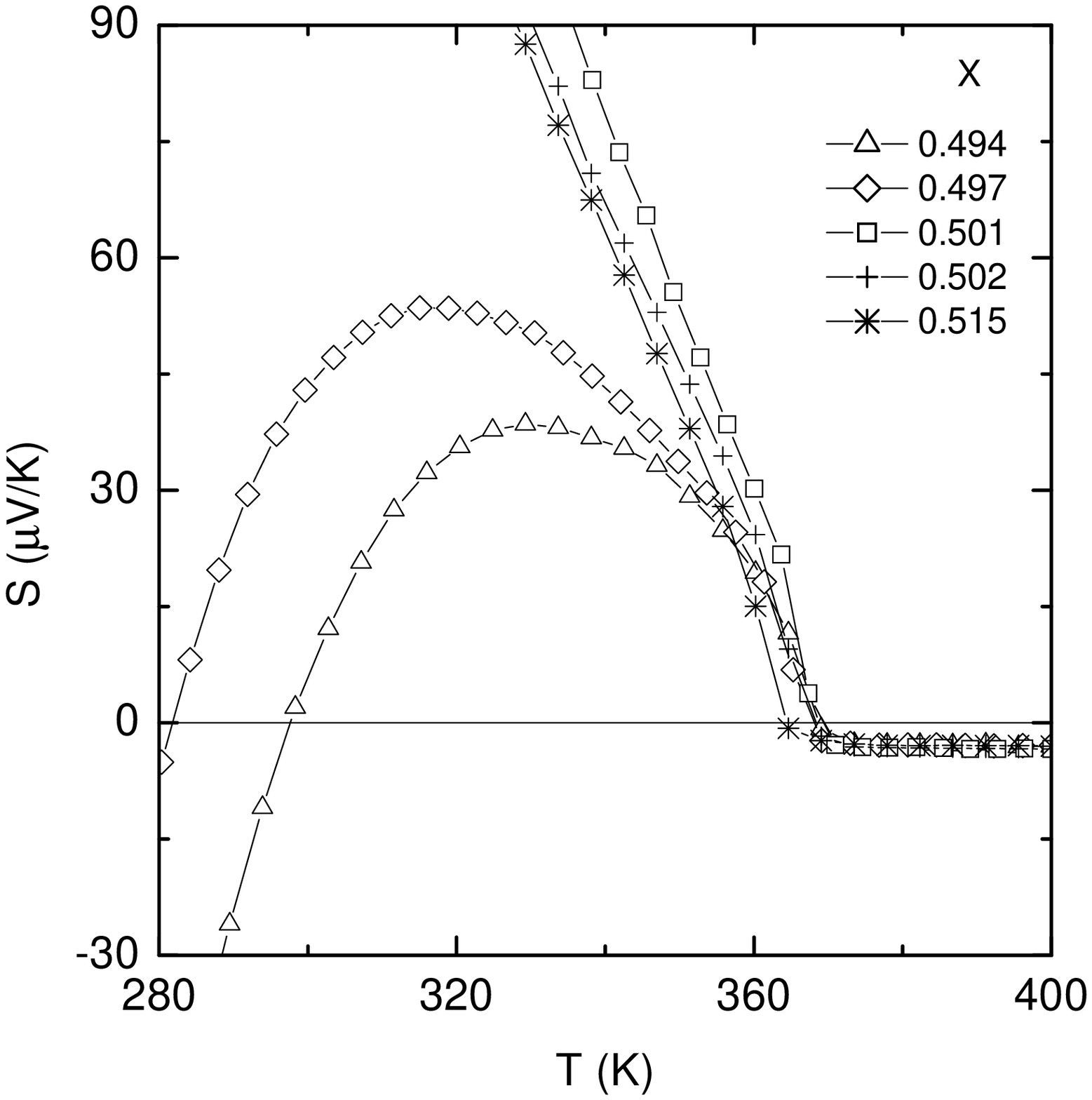}
\caption{Temperature dependences of the thermoelectric power in
GdBaCo$_{2}$O$_{5+x}$ near the metal-insulator transition.}
\end{figure}

The metal-insulator transition at $T_{\text{MIT}}\approx 360$ K has a
spectacular manifestation in the thermoelectric properties (Figs. 10 and
11), owing to the high sensitivity of $S(T)$ to details of the
electronic structure. As can be seen in Fig. 11, at high temperatures
$T\geq 370$ K, $S$ is very small ($\approx -4$ $\mu$V/K), as is usually
the case with ordinary metals, and virtually independent of temperature.
Moreover, it does not change with doping either, implying that the Fermi
level is positioned in the middle of a wide conduction band. Upon
cooling the sample below 370 K, however, the thermoelectric power
abruptly increases, indicating that the conduction band splits and a
band gap opens near the Fermi level. Depending on the doping, the Fermi
level gets trapped inside the newly created valence or conduction band,
determining which type of carriers -- electrons or holes
-- will prevail. Immediately after the transition, $S$ appears to be positive
even in slightly electron-doped samples ($x<0.5$), indicating that while
both electrons and holes are contributing to charge transport at these
temperatures, the holes have considerably higher mobility. At lower
temperatures, carrier excitations over the gap become suppressed, letting
only one type of carriers survive; consequently, at $T < 250-280$ K the
sign of $S$ is uniquely determined by the doping level ($x<0.5$ or
$x>0.5$).

\begin{figure}[!t]
\includegraphics*[width=7.9cm]{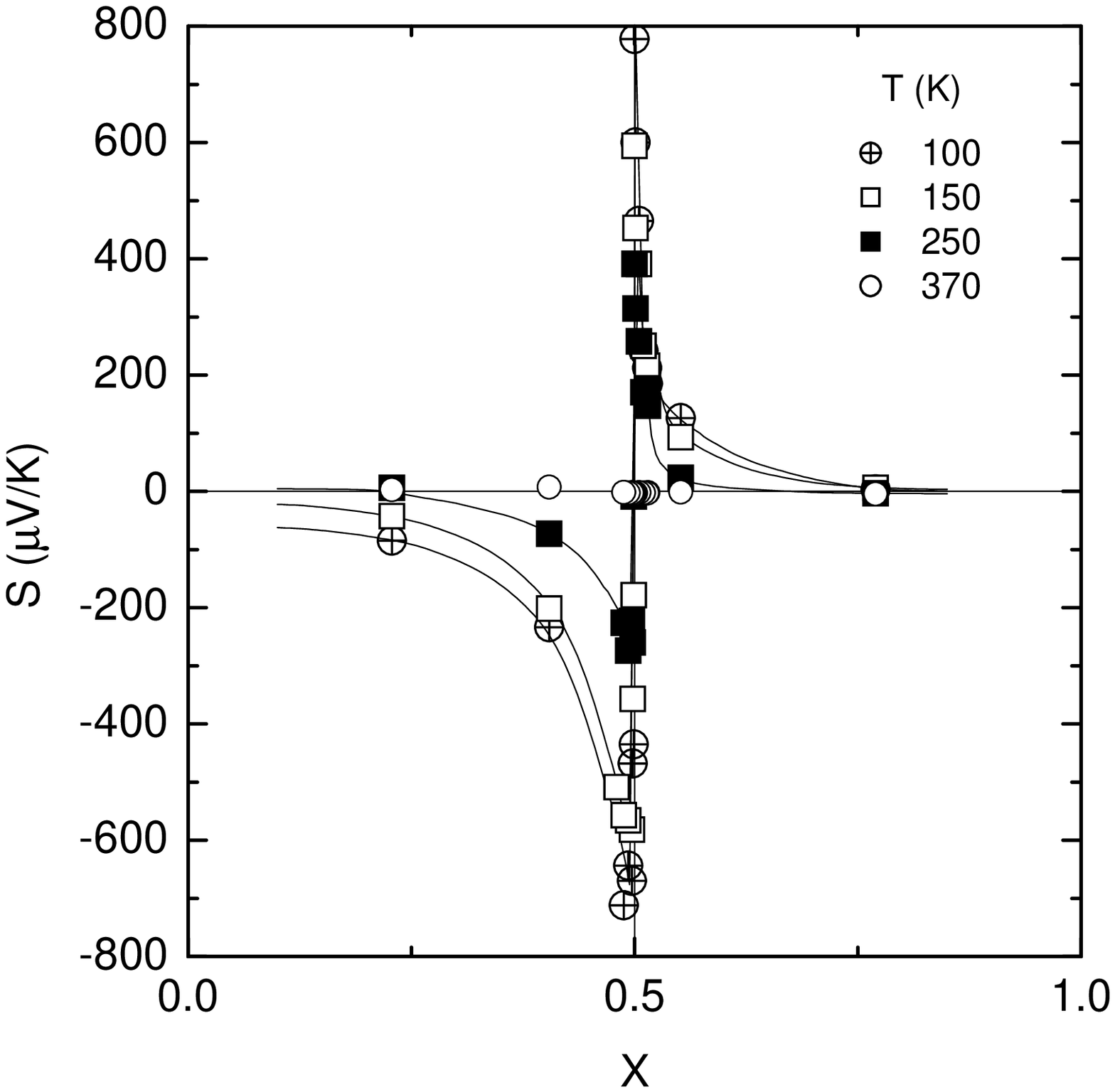}
\caption{Doping dependences of the thermoelectric power in
GdBaCo$_{2}$O$_{5+x}$ at several temperatures.}
\end{figure}

One may wonder whether and how these two transport properties --
resistivity and thermoelectric power -- correlate in
GdBaCo$_{2}$O$_{5+x}$. In conventional doped semiconductors, for
instance, such correlation is rather straightforward if one type of
carriers dominates. Namely, the resistivity and thermoelectric power are
both governed by the activation energy of carriers, $\rho \propto
\exp(\Delta/k_{B}T)$ and $S \approx A \pm (k_{\text{B}}/e)
(\Delta/k_{\text{B}}T)$, and thus are related roughly as $S \approx C
\pm (k_{\text{B}}/e)\ln\rho$. This qualitative relation implies that
upon doping a semiconductor, its thermoelectric power should decrease
(in absolute value) by roughly 200 $\mu$V/K per order-of-magnitude
reduction in resistivity. In fact, this trend is not restricted to
conventional semiconductors, but holds also in many other non-metallic
systems, including doped Mott insulators, \cite{Jonker_P, Jonk_cup,
FeCrSe} though the slope $\partial S/\partial (\ln\rho)$ may differ from
$k_{\text{B}}/e\approx 86.2$ $\mu$V/K.

\begin{figure}[!b]
\includegraphics*[width=6.5cm]{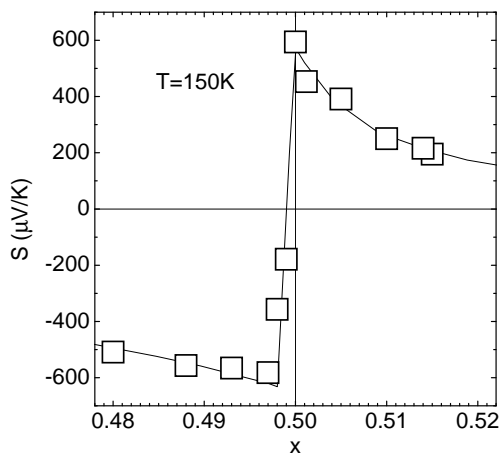}
\caption{Thermoelectric power at $T=150$ K as a function of doping in
the vicinity of $x=0.5$.}
\end{figure}

If this $S$-vs.-$\rho$ correlation were also working in the case of
GdBaCo$_{2}$O$_{5+x}$, its thermoelectric power would show a step-like
decrease in magnitude upon increasing $x$, following the asymmetric
resistivity curves ($\ln\rho$ vs. $x$) shown in Fig. 7. However, Fig.
12, which presents $S$ at several temperatures as a function of the
oxygen concentration, makes it immediately clear that these expectations
fail. Instead, the doping dependence of thermoelectric power turns out
to be fairly symmetric with respect to the parent, $x=0.5$, composition:
$S(x)$ exhibits a spectacular divergence with approaching $x=0.5$, where
it reaches large absolute values and changes its sign upon crossing this
peculiar doping level. It is worth noting that $S(x)$ switches from
being negative (electron-like) to positive (hole-like) within a very
narrow but discernible region around $x=0.5$ (its expanded view is given
in Fig. 13); this slight thermally-induced smearing gives evidence for a
continuous evolution in the CoO$_2$-plane doping. When the oxygen
concentration deviates from $x=0.5$, $S(x)$ smoothly reduces its
magnitude, eventually approaching a small constant value (Fig. 12). The
$S(x)$ curves keep this singular and fairly symmetric behavior
regardless of temperature, except for the high-temperature ``metallic''
region ($T\geq370$ K), where the thermoelectric power is small and
virtually doping independent (Fig. 12). In fact, the striking contrast
in the doping dependences of two transport properties -- symmetric
$S(x)$ and asymmetric $\rho(x)$ -- is one of the most intriguing feature
in the thermoelectric behavior of GdBaCo$_{2}$O$_{5+x}$.

\begin{figure}[!t]
\includegraphics*[width=6.5cm]{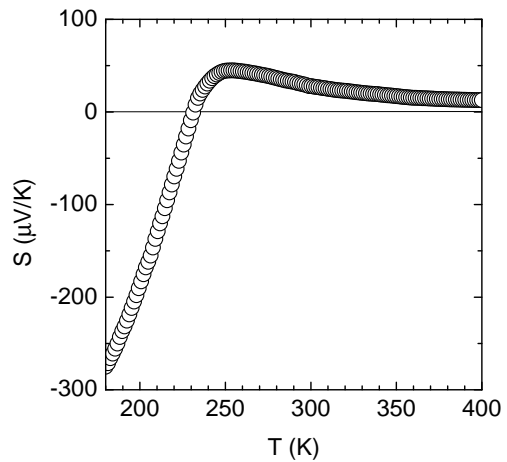}
\caption{Temperature dependences of the thermoelectric power in
GdBaCo$_{2}$O$_{5+x}$ for $x=0$.}
\end{figure}

The smooth evolution of $S$ with decreasing oxygen content below $x=0.5$
shown in Fig. 12 breaks down, however, upon approaching the limit of
$x=0$ -- another peculiar composition corresponding to the 50\% electron
doping. This composition is known for the long-range charge ordering
that sets in below $\approx 210-220$ K, as was observed in
YBaCo$_{2}$O$_{5.0}$ and HoBaCo$_{2}$O$_{5.0}$.\cite{Vogt,Suard} A
similar charge ordering presumably takes place in GdBaCo$_{2}$O$_{5.0}$
as well, and it has a clear manifestation in the behavior of
thermoelectric power, whose magnitude quickly increases below 230-240 K
(Fig. 14). The long-range charge ordering at 50\%-electron doping and
its impact on the charge transport, being interesting in their own
right, need a detailed study, which however goes beyond the scope of the
present paper.

\subsection{Magnetization}

\subsubsection{General features}
\label{sec:magG}

Broadly speaking, it would be quite natural for cobalt oxides to possess
magnetic properties more complex than those of isostructural compounds
based on most other transition metals. The reason for this additional
complexity is that cobalt ions in a given valence state can have more
than one allowed {\it spin} state; for example, Co$^{3+}$ ions can
acquire a low-spin (LS:~~$t^6_{2g}$, $e^0_g$; $S=0$), intermediate-spin
(IS:~~$t^5_{2g}$, $e^1_g$; $S=1$), or high-spin (HS:~~$t^4_{2g}$,
$e^2_g$; $S=2$) state.\cite{Martin, Maignan, cubic_Co, BiCo2201, Troy1,
Troy2, OpticalSm, Suard, Vogt, Moritomo, Akahoshi, Kusuya, Respaud,
GBC_PRL, Moritomo2, Frontera} These states appear to be located quite
close to each other in energy, bringing about a possibility of the
spin-state transitions/crossovers upon changing temperature or lattice
deformation.

Keeping in mind a possibility of wide-range charge doping in
GdBaCo$_{2}$O$_{5+x}$, which affects both the valence state of Co ions
and the magnetic interactions, structural ordering phenomena, and a
possible nanoscopic phase separation, in addition to the spin-state
degree of freedom, one might expect the magnetic phase diagram in this
system to be extremely complicated or even messy. In reality, the
magnetic behavior of GdBaCo$_{2}$O$_{5+x}$ turns out to be indeed rich,
but still following fairly simple empirical rules. As will be discussed
below, the possible spin arrangements in GdBaCo$_{2}$O$_{5+x}$ appear to
be dramatically simplified by an exceptionally strong spin anisotropy,
which essentially pins the spin orientation.\cite{GBC_PRL} This
fortunate feature turns GBCO into a good model system, yet it also
emphasizes the necessity to perform all the magnetic measurements on
single crystals only.

\begin{figure}
\includegraphics*[width=8.6cm]{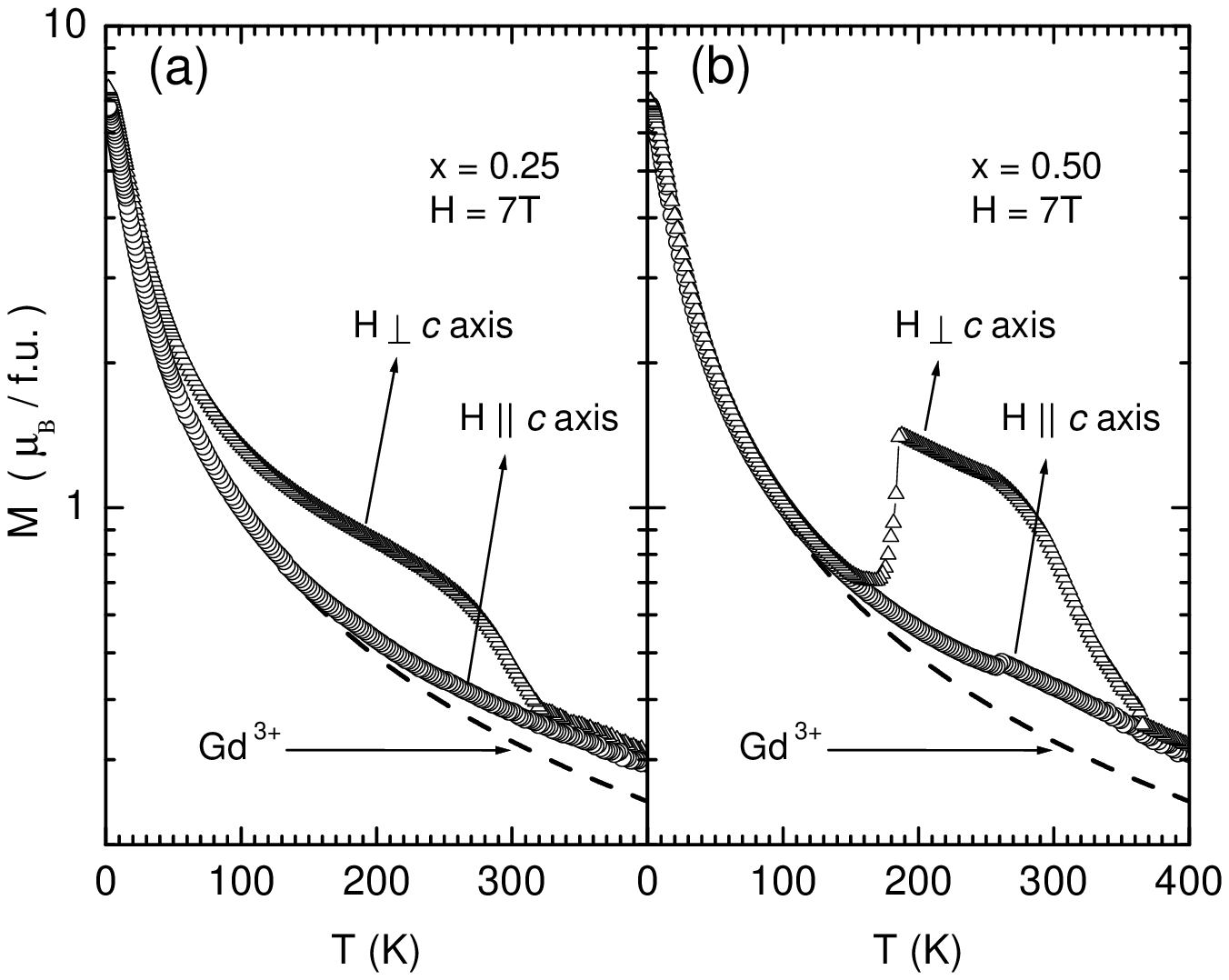}
\caption{Raw magnetization $M(T)$ data obtained for GdBaCo$_{2}$O$_{5+x}$
single crystals with $x=0.25$ (a) and $x=0.50$ (b) at the magnetic field
$H=7$ T applied parallel or perpendicular to the $c$ axis. Dashed lines
show the Curie-Weiss contribution of Gd$^{3+}$ ions
($\mu_{\text{eff}}=7.94\,\mu_{\text{B}}$; $\theta = 0$ K).}
\end{figure}

Figure 15 illustrates the method used to determine the anisotropic
magnetization contribution coming from the cobalt ions. For every oxygen
composition, we measured the magnetization $M(T)$ of
GdBaCo$_{2}$O$_{5+x}$ crystals with the magnetic field $H=0.01; 0.1; 1;
7$ T applied along or transverse to the CoO$_2$ planes. From the overall
magnetization we then subtracted an isotropic contribution of Gd$^{3+}$
ions (shown by dashed lines in Fig. 15), assuming their ideal
paramagnetic behavior with spin $S=7/2$; the latter appeared to be a
good approximation since we found no sign of Gd$^{3+}$ spin ordering
down to $T=1.7$ K in any of the crystals. Hereafter, all the
magnetization data will be presented after subtracting the contribution
of Gd$^{3+}$ ions.

When the magnetic field is applied along the CoO$_2$ planes (${\bf
H}$$\,\parallel\,$$ab$), their magnetization exhibits several
doping-dependent anomalies, which include a ferromagnetic-like behavior
at moderate temperatures (Fig. 15). In sharp contrast, the $c$-axis
magnetization (${\bf H}$$\,\parallel\,$$c$) appears to be always smaller
and virtually featureless throughout the whole doping range, except for
a weak step-like feature associated with the metal-insulator transition
at $T\approx 360$ K for $x\sim 0.5$, which is observed for any field
orientation. This persistent anisotropy clearly indicates that
regardless of the oxygen content, the cobalt spins in
GdBaCo$_{2}$O$_{5+x}$ are strongly confined to the CoO$_2$ planes. As
can be seen in Fig. 15, even when cobalt ions develop a fairly strong
ferromagnetic moment along the $ab$ plane ($\sim 1\,\mu_{\text{B}}$ per
formula unit at $T\approx 200-260$ K), the 7 T field appears to be far
too weak to overcome the spin anisotropy and to turn this ferromagnetic
moment towards the $c$-axis. In the following, we will focus on the
in-plane magnetic behavior, since the $c$-axis magnetization is governed
simply by a weak field-induced rotation of Co moments out of the $ab$
plane, that is, by a competition between the Zeeman energy and a
single-ion spin anisotropy.

Another general feature of the magnetization in GdBaCo$_{2}$O$_{5+x}$ is
a pronounced thermo-magnetic irreversibility. Regardless of doping, the
data taken upon cooling the sample in magnetic field (FC) differed from
those obtained on heating after the sample was cooled in zero field
(ZFC), as illustrated in Fig. 16. As the magnetic field increases, the
ZFC curve approaches the FC one, yet the difference remains well
discernible even at $H=7$ T.

\begin{figure}
\includegraphics*[width=8.6cm]{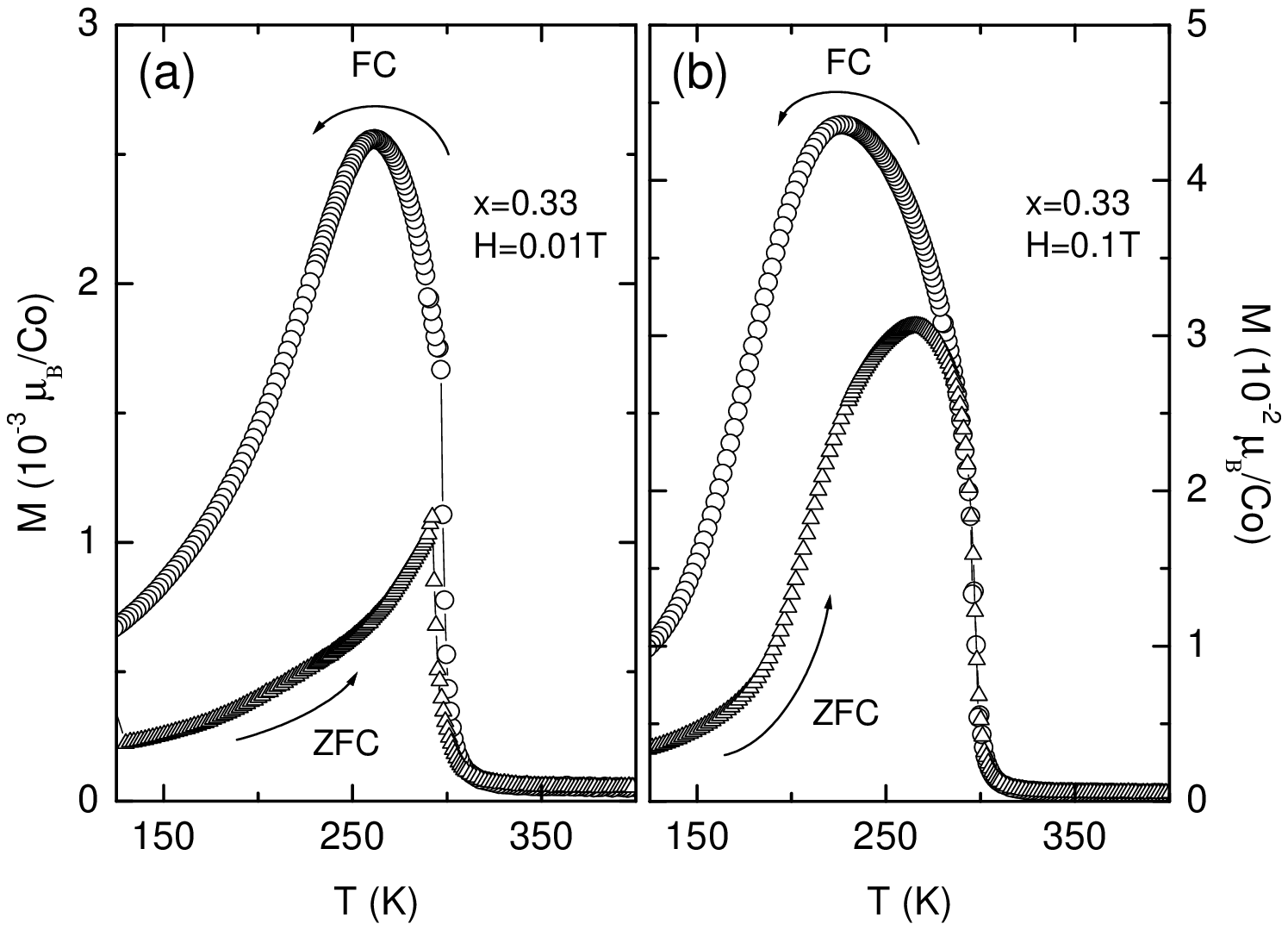}
\caption{Irreversible behavior of the magnetization in
GdBaCo$_{2}$O$_{5+x}$ with $x=0.33$ (the contribution of Gd$^{3+}$ ions
has been subtracted). Measurements were performed either on cooling the
crystal in magnetic field (FC) or on heating after the crystal was
cooled down to 2 K in zero magnetic field (ZFC). The magnetic field
$H=0.01$ T (a) or $H=0.1$ T (b) was applied along the $ab$ plane.}
\end{figure}

Quite often such magnetic irreversibility observed in transition-metal
oxides is attributed, sometimes without due care, to the formation of a
spin-glass state. It should be noted, however, that the idea of a spin
glass implies that no long-range (or intermediate-range) order is
developed in the spin system.\cite{SG} We have found that in the case of
GdBaCo$_{2}$O$_{5+x}$, the magnetic irreversibility shows up {\it only}
below the onset of a ferromagnetic-like behavior (Fig. 16) and thus is
most likely associated with a conventional ferromagnetic domain
structure or with a metamagnetic transition. It should not be surprising
also if the FM domain structure in GBCO appears to be very robust: one
may just imagine how difficult it should be for a FM domain composed of
Ising-like spins\cite{GBC_PRL} to change the direction of magnetization.
Whatever the doping level in GdBaCo$_{2}$O$_{5+x}$ was, we never
observed a conventional spin-glass behavior, and thus we suggest that
all magnetic irreversibilities in this system are related to the spin
rearrangement within a spin-ordered state.

\subsubsection{Evolution with doping}

\begin{figure}[!b]
\includegraphics*[width=6cm]{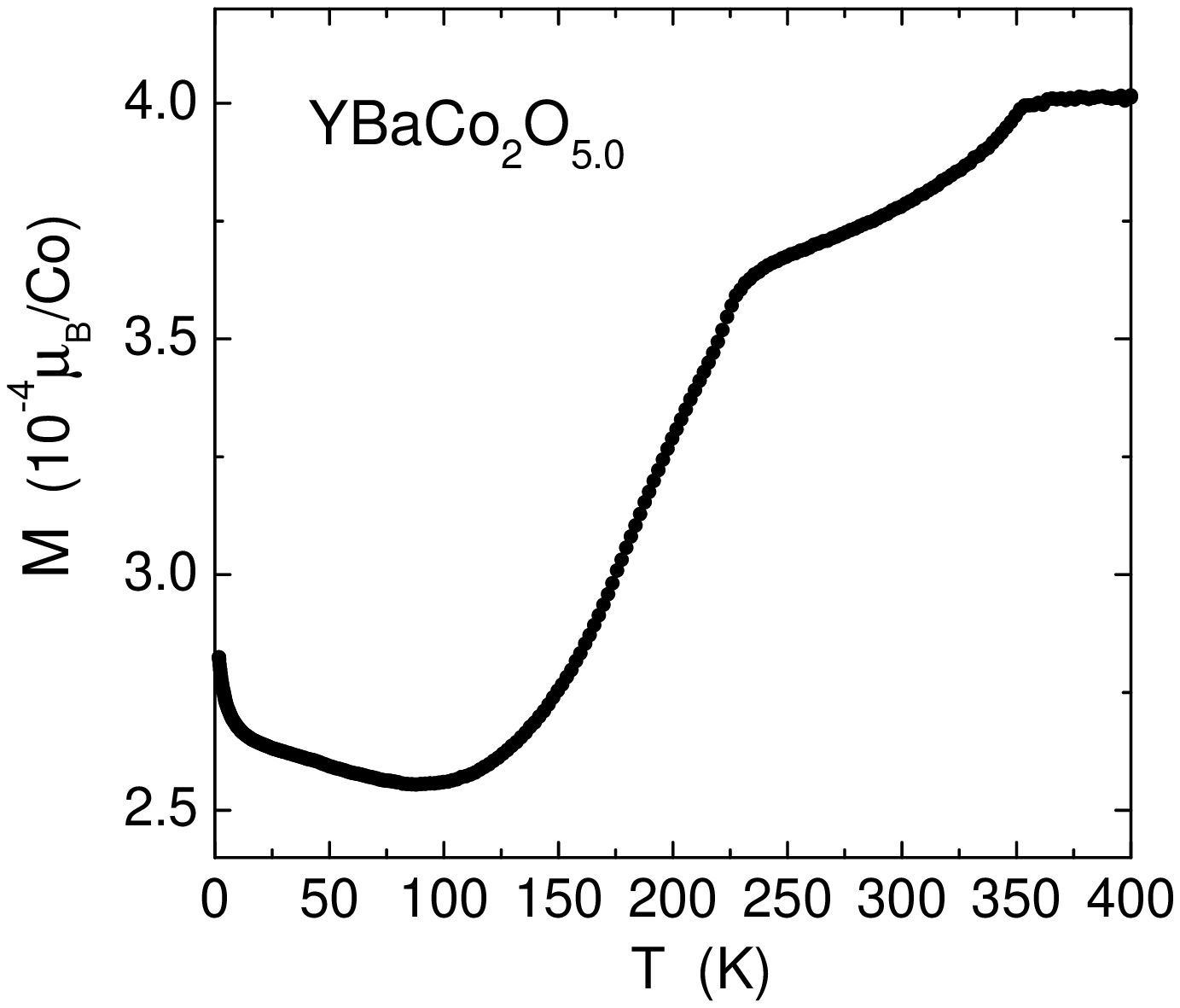}
\caption{Magnetization of an YBaCo$_{2}$O$_{5+x}$ ceramic
measured at $H=0.1$ T.}
\end{figure}

\begin{figure*}[!]
\includegraphics*[width=0.8\linewidth]{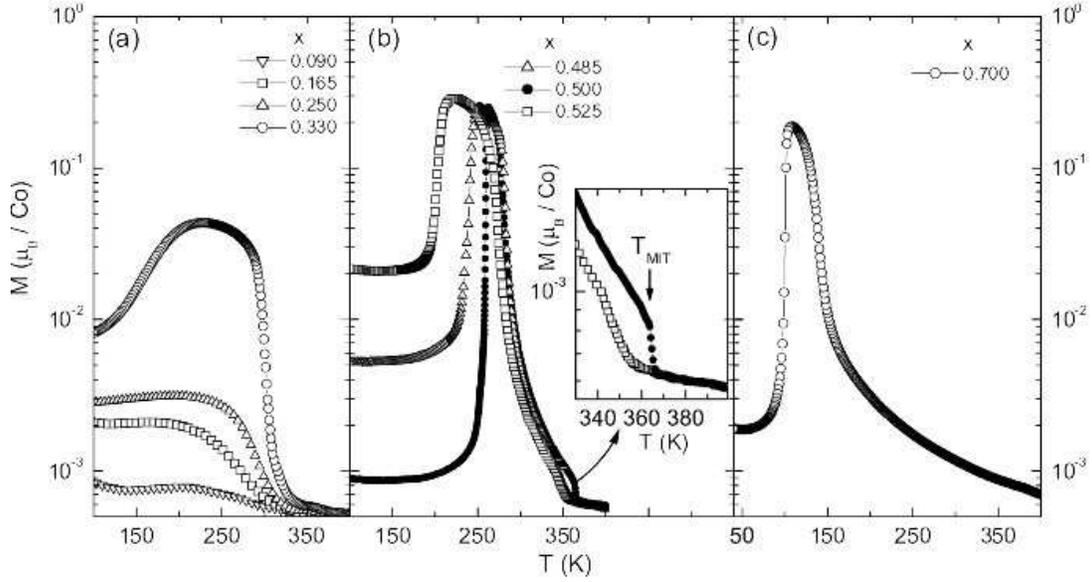}
\caption{Evolution of the magnetization behavior in GdBaCo$_{2}$O$_{5+x}$
with doping (the contribution of Gd$^{3+}$ ions has been subtracted).
Selected curves are shown for three distinct oxygen-concentration
regions: $0<x<0.45$ (a), $0.45<x<0.55$ (b), and $x>0.55$ (c). The $M(T)$
data are taken on cooling in a magnetic field of 0.1 T applied along the
$ab$ plane. The inset in (b) shows an expanded view of the
metal-insulator transition being also a spin-state transition.}
\end{figure*}

At $x=0.0$, GdBaCo$_{2}$O$_{5+x}$ exhibits an antiferromagnetic behavior
as was previously reported for other RBaCo$_{2}$O$_{5.0}$ compounds
studied by magnetization measurements and neutron powder
diffraction;\cite{Suard, Akahoshi, Vogt} the latter identified a
$G$-type AF ordering for spins of the cobalt ions [labeled as AF(1)
hereafter]. It becomes increasingly more difficult to determine the
magnetization component coming from the Co ions as $x$ approaches zero,
because the contribution from the Gd ions becomes dominant in the
temperature range of interest; therefore, it is useful to look at the
$M(T)$ behavior of YBaCo$_{2}$O$_{5+x}$ -- the only known isostructural
compound with non-magnetic R ions\cite{YLaLu} (Fig. 17). The
magnetization of YBaCo$_{2}$O$_{5.0}$, being in rough agreement with the
evaluated contribution of cobalt ions to the magnetization of
GdBaCo$_{2}$O$_{5.0}$, is quite small and shows two anomalies at
$\approx 340-350$ K and $\approx 220$ K, which were attributed by Vogt
{\it et al.} to the onset of an AF spin order and a unidirectional
charge order, respectively.\cite{Vogt} It is worth noting that the same
types of ordering at virtually the same temperatures were also found in
HoBaCo$_{2}$O$_{5.0}$ (Ref.~\onlinecite{Suard}) and
NdBaCo$_{2}$O$_{5.0}$ (Ref.~\onlinecite{Soda}), which gives evidence
that this spin and charge ordering is governed solely by the physics of
CoO$_2$ planes.

When the antiferromagnetic GdBaCo$_{2}$O$_{5.0}$ is doped with oxygen,
it immediately causes an increase of magnetization at temperatures below
$\approx 300-320$ K [Fig. 18(a)]; this looks like a new
ferromagnetic-like component emerging and growing in a rough proportion
to the oxygen concentration up to $x\approx 0.3$. To our surprise, the
onset temperature of this FM behavior remains constant, ignoring the
growth of the overall FM moment. Apparently, such behavior would
certainly indicate a formation of a distinct chemical phase, if the
structural data did not demonstrate that all the crystals were perfectly
homogeneous on a macroscopic scale (down to the scale resolved by
conventional X-ray technique).

One might speculate that the doped oxygen induces canting of
antiferromagnetically ordered cobalt spins which then mimic the
ferromagnetic behavior, as happens in many other canted
antiferromagnets, such as La$_2$CuO$_4$ \cite{LCO}, where a slight spin
rotation brings about a weak ferromagnetism. However, a rather large
magnitude of FM moment, exceeding, for example, in the $x=0.33$ sample
$\sim 0.3\,\mu_{\text{B}} /$Co at $H=7$ T, is apparently inconsistent
with such weak spin canting.

The remaining possibility is that the ferromagnetic response originates
from nanoscopic FM droplets imbedded in the AF(1) matrix; they may be
associated with adjacent oxygen clusters in GdO$_x$ planes or just with
nanoscopic regions of CoO$_2$ planes enriched with charge carriers.
These FM clusters should be at least several unit cells in size to
survive thermal fluctuations and to produce more than just a
paramagnetic response, yet small enough to avoid being detected as a
macroscopic phase. Indeed, this hypothesis finds support from the field
dependences of the magnetization measured in moderately doped
GdBaCo$_{2}$O$_{5+x}$ ($0<x<0.3$), which exhibit a curious combination
of ferromagnetic features (e. g., thermomagnetic irreversibility) with
superparamagnetic ones. For example, in the $x=0.165$ crystal, the
$M(H)$ curves appear to be almost perfectly linear, that is, $M/H$ is
independent of $H$, without any sign of saturation up to 7 T [Fig.
19(a)]. In fact, this is a canonical behavior of a superparamagnet -- a
paramagnetic system, where the role of individual spins is played by the
moments of FM clusters. A slow saturation appears in the $M(H)$ curves
at higher oxygen concentrations (as an example, Fig. 19(b) shows $M$/$H$
data for $x=0.330$), indicating that FM domains grow and probably also
start interacting with each other.

\begin{figure}[!b]
\includegraphics*[width=8.6cm]{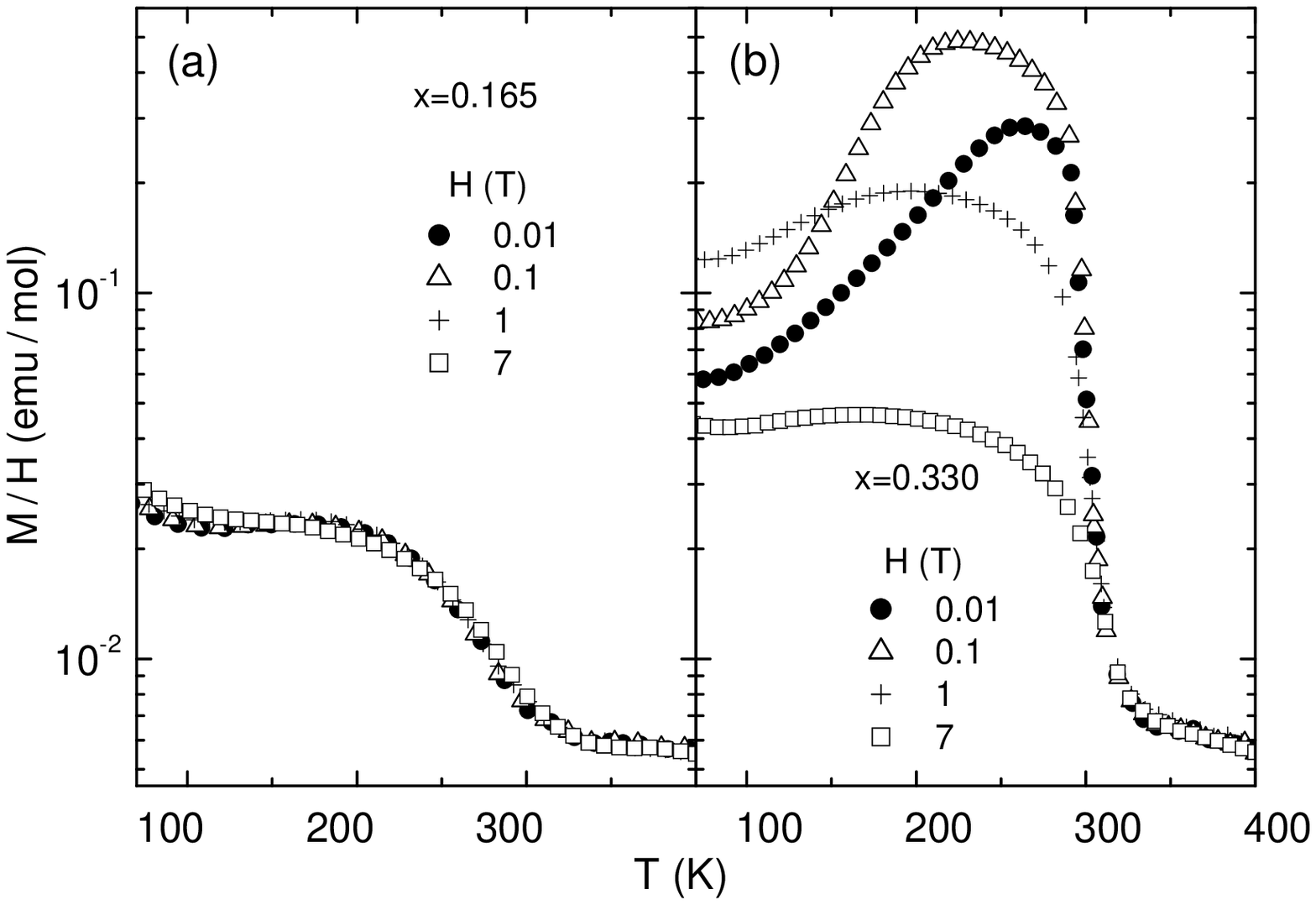}
\caption{$M/H$ in GdBaCo$_{2}$O$_{5+x}$ measured at several magnetic
fields in a range of $0.01-7$ T for two oxygen concentrations, $x=0.165$
(a) and $x=0.330$ (b). The data are taken on cooling in a magnetic field
applied along the $ab$ plane.}
\end{figure}

Interestingly, the FM ordering introduced by oxygen seems to be unstable
at low temperatures and the magnetization increase starting below
300-320 K smoothly gives way to an AF-like behavior as the temperature
is further reduced [Figs. 18(a) and 19(b)]. As can be seen in Fig.
19(b), a strong magnetic field can affect this FM-AF competition,
suppressing the low-temperature magnetization drop: at $H=0.1$ T, $M$
diminishes by more than 5 times with decreasing temperature from 200 K
to 100 K, while at $H=7$ T it remains almost constant.

The sequence of magnetic transitions becomes most clear in the
oxygen-concentration range near $x\approx0.50$ [Fig. 18(b)]. For these
compositions, a net ferromagnetic moment appears in the cobalt
sublattice below $\approx 300$ K, and then suddenly vanishes upon
further decreasing temperature (at $\approx 260$ K for $x=0.500$),
indicating successive PM-FM-AF transitions. The FM state shows up only
in a narrow temperature window, which has the smallest width of less
than 40 K for $x=0.500$. Whenever the oxygen concentration deviates from
the parent composition $x=0.500$, be it on the electron-doped or the
hole-doped side, the FM phase becomes more stable and the FM-AF
transition shifts to lower temperatures [Fig. 18(b)]. Also, the
low-temperature magnetization in both cases shows notably higher values,
indicating a deviation from a pure AF spin order that is realized in
GdBaCo$_{2}$O$_{5.50}$.

Another important feature of the magnetization behavior in the range of
$0.45<x<0.55$ is a step-like change of both the magnetization and its
slope at $T\approx 360$ K [inset in Fig. 18(b)], exactly at the
temperature of the metal-insulator transition observed in our
resistivity and thermoelectric-power measurements (Figs. 6 and 11). This
change takes place in the paramagnetic region and is caused by the
spin-state transition of Co$^{3+}$ ions as has been concluded based on
the Curie-Weiss fitting of the PM susceptibility and structural studies
of GdBaCo$_{2}$O$_{5+x}$ and isostructural compounds.\cite{Respaud,
Frontera, Martin, Moritomo, GBC_PRL} Similar to the behavior observed in
resistivity (Fig. 6), any deviation in stoichiometry from $x=0.50$ blurs
the spin-state/metal-insulator transition and shifts it towards lower
temperatures [inset in Fig. 18(b)].

\begin{figure}
\includegraphics*[width=8.6cm]{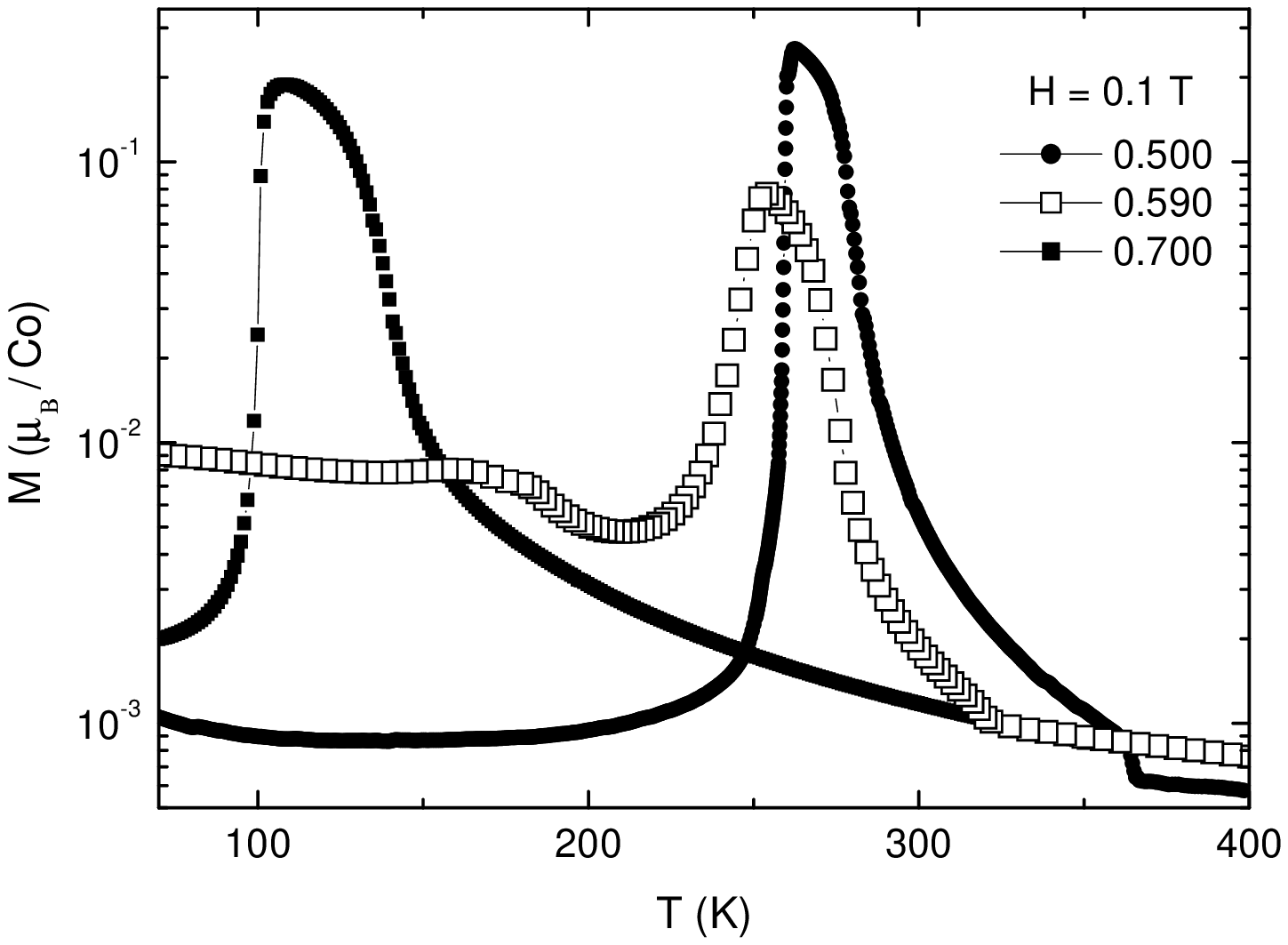}
\caption{An example of the phase separation in a GdBaCo$_{2}$O$_{5+x}$
crystal with $x=0.59$, which is manifested in the magnetization behavior
as the presence of two different magnetic phases typical for neighboring
compositions of $x\approx 0.5$ and $x\approx 0.7$. Magnetization data
for crystals with $x=0.500$ and $x=0.700$ are also shown for
comparison.}
\end{figure}

As the oxygen content in GdBaCo$_{2}$O$_{5+x}$ exceeds $\approx 0.55$,
its magnetic behavior notably changes. In this moderately hole-doped
region, only crystals with $x\approx 0.70$ give an impression of being
homogeneous [Fig. 18(c)]. For this composition, the FM order develops at
much lower temperatures than for $x<0.55$, namely, below 150 K,
suggesting a different spin arrangement of Co$^{3+}$ and Co$^{4+}$ ions;
the FM ordering however is still followed by a sharp FM--AF transition
at $\sim 100$ K [Fig. 18(c)]. For other oxygen concentrations, the
$M(T)$ curves clearly demonstrate a superposition of two magnetic
phases, corresponding to $x\approx 0.5$ and $x\approx 0.7$ (Fig. 20).
What distinguishes this composition region from the electron-doped
samples ($x<0.45$) is that here the phase separation (taking place in
macroscopically homogeneous crystals\cite{homogen}) is also well seen in
the structural data, see Sec.\ \ref{sec:Str}, which implies a fairly
large size of magnetic and crystallographic domains. Since the Coulomb
interaction would prevent a formation of large charged domains, the
phase separation should also involve a macroscopic redistribution of
oxygen ions over the crystal. One might further speculate that the
homogeneous state with $x\approx 0.7$ corresponds to an ordered
``Ortho-III'' phase where two filled oxygen chains in GdO$_x$ layers
alter with one empty chain (O-O-X-O-O-X), in contrast to the $x=0.5$
phase exhibiting a one-to-one alteration (O-X-O-X). It would be
interesting to search for the superlattice peaks corresponding to this
putative ``Ortho-III'' structure using high-intensity x-ray
measurements. Given a very high oxygen mobility, the oxygen ions could
easily form ordered domains when crystals were cooled down after
annealing. More structural studies are however necessary to build a
conclusive picture.

\subsubsection{Detwinned GdBaCo$_{2}$O$_{5.50}$ crystals}

\begin{figure}[!t]
\includegraphics*[width=8.6cm]{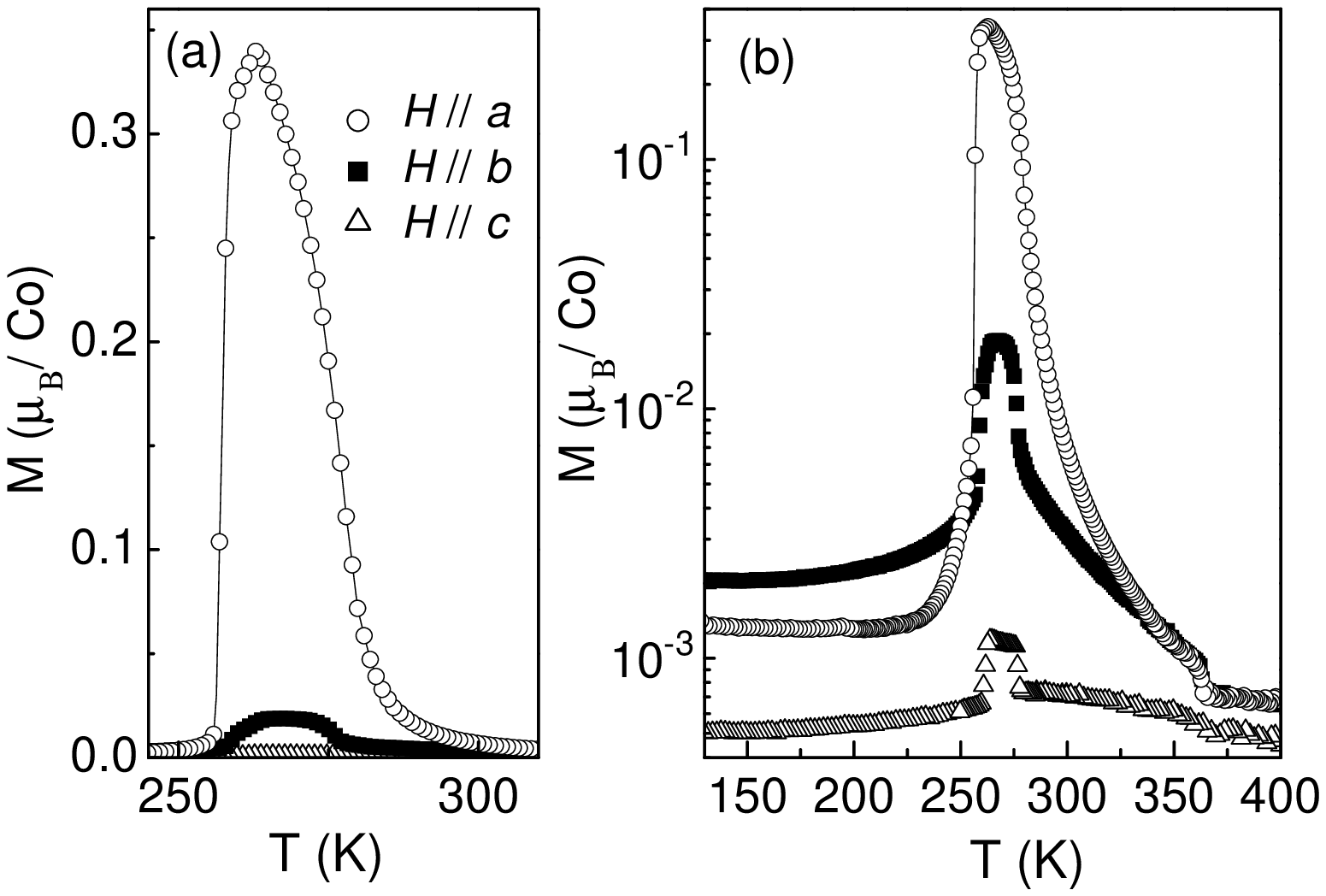}
\caption{Magnetization of a detwinned GdBaCo$_{2}$O$_{5.50}$ crystal
in a linear (a) and logarithmic scale (b); measurements are done in
$H=0.1$ T applied along one of the crystal axes (the contribution of
Gd$^{3+}$ ions has been subtracted).}
\end{figure}

As it becomes clear from the data presented above, the magnetic
properties of the parent GdBaCo$_{2}$O$_{5.50}$ composition are most
important for understanding the overall magnetic behavior of
GdBaCo$_{2}$O$_{5+x}$. Since the crystal structure at $x=0.50$ is
orthorhombic (see Sec.\ \ref{sec:Str}), we had to detwin crystals in
order to obtain single-domain samples for an accurate and detailed
study. The magnetization of a detwinned GdBaCo$_2$O$_{5.50}$ single
crystal (a fraction of misoriented domains $\sim 4$\%) measured along
the $a$, $b$, and $c$ axes reveals a remarkable anisotropy of the spin
system [Fig. 21(a)]: In the FM state that shows up in a narrow
temperature window below 300 K, the net FM moment appears only along one
of the orthorhombic axes, namely, along the $a$ axis. This behavior
suggests that the cobalt spins in GdBaCo$_{2}$O$_{5.50}$ are not only
strongly confined to the CoO$_2$ planes (Sec.\ \ref{sec:magG}), but are
also pinned to one of the in-plane directions; in other words, the spins
system appears to be {\it Ising-like}. Note that a small magnetization
along the $b$ axis in Fig. 21 comes mostly from residual misoriented
domains.

\begin{figure*}[!t]
\includegraphics*[width=0.8\linewidth]{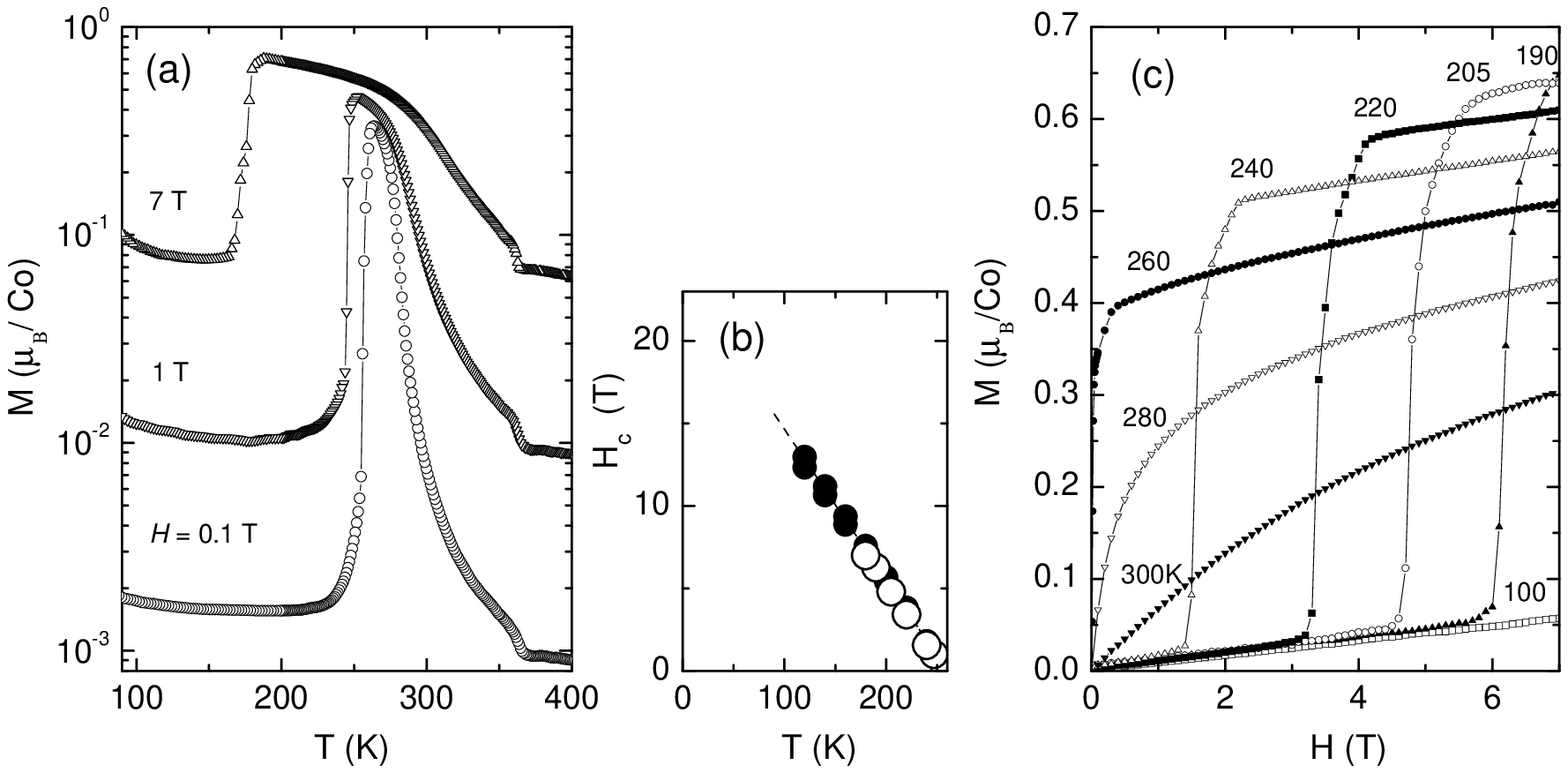}
\caption{(a) $a$-axis magnetization of GdBaCo$_{2}$O$_{5.50}$ measured
in different magnetic fields. (b) AF-FM phase boundary determined from
the $M(T)$ and $M(H)$ data (open circles) and magnetoresistance
measurements (solid circles). (c) Isothermal magnetization $M(H)$
measured upon increasing magnetic field ${\bf H}$$\,\parallel\,$$a$ at
several temperatures; the AF-FM transition exhibits a weak hysteresis
(not shown).}
\end{figure*}

The logarithmic plot in Fig. 21(b) provides additional details. As we
have already discussed, the $c$-axis magnetization in
GdBaCo$_{2}$O$_{5.50}$ is almost featureless similar to all other
compositions (a small hump in the FM region is several hundred times
smaller than for ${\bf H}$$\,\parallel\,$$a$; moreover, it may come in
part from imperfect alignment of the crystal in the magnetometer). The
behavior of the in-plane magnetization is more interesting: In the FM
region, $M$ is much larger for ${\bf H}$$\,\parallel\,$$a$ than for
${\bf H}$$\,\parallel\,$$b$, but the situation abruptly turns over as
the crystal enters the AF region. Given that the transverse
susceptibility of an antiferromagnet exceeds the longitudinal one, we
can conclude that the cobalt spins keep their spin easy axis ($\parallel
a$) upon the FM-AF transition. In other words, the cobalt spins are
aligned along the $a$ axis in the FM state and they keep being aligned
along the $a$ axis in the AF state; what happens upon the FM-AF
transition is that the spins in one of sublattices just flip by
$180^{\circ}$.

The balance of FM and AF ordering in GdBaCo$_{2}$O$_{5.50}$ turns out to
be quite delicate, so that it can be easily affected by magnetic fields
applied along the spin-easy $a$ axis, which stabilize the FM state and
shift the FM-AF transition to lower temperatures (Fig. 22). We find that
the magnetic field ${\bf H}$$\,\parallel\,$$a$ required to overcome the
AF coupling grows roughly linearly with cooling, from zero at $T\approx
260$ K up to $\sim 20$ T at $T=0$. It is worth noting that the AF-FM
transition remains very sharp even in the temperature range close to 260
K, where rather weak fields ($\sim 1.5$ T at 240 K, i.e.
$\mu_{\text{B}}H\ll kT$) are capable of recovering the FM order [Figs.
22(a) and 22(c)], which shows that thermal fluctuations are irrelevant
here. This behavior clearly indicates that the observed AF-FM switching
is a metamagnetic transition, that is, it occurs {\it within} the
ordered spin state and is governed by the relative reorientation of
weakly-coupled spin sublattices.

Whatever the temperature, the 7-T field ${\bf H}$$\,\parallel\,$$b$ or
${\bf H}$$\,\parallel\,$$c$ appears to be too week to compete with the
spin anisotropy; it neither can turn the FM moment towards the $b$ or
$c$ axis in the FM region, nor can it induce the AF-FM transition at
$T\leq 260$ K. The only impact of ${\bf H}$$\,\parallel\,$$bc$ is
therefore to cause a partial tilting of the cobalt spins from their easy
$a$ axis, thus giving linear $M(H)$ curves in both FM and AF regions
(Fig. 23). The spin anisotropy between the $c$ axis and the $ab$ plane
is several times stronger than the in-plane anisotropy; consequently,
the slope of the $M(H)$ curves for ${\bf H}$$\,\parallel\,$$c$ is
several times smaller than that shown in Fig. 23.

\begin{figure}[!b]
\includegraphics*[width=5.6cm]{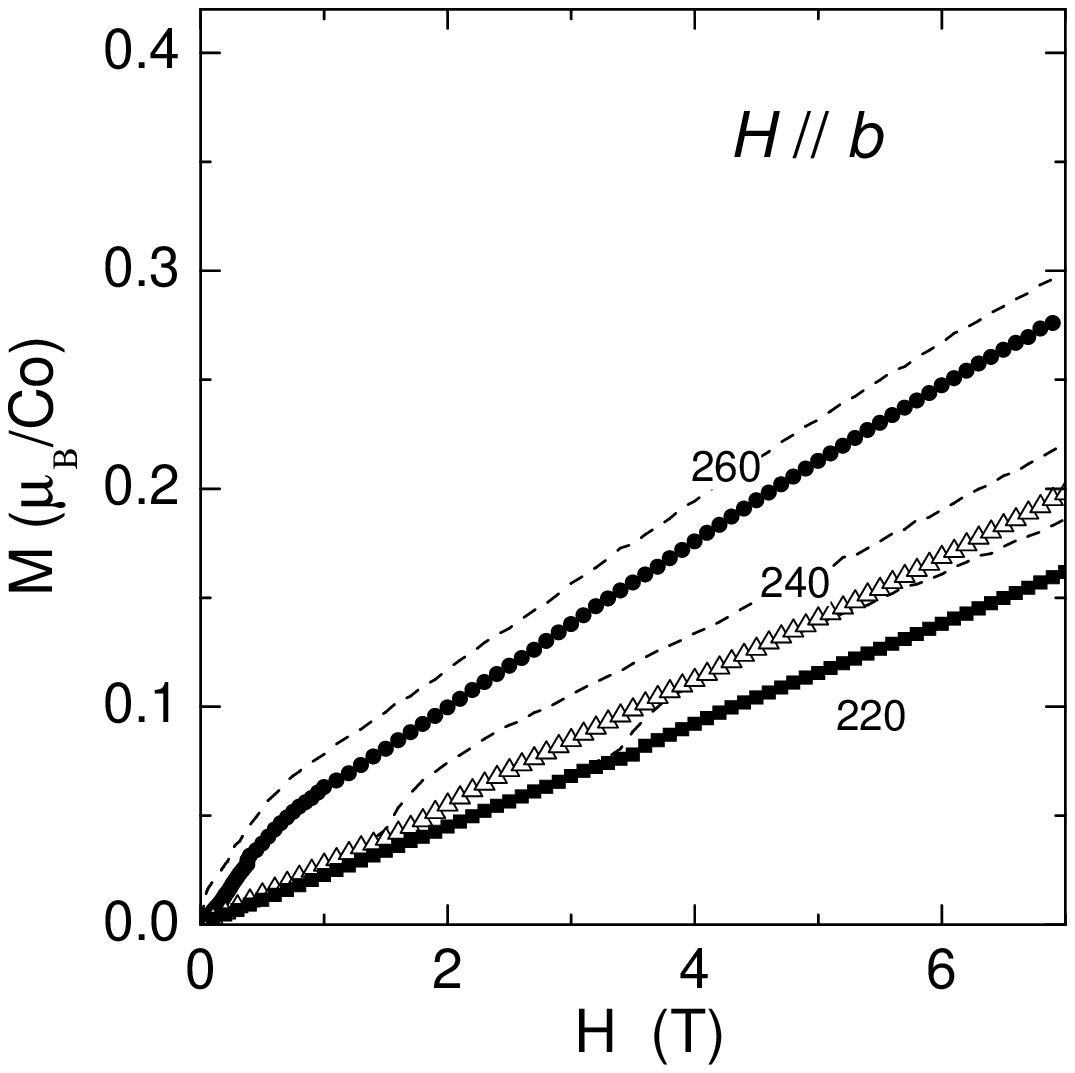}
\caption{Isothermal magnetization of GdBaCo$_{2}$O$_{5.50}$ for
${\bf H}$$\,\parallel\,$$b$. Dashed lines indicate the experimental
data, and symbols show $M(H)$ after subtracting the contribution from
misoriented domains (4\% of the total amount).}
\end{figure}

It is interesting to examine the magnitude of the moment that shows up
in the FM state, which can be easily done using the $M(H)$ data in Fig.
22(c). The FM moment turns out to exceed $0.6\,\mu_{\text{B}}/$Co at
$T=200$ K, and a rough extrapolation to $T=0$ suggests a saturated
magnetic moment of $\approx 1\,\mu_{\text{B}}/$Co (Fig. 24). It is worth
noting that {\it polycrystalline} RBaCo$_{2}$O$_{5.5}$ samples were
reported to demonstrate noticeably smaller FM moments.\cite{Troy1,
Respaud, Akahoshi} This apparent discrepancy originates from the Ising
spin anisotropy that prevents moments from being seen along the $b$ and
$c$ axes; a detwinned single crystal is clearly necessary to make the
true FM moment of $\approx 1\,\mu_{\text{B}}/$Co visible.

\begin{figure}[!t]
\includegraphics*[width=5.8cm]{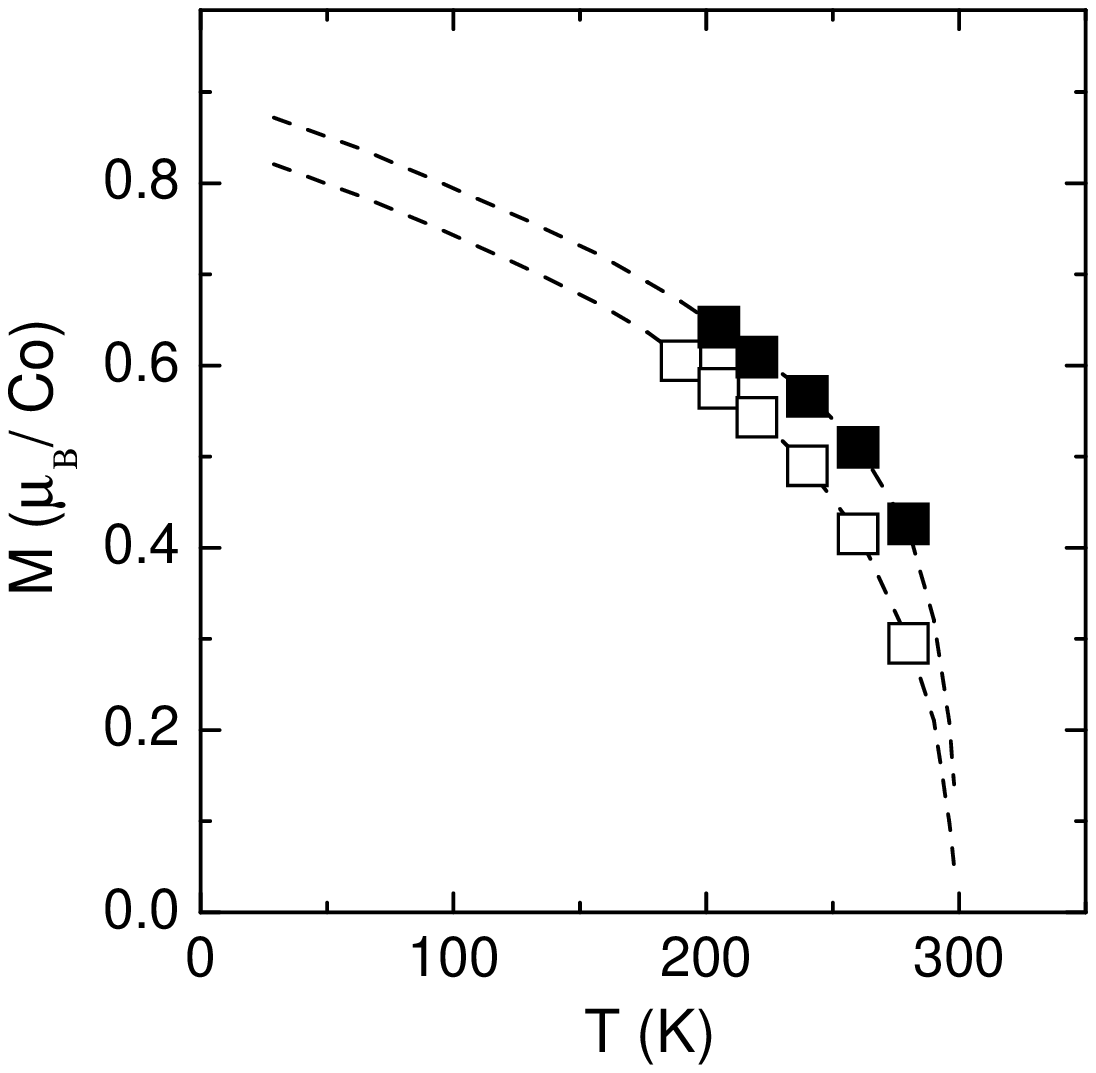}
\caption{Net magnetic moment along the $a$ axis measured at
$H=7$ T (solid squares), and the extrapolated FM moment at $H=0$ (open
squares). Dashed lines are a guide to the eye.}
\end{figure}

While this value is apparently too large to be accounted for by any
spin-canting picture, it agrees well with the expectation for a simple
FM order, if the Co$^{3+}$ ions in this temperature range possess a 1:1
mixture of low-spin $S=0$ and intermediate-spin $S=1$ states. In fact,
the same conclusion on the spin states of Co$^{3+}$ ions has been also
reached based on the Curie-Weiss fitting of the PM
susceptibility\cite{Respaud, Martin, GBC_PRL} in the temperature range
300-360 K (Figs. 18 and 22), as well as based on the structural
data.\cite{Frontera}

\subsection{Magnetoresistance}
\label{sec:MR}

\begin{figure}[!b]
\includegraphics*[width=8.6cm]{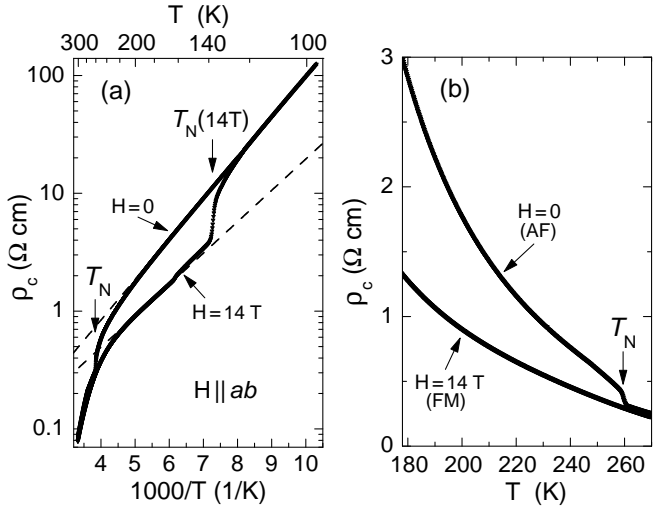}
\caption{(a) Temperature dependence of $\rho_c$ measured at $H=0$ and
14 T applied along the $ab$ plane. The dashed lines show simple
activation fits $\rho_c \propto \exp(\Delta/T)$ for both the AF and FM
states, and $T_{\text{N}}$ indicates the temperature of the FM-AF
transition that is shifted to lower temperatures by the applied field.
(b) A linear-scale view of the high temperature region, illustrating a
step-like increase of $\rho_c$ at $T_{\text{N}}$.}
\end{figure}

\begin{figure*}[!]
\includegraphics*[width=0.95\linewidth]{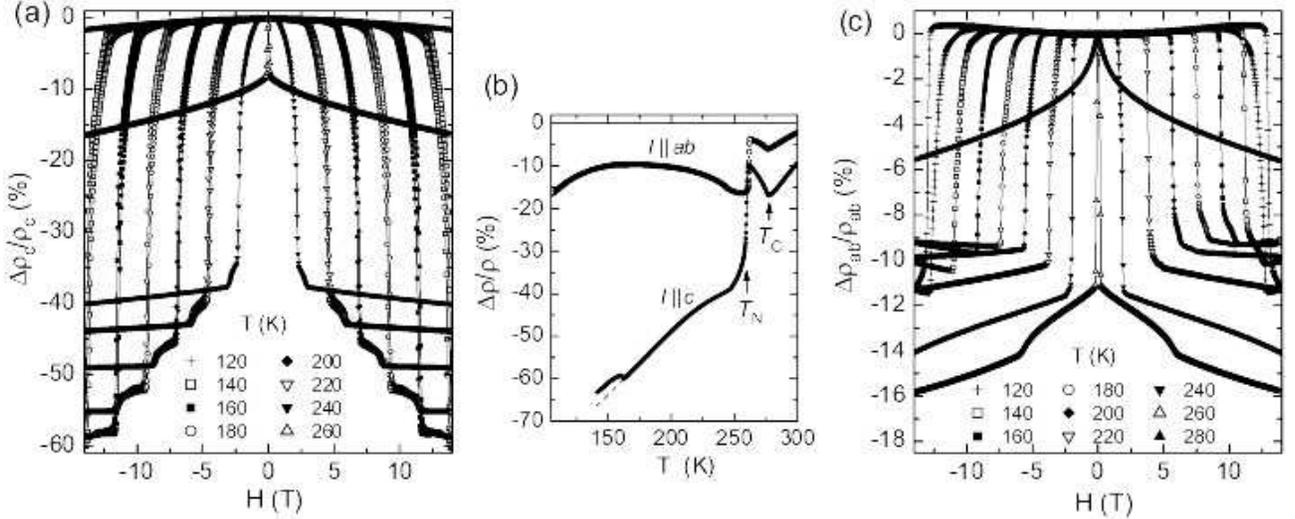}
\caption{Magnetoresistance of twinned GdBaCo$_{2}$O$_{5.5}$ crystals:
field dependences of $\Delta\rho_c/\rho_c$ (a) and
$\Delta\rho_{ab}/\rho_{ab}$ (c) measured at several temperatures for
${\bf H}$$\,\parallel\,$$ab$. For clarity, in panel (c) the hysteresis
is shown only for $T=120$ K; for other temperatures, the data are given
for sweeping the field one-way from -14 T to +14 T. (b) Temperature
dependence of $\Delta\rho_c/\rho_c$ and $\Delta\rho_{ab}/\rho_{ab}$
measured at $H=14$ T applied along the $ab$ plane.}
\label{fig26}
\end{figure*}

In compounds with competing magnetic orders, a magnetic field that
favors one kind of ordering often causes also a large magnetoresistance;
GdBaCo$_{2}$O$_{5+x}$ is no exception. The charge transport in this
system turns out to be very sensitive to both the FM and AF ordering,
and magnetic fields readily induce a giant magnetoresistance by
affecting the subtle AF-FM balance.\cite{Respaud, GBC_PRL} As an
example, Fig. 25 shows the $c$-axis resistivity of a twinned
GdBaCo$_{2}$O$_{5.50}$ crystal measured at $H=0$ and 14 T. In zero
field, the FM-AF transition at $\approx 260$ K brings about a step-like
increase of the resistivity $\rho_c$ [Fig. 25(b)]; a similar, albeit
smaller, step is observed in $\rho_{ab}$ as well. A 14-T field applied
along the $ab$ plane shifts the magnetic transition towards lower
temperatures and wipes out the resistivity increase, thus causing the
resistivity to drop by up to several times.

As can be seen in Fig. 25(a), $\rho_c$ grows roughly exponentially upon
cooling below $\sim 230$ K regardless of the applied field; the charge
carriers, therefore, have different activation energies in the FM and AF
states, and the MR originates from the reduction in this activation
energy. As soon as the magnetic field becomes insufficient to maintain
the FM order, the system switches into the AF state [at
$T=T_{\text{N}}(14\,$T$)$], and the resistivity jumps to its zero-field
value [Fig. 25(a)].

Unfortunately, we did not succeed in detwinning the crystals with
already prepared electrical contacts, and thus the measurements were
carried out on twinned crystals, which does not allow us to analyze the
MR quantitatively. The role of twins is less critical for $\rho_c(T)$,
since we find the orthorhombic domains to always go through the whole
crystal from one face to another. Consequently, the measuring current
does not cross domain boundaries, and the observed magnetoresistance
$\Delta\rho_c/\rho_c$ is just reduced from its true value by some
factor, since not all domains are affected by the magnetic field. Upon
measuring $\rho_{ab}$, however, the current flowing in the $ab$ plane
has to pass through both kinds of orthorhombic domains which form a
striped structure. The measured $\Delta\rho_{ab}/\rho_{ab}$ thus depends
not only on the ratio of domains, but also on the yet unknown anisotropy
$\rho_a/\rho_b$.

In spite of this complication, the qualitative behavior of the MR
remains clear. At $T<T_{\text{N}}\,$$\approx\,$260 K, the in-plane
magnetic field induces an abrupt decrease in both $\rho_c$ and
$\rho_{ab}$ (Fig. 26), which occurs at exactly the same field as the
AF-FM transition in magnetization, leaving no doubts about its origin.
At higher temperatures, the resistivity changes gradually, again
resembling the magnetization behavior in Fig. 22(c). In the latter case,
the MR seems to originate from the field-induced stabilization of the FM
order; the magnetic field suppresses critical FM fluctuations, thus
facilitating the charge motion.

The temperature dependences of $\Delta\rho_{ab}/\rho_{ab}$ and
$\Delta\rho_c/\rho_c$ in Fig. 26(b) nicely illustrate the correlation of
the MR with the magnetic behavior. Upon cooling, we first observe a
clear peak at $\sim 278$ K in the MR, which can be associated with the
FM transition (the peak position coincides with the peak in $dM/dT$ in
Fig. 21). Apparently, the FM fluctuations that are strongest near
$T_{\text{C}}$ frustrate the charge motion, providing contribution of
about 20\% to resistivity, and this contribution is removed when a
strong magnetic field is applied. It is worth noting, that this negative
MR peak near $T_{\text{C}}$ closely resembles the MR behavior of such
ferromagnetic oxides as cubic La$_{0.5}$Ba$_{0.5}$CoO$_3$
(Ref.~\onlinecite{Fauth}) and La$_{1-x}$Sr$_x$MnO$_3$
(Ref.~\onlinecite{Imada}), where the resistivity change is usually
explained by the double-exchange mechanism.\cite{Double} Upon further
cooling, the MR abruptly increases at $T=T_{\text{N}}$. The origin of
this MR is obvious: Without field, the AF ordering enhances the
resistivity, while the applied magnetic field prevents establishing the
AF order. As can be seen in Fig. 26(b), the $c$-axis MR quickly gains
strength with cooling. At 150 K, $\rho_c$ drops by three times at 14 T,
and the ratio $\rho_c(0,T)/\rho_c(H,T)$ would keep growing to much
larger values with lowering temperature if fields exceeding $H_c(T=0)$
were applied.\cite{Respaud} The in-plane MR $\Delta\rho_{ab}/\rho_{ab}$
seems to be by several times smaller; the magnitude and the temperature
dependence of $\Delta\rho_{ab}/\rho_{ab}$ should, however, be taken with
a grain of salt, since, as we mentioned above, a detwinned crystal
should be measured to determine the true MR values.

\begin{figure}[!b]
\includegraphics*[width=8.6cm]{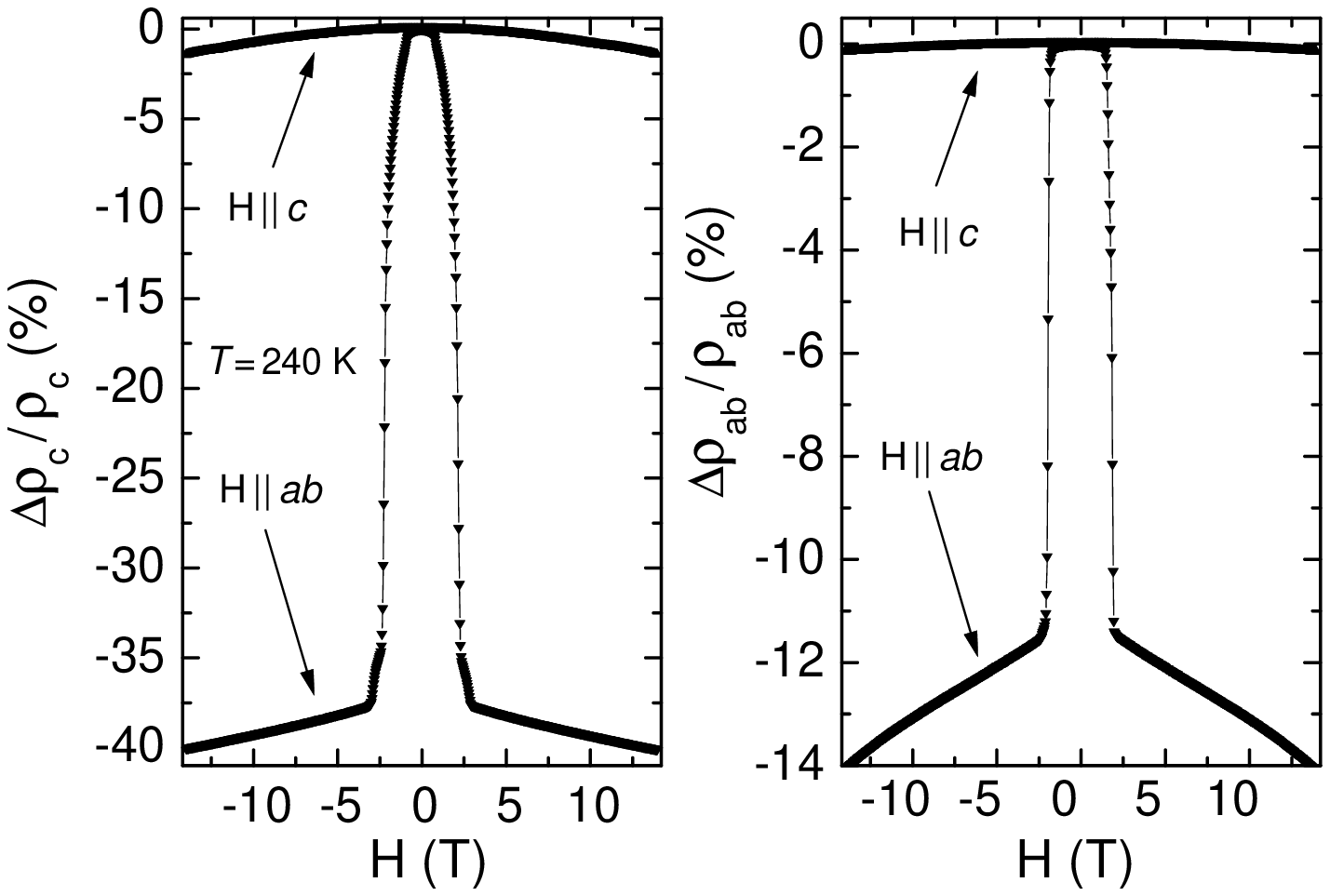}
\caption{Dependence of the magnetoresistance on the field direction:
$\Delta\rho_c/\rho_c$ (left) and $\Delta\rho_{ab}/\rho_{ab}$ (right)
measured at $T=240$ K with the magnetic field applied along or
transverse to the $ab$ plane.}
\end{figure}

The MR anisotropy with respect to the magnetic-field direction can
provide interesting information on the cobalt-spin anisotropy. In
general, when an {\it Ising-like} antiferromagnet is subjected to an
increasingly strong magnetic field, it eventually turns into a FM state,
whatever the field direction is. If the field is applied along the
spin-easy $a$ axis, the spin-flip transition is abrupt, while for the
transverse direction, the spins rotate gradually and much higher fields
are required to align them. In the latter case, the rotation angle is
roughly $\sin\alpha \sim g\mu_{\text{B}}H/2J_{a-c}$, where $J_{a-c}$ is
the spin anisotropy energy. It is reasonable to expect that the MR in
GdBaCo$_{2}$O$_{5+x}$, being roughly proportional to the magnetization,
will change with the field as $\propto g\mu_{\text{B}}H/2J_{a-c}$,
tending to reach its saturation value at $H\sim
2J_{a-c}/g\mu_{\text{B}}$. Thus, the anisotropy of magnetoresistance in
addition to that of magnetization may be used to probe how strongly the
spins are coupled with the crystal lattice.

An experimental study of GdBaCo$_{2}$O$_{5.50}$ crystals reveals that
the MR anisotropy is surprisingly large and the 14-T field ${\bf
H}$$\,\parallel\,$$c$ can do nothing comparable to the MR caused by a
much less field ${\bf H}$$\,\parallel\,$$ab$ (Fig. 27). Indeed, the MR
is barely seen for ${\bf H}$$\,\parallel\,$$c$, especially
$\Delta\rho_{ab}/\rho_{ab}$ [Fig. 27(b)], which is as small as 0.14\% at
14 T, while for the in-plane field it readily reaches a two orders of
magnitude larger value. This indicates that magnetic fields in the 100-T
range would be necessary to rotate the cobalt spins in
GdBaCo$_{2}$O$_{5.50}$ from the $a$ to the $c$ axis, overcoming the
remarkable spin anisotropy. It is worth noting, that such
spin-anisotropy energy, roughly estimated to be of the order of 10
meV/Co ($\mu_{\text{B}}H$), is extremely large compared with any known
magnetic material\cite{typical_MA,largest_MA} and is close to the giant
magnetic anisotropy observed in nanoclusters.\cite{largest_MA} This huge
spin anisotropy can also account for a considerable anisotropy in
susceptibility, $\chi_{ab}/\chi_c>1$, that survives even in the
paramagnetic state up to 400 K.

\subsection{Phase diagram of GdBaCo$_{2}$O$_{5+x}$}
\label{sec:phasedia}

An empirical phase diagram of GdBaCo$_{2}$O$_{5+x}$ based on the
obtained structural, transport, and magnetic data is sketched in Fig. 28
as a function of the oxygen concentration. At the lowest oxygen content,
$x=0$, GdBaCo$_{2}$O$_{5+x}$ is an antiferromagnetic insulator [AFI(1)];
its low-temperature phase should most likely correspond to a
charge-ordered $G$-type antiferromagnet, by analogy with other
isostructural compounds studied using neutron diffraction.\cite{Vogt,
Suard, Soda}

\begin{figure}[!b]
\includegraphics*[width=8.6cm]{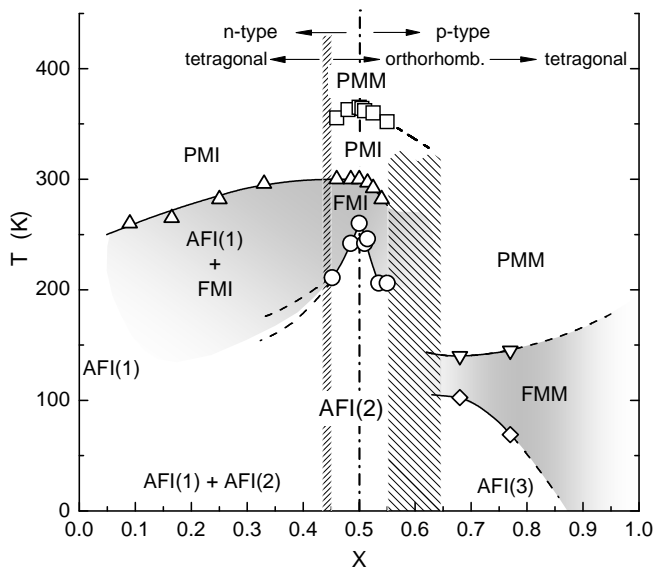}
\caption{Phase diagram of GdBaCo$_{2}$O$_{5+x}$, including regions
of a paramagnetic metal (PMM), paramagnetic insulator (PMI),
ferromagnetic metal (FMM), ferromagnetic insulator (FMI), and
antiferromagnetic insulator (AFI).}
\end{figure}

An increase in the oxygen content $x$ is found to immediately result in
the formation of isolated ferromagnetic clusters imbedded in the
antiferromagnetic [AFI(1)] matrix. These clusters demonstrate a
paramagnetic behavior above $T_{\text{C}}=250-300$ K (marked by open
triangles in Fig. 28), and a curious combination of superparamagnetic
and ferromagnetic features below $T_{\text{C}}$; namely, the
magnetization does not saturate in magnetic fields up to 7 T (as in
canonical superparamagnets), yet the FM moments exhibit an extremely
strong coupling with the crystal lattice, which is manifested in the
magnetic anisotropy and thermo-magnetic irreversibility.

With further increasing oxygen content in GdBaCo$_{2}$O$_{5+x}$ (up to
$x=0.45$), the FM clusters smoothly grow in size, reaching percolation,
and develop at low temperatures an intrinsic instability towards a new
type of AF ordering [AFI(2)], which is the ground state of the parent
GdBaCo$_{2}$O$_{5.50}$ compound.\cite{GBC_PRL} The AFI(2) order emerges
from the FM order in clusters and differs essentially from the AFI(1)
state that is realized at $x=0$; for instance, the AFI(2) ordering can
be easily suppressed by a magnetic field, while AFI(1) is quite robust.
One can consider GdBaCo$_{2}$O$_{5+x}$ over a wide composition region of
$0<x<0.45$ as being composed of nanoscopic phases mixed together: At
high temperatures, the system behaves as a paramagnetic insulator; then,
below $T_{\text{C}}$, as a mixture of FM and AF insulating phases; and
finally at low temperatures, it evolves into a mixture of two distinct
AF phases [AFI(1) and AFI(2)]. The resistivity in this composition range
shows a robust insulating behavior without any noticeable change with
increasing $x$. The room-temperature crystallographic structure retains
a macroscopically tetragonal symmetry with smoothly changing lattice
parameters.

The region with the richest behavior, $0.45\leq x\leq 0.55$, occupies
the center of the phase diagram around the ``parent'' composition
$x=0.50$. In this doping range, the crystal structure is macroscopically
orthorhombic up to rather high temperatures due to the ordering of
oxygen into alternating full and empty chains that run along the $a$
axis (at $T=260^{\circ}$C, we still observed a twin structure related to
the oxygen ordering). Owing to this structural order, the broad magnetic
and transport features emerging already at $x=0.3-0.4$ come in focus
here, and GdBaCo$_{2}$O$_{5+x}$ demonstrates a series of sharp phase
transitions upon cooling: first, from a paramagnetic metal (PMM) to a
paramagnetic insulator (PMI) at $T_{\text{MIT}}$ (open squares in Fig.
28), which is also accompanied with a spin-state transition, then to a
ferromagnetic insulator (FMI) at $T_{\text{C}}$, and, finally, to an
antiferromagnetic insulator [AFI(2)] at $T_{\text{FM-AF}}$ (open
circles).

The ``parent'' composition GdBaCo$_{2}$O$_{5.50}$ is a metal at high
temperatures but turns into a semiconductor upon cooling below the
metal-insulator transition. In fact, the $x=0.50$ composition represents
a kind of borderline that divides the central region of the phase
diagram into two roughly symmetric parts which correspond to
electron-doped and hole-doped semiconducting states below
$T_{\text{MIT}}$. All transport properties show anomalies in their
doping dependences at $x=0.50$, indicating that the density of doped
carries smoothly goes to zero with approaching this point from either
side. As can be seen in Fig. 28, all the transition temperatures also
exhibit maximum values exactly at $x=0.5$. The only property
demonstrating a striking asymmetry with respect to $x=0.5$ is the
resistivity which is much smaller for the hole-doped side ($x>0.5$).

When the oxygen concentration exceeds $x=0.55$, GdBaCo$_{2}$O$_{5+x}$
develops a new phase possessing considerably lower temperatures of the
ferromagnetic transition ($T_{\text{C}}<150$ K) and the FM-AF transition
($T_{\text{FM-AF}}\leq100$ K). Until $x$ reaches $\approx 0.7$, this new
phase is mixed (on a mesoscopic scale) with the $x\approx 0.5$ phase,
and only at $x\geq 0.68-0.70$ the system again recovers its homogeneity.
Surprisingly, in the $x\approx 0.7$ region, the series of successive
PM-FM-AF transitions, albeit happening at reduced temperatures, still
remains remarkably similar to that at $x\approx 0.5$, in spite of very
different transport properties and different spin states of cobalt ions
in these phases.

In general, GdBaCo$_{2}$O$_{5+x}$ seems to have just a few stable phases
such as $x=0$, $x\approx 0.5$, and $x\approx 0.7$, while intermediate
compositions always tend to phase separate on a nanoscopic or mesoscopic
scale.

Although the compositions with $x>0.77$ are hard to achieve in
GdBaCo$_{2}$O$_{5+x}$, we can extrapolate the phase boundaries to higher
doping, as shown in Fig. 28, using an analogy with the cubic
La$_{0.5}$Ba$_{0.5}$CoO$_{3-\delta}$. Our measurements have shown that
the latter compound is a ferromagnet with $T_{\text{C}}\approx 200$ K
(only weakly dependent on $\delta$) with no sign of reentrant AF
behavior at low temperatures.

\section{DISCUSSION}

It is natural to start discussing the GdBaCo$_{2}$O$_{5+x}$ compound
beginning with the ``parent'' $x=0.50$ composition, whose properties
constitute a basis for understanding the transport and magnetic behavior
of this layered cobalt oxide over the entire doping range.

\subsection{Magnetic structure of GdBaCo$_{2}$O$_{5.50}$}

By now, there were several attempts to elucidate the magnetically
ordered states in isostructural RBaCo$_{2}$O$_{5+x}$ (R = Y, Nd, Tb, Ho)
compounds using neutron scattering (for GdBaCo$_{2}$O$_{5+x}$, such a
study is precluded by a large absorption of neutrons by Gd).\cite{Vogt,
Suard, Soda, Fauth2, Neutron2, Neutron3, Neutron4} In the case of
samples with the lowest oxygen concentration $x=0$, whose preparation is
quite straightforward, the neutron diffraction has indeed provided
reproducible data on the charge and spin ordering, being in good
agreement with the magnetization and transport measurements.\cite{Vogt,
Suard, Soda} In contrast, much more complicated and sample-dependent
diffraction patterns were obtained for compositions near $x\approx 0.5$,
bringing about controversial models of the magnetic order in this region
of the phase diagram.\cite{Fauth2, Neutron2, Neutron3, Neutron4} This
apparent controversy, however, is in fact not very surprising given the
remarkable sensitivity of the magnetic and structural properties of
RBaCo$_{2}$O$_{5+x}$ to even a slight modification in the oxygen
concentration, that was revealed in the present study; note also a
strong tendency to phase separation, which should inevitably take place
in all the samples except for those located in a few narrow composition
regions (see Fig. 28). As we have shown here, the actual composition of
so-called ``as-grown'' and ``oxygen-annealed'' samples usually used in
most of the previous studies may deviate a lot from the required
$x=0.50$, and a rather delicate technique should be developed to tune
the oxygen content precisely to this peculiar point.

To the best of our knowledge, none of the models, that have been
suggested for the $x=0.50$ composition based on the neutron-scattering
data, appears to be capable of explaining the entire
experimentally-observed magnetic behavior. For example, a G-type AF
order was proposed by Fauth {\it et al.}\cite{Fauth2} for temperatures
where a rather large FM moment shows up in our magnetization
measurements. On the other hand, the model suggested by Soda {\it et
al.},\cite{Neutron2} being quite close to that one reported recently by
us,\cite{GBC_PRL} indeed captures such gross features as the FM and AF
states and a possibility of easy and abrupt switching between these two;
nevertheless, it still implies a kind of spiral spin order for the AF
state, which is clearly inconsistent with the Ising-like magnetic
behavior observed in the magnetization of detwinned
crystals.\cite{GBC_PRL}

In view of the above mentioned problems with direct methods, we should
try to reconstruct the magnetic structure of GdBaCo$_{2}$O$_{5.50}$
relying mostly on the magnetization data obtained on single crystals
with precisely tuned stoichiometry. Fortunately, this task is simplified
by the discovered strong uniaxial anisotropy of cobalt spins which
dramatically narrows down the range of possible spin arrangements. Let
us first summarize the robust facts established for
GdBaCo$_{2}$O$_{5.50}$ as follows:

(i) The magnetic moments of Co$^{3+}$ ions in GdBaCo$_{2}$O$_{5.50}$
exhibit an Ising-like behavior, being exceptionally strongly confined to
the CoO$_2$ ($ab$) planes (with the anisotropy energy of the order of 10
meV per cobalt ion) and to a lesser extend, though still strongly, to
the orthorhombic $a$ axis. Note also that the confinement of cobalt
spins to the $ab$ plane appears to be a generic feature of
GdBaCo$_{2}$O$_{5+x}$ regardless of the oxygen content and thus
regardless of the valence and spin states of cobalt ions (Sec.\
\ref{sec:magG}). One might wonder whether there can be another
interpretation of the observed magnetic anisotropy: In magnetically
ordered states, for example, the magnetization may indeed be larger for
a direction transverse to the spin-easy axis, which is usually the case
with antiferromagnets and canted antiferromagnets. However, a
possibility of the spin-easy axis to be parallel to the $c$ axis is
immediately ruled out by a considerable anisotropy $\chi_{ab}>\chi_c$
that survives up to high temperatures in the {\it paramagnetic} state.
In turn, by analyzing various canted AF configurations within the $ab$
plane, regardless of whether they are allowed by symmetry or not, we
find that non of them can account for the observed large FM moment,
which would require a large canting angle of cobalt spins, and at the
same time for a strong in-plane anisotropy, where the FM moment appears
only along the $a$ axis (Fig. 21). For example, the spin structure
suggested in Ref.~\onlinecite{Neutron2} with a canting angle of $\sim
\pi/4$ would inevitably result in a magnetic behavior to be essentially
isotropic within the $ab$ plane.

(ii) The spontaneous FM state below $T_{\text{C}}$ and the FM state
induced by a magnetic field ${\bf H}$$\,\parallel\,$$a$ at
$T<T_{\text{FM-AF}}$ represent the same phase with a net FM moment
smoothly growing upon cooling, tending to $\approx 1\,\mu_{\text{B}}/$Co
at $T=0$ (Fig. 24). After the magnetic field stabilizes the FM order, no
anomaly can be seen in the magnetization or resistivity at the
zero-field FM-AF transition temperature.

(iii) The FM-AF switching in GdBaCo$_{2}$O$_{5.50}$ is a metamagnetic
transition: It is induced by a relative reorientation of weakly coupled
spin sublattices, while within each sublattice the spins are kept
ordered by a much stronger interaction. This remarkable hierarchy of
spin interactions follows from the fact that the AF-FM transition
remains sharp even when it is induced by a very weak field
$\mu_{\text{B}}H\ll kT$ (Fig. 22).

Based on the above listed observations (i)-(iii), one can quite easily
sort out the spin structures that may be realized in
GdBaCo$_{2}$O$_{5.50}$. The point (i) indicates that the cobalt spins in
this compound should form a collinear structure aligned along the $a$
axis; this conclusion agrees with all the neutron-scattering
studies,\cite{Vogt, Suard, Soda, Fauth2, Neutron3} except for
Ref.~\onlinecite{Neutron2}. However, if we consider a uniform FM order
at $T<T_{\text{C}}$, the observed net FM moment of $\approx
1\,\mu_{\text{B}}/$Co [see (ii)] appears to be inconsistent with the
allowed spin states of Co$^{3+}$ ions (0, $2\,\mu_{\text{B}}$, and
$4\,\mu_{\text{B}}$ for LS, IS, and HS states, respectively). Although
the magnetic moments of Co$^{3+}$ ions could be modified from their
spin-only values by the orbital moments, which at a first glance may
resolve this contradiction, a homogeneous FM state is also clearly
inconsistent with the FM-AF switching (iii). We thus inevitably arrive
at a conclusion that the magnetic structure of GdBaCo$_{2}$O$_{5.50}$
should correspond to a ferrimagnet composed of two or more antiparallel
FM sublattices; note that one of the sublattices may carry zero moment
if it involves the LS cobalt ions. Apparently, the net FM moment of
$\approx 1\,\mu_{\text{B}}/$Co -- a difference between the sublattice
moments -- may be accounted for if the Co$^{3+}$ ions constituting the
two ferrimagnetic sublattices are either in the LS and IS states, or in
the IS and HS states. One can construct more complex spin structures by
increasing the number of magnetic sublattices; for instance, a
zero-moment sublattice may be composed of LS Co$^{3+}$ ions, or
antiferromagnetically ordered IS(HS) ions.

\begin{figure*}[!t]
\includegraphics*[width=1.0\linewidth]{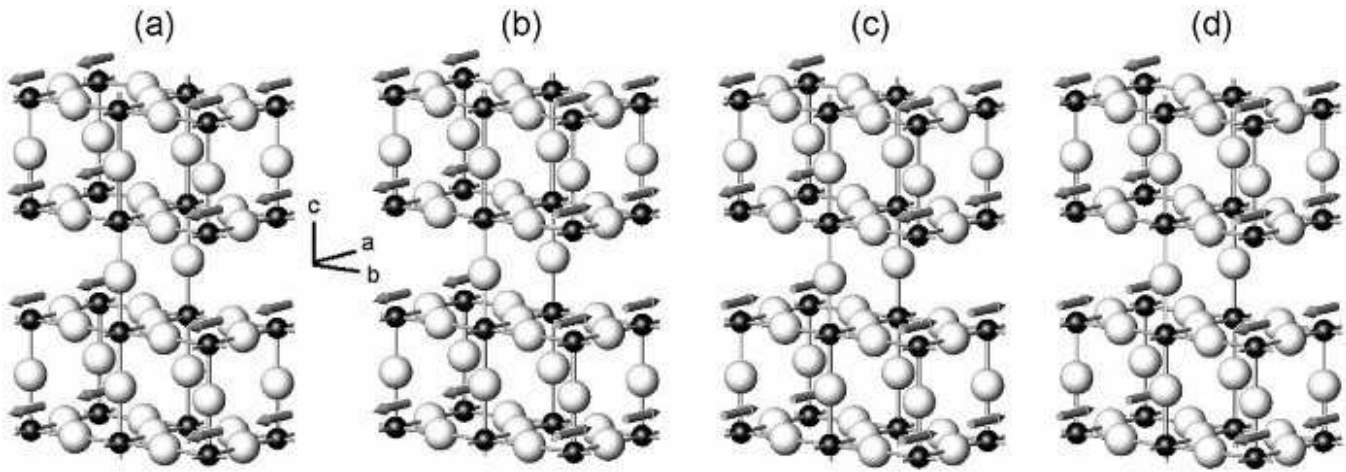}
\caption{A sketch of the magnetic structure of GdBaCo$_{2}$O$_{5.50}$ in
the FM state (a) and possible AF states (b)-(d); Ba and Gd ions are
omitted for clarity. At $T<T_{\text{MIT}}$, the Co$^{3+}$ ions in
octahedral environment are considered to have a low-spin ($S=0$) state,
while those in pyramidal environment have an intermediate ($S=1$) state
(the spin direction is indicated by arrows).}
\end{figure*}

Now we can try to map the ferrimagnetic sublattices onto the actual
crystal structure, keeping in mind the following important conditions.
First, the ferrimagnetic sublattices should be strongly coupled with
each other, since, according to Respaud {\it et al.}, even a 35-T field
fails to induce a ferrimagnetic-ferromagnetic transition.\cite{Respaud}
Another explanation for this robustness is that only one of the
sublattices actually possesses a non-zero FM moment; in that case, no
decoupling transition should be expected. The second point is that
besides strong FM bonds, each 3D ferrimagnetic sublattice should also
contain planes of very weak magnetic coupling to allow for an easy
folding of the magnetic unit cell upon switching into the AF state. One
can see in Fig. 1, that the crystallographic planes responsible for weak
magnetic coupling can naturally be the GdO$_x$ plane and/or an $ac$
plane passing between the pyramidal and octahedral cobalt ions.
Apparently, such kind of effective decoupling of neighboring Co$^{3+}$
ions can hardly be possible unless either of the two $ac$ CoO$_2$ planes
-- composed of cobalt ions in the pyramidal or octahedral environment --
is essentially non-magnetic. Consequently, the other {\it half} of
cobalt ions should provide an average moment of $\approx
2\,\mu_{\text{B}}/$Co, which corresponds to the IS state or an $\approx
\,$1:1 ratio of LS and HS states. Note that the existence of a
non-magnetic sublattice also gives the best account for the high-field
magnetization data of Respaud {\it et al.}.\cite{Respaud}

An important information on the magnetic states of Co$^{3+}$ ions in
GdBaCo$_{2}$O$_{5.50}$ has been also obtained from the Curie-Weiss
fitting of the PM susceptibility\cite{Respaud, Martin, GBC_PRL} in the
temperature range 300-360 K (Figs. 18 and 22), which better agrees with
the presence of 50\% of ions in the IS state, while the same fitting
above $T_{\text{MIT}}\approx 360$ K points to a 1:1 mixture of IS and HS
states. Although our magnetization experiments can hardly distinguish
which cobalt ions alter their spin state at the spin-state transition at
$T_{\text{MIT}}$ and whether the IS state is realized in pyramidal or
octahedral positions, a structural analysis of the oxygen coordination
by Frontera {\it et al.} suggests that it is the cobalt ions in
octahedral positions that participate in the spin-state (LS--HS)
transition, while the cobalt ions in pyramidal sites keep their IS state
regardless of temperature.\cite{Frontera}

We have eventually reached the most likely picture of the magnetic
ordering in GdBaCo$_{2}$O$_{5.50}$, which is illustrated in Fig. 29. The
alternating filled and empty oxygen chains shown in Fig. 29(a) create
two types of structural environment, octahedral and pyramidal, for
Co$^{3+}$ ions; the former favors the non-magnetic LS ground state,
while the latter makes the IS state preferable. At high temperatures,
the entropy keeps the cobalt ions in octahedral positions in the HS
state, and the compound behaves as a paramagnet composed of IS and HS
Co$^{3+}$ ions (not to mention Gd$^{3+}$ ions which are always
paramagnetic). Upon cooling below $T_{\text{MIT}}\approx 360$ K, the
octahedral cobalt ions cooperatively switch into the LS state, and thus
the CoO$_2$ planes develop a spin-state order consisting of alternating
rows of Co$^{3+}$ ions in the LS and IS states. Consequently, the
magnetic cobalt ions form 2-leg ladders extended along the $a$ axis and
separated from each other by non-magnetic CoO$_2$ $ac$ layers along the
$b$ axis and by paramagnetic GdO$_{0.5}$ layers along the $c$ axis, as
sketched in Fig. 29(a).

Empirically, the spin interaction in ladders formed below
$T_{\text{MIT}}$ turns out to be ferromagnetic, and eventually the
ladders establish a FM order below $T_{\text{C}}\approx 280-300$ K with
spins aligned strictly along the $a$ axis [Fig. 29(a)]. Owing to the
reduced dimensionality (quasi-1D/2D) of the ladders, the FM order
develops quite gradually upon cooling, being subject to strong thermal
fluctuations. The strength of the FM interaction in ladders,
$J/k_{\text{B}}$, can be estimated from the Curie temperature: The
molecular-field theory for 3D systems \cite {J_Tc} would give
$J/k_{\text{B}}\sim 150$ K for $T_{\text{C}} \sim 300$ K; however, for
the quasi-1D/2D magnetic ordering, $J$ should be roughly twice as large
for the same $T_{\text{C}}$ to be reached.\cite {Fisher} It is worth
noting that GdBaCo$_{2}$O$_{5.50}$ is an insulator and thus such
mechanism of the FM ordering as the double exchange,\cite{Double} which
is common in metallic oxides,\cite{Imada} is irrelevant here. In
insulators, the magnetic ordering is known to be predominantly caused by
the superexchange interaction, whose sign for each pair of ions can be
estimated using Goodenough-Kanamori rules.\cite{Anderson, Goodenough,
Kanamori, Weihe} To account for the FM spin order in ladders, we need to
assume some kind of orbital ordering among the IS Co-ions.\cite{GBC_PRL}
We can even speculate that the superstructure reflections along the $a$
axis often observed in the neutron scattering are brought about by such
orbital ordering.

Whether a macroscopic magnetic moment will emerge or not after the
FM-ordered ladders are formed depends on their relative orientation,
which can be either ferro- or antiferromagnetic due to the Ising-like
nature of the spins. In a narrow temperature range below $T_{\text{C}}$,
the weak effective interaction between ladders turns out to be
ferromagnetic and a net FM moment shows up; upon further cooling below
$T_{\text{FM-AF}}$, however, it gradually changes sign, resulting in the
AF ground state. The ladder stacking may become antiferromagnetic along
the $b$ axis [Fig. 29(b)], inducing doubling of the magnetic unit cell
along that axis, or along the $c$ axis [Fig. 29(c)], or both [Fig.
29(d)].

Quite naturally, the inter-ladder coupling across non-magnetic layers is
weak, and switching from the ground-state AF order to the field-induced
FM one can be induced by a fairly weak magnetic field $H_c$ that depends
on temperature. To understand the role of temperature, one should
consider the thermally-excited states: upon heating, a certain amount of
LS Co$^{3+}$ ions in octahedral positions become Co$^{2+}$, Co$^{4+}$,
or change their spin state. Whatever the case, each excited ion acquires
a non-zero spin, providing an additional bridge between spin-ordered FM
ladders. In this respect, it is worth noting that the AF ordering can be
suppressed not only by magnetic fields or thermal excitations, but also
by changing the oxygen stoichiometry: any deviation from $x=0.5$ also
introduces Co$^{2+}$ or Co$^{4+}$ ions and shifts the AF-FM phase
transition to lower temperature [see Fig. 18(b) and Fig. 28].

We can conclude that the GdBaCo$_{2}$O$_{5.50}$ compound with precisely
tuned stoichiometry represents a very interesting magnetic system of
weakly interacting FM ladders with Ising-like moments.

\subsection{Electronic structure of GdBaCo$_{2}$O$_{5.50}$}
\label{sec:elstruct}

\begin{figure}[!t]
\includegraphics*[width=8.6cm]{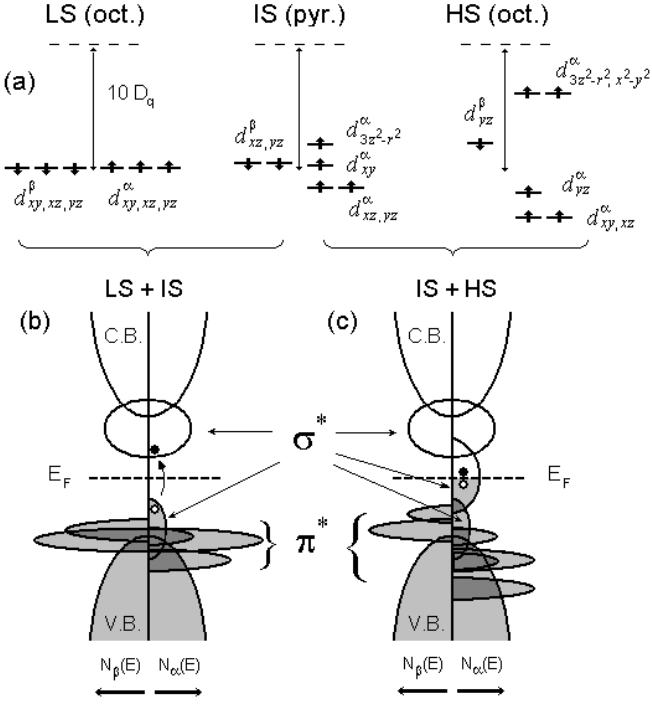}
\caption{Schematic representation of the electronic structure of
GdBaCo$_{2}$O$_{5.50}$. (a) Electronic levels of a Co$^{3+}$ ion in the
LS state (octahedral environment); IS state (pyramidal environment); and
HS state (octahedral environment). The electronic band structure of
CoO$_2$ planes below the metal-insulator/spin-state transition
originates from a superposition of the LS(oct.) and IS(pyr.) states (b),
while above the transition it originates from IS(pyr.) and HS(oct.)
states (c).}
\end{figure}

As follows from the magnetic and structural data of
GdBaCo$_{2}$O$_{5.50}$ discussed above, the metal-insulator transition
at $T_{\text{MIT}}\approx 360$ K is accompanied with a change in the
spin-state of Co$^{3+}$ ions: Upon heating across $T_{\text{MIT}}$, the
Co$^{3+}$ ions in octahedral positions switch their spin state from the
LS to HS state, while those in pyramidal positions keep their IS state
at all temperatures. Figure 30(a) shows the electronic energy levels of
Co$^{3+}$ ions for the relevant states, namely, the LS and HS states in
octahedral environment, and the IS state in pyramidal environment. This
level structure is evaluated in the framework of the ionic model that
takes into account the crystal-field (CF) energy splitting ($10D_q$) and
on-site electrons interactions (the intra-atomic exchange energy
$J_{\text{H}}$ and the Coulomb repulsion energies $U$ and $U^\prime$,
defined following Kanamori\cite{Kanamori, Pouchard}).

In the octahedral environment ($O_h$ symmetry), the crystal field splits
the 3d electron energy levels of a Co$^{3+}$ ion into three $t_{2g}$
orbitals ($d_{xy}$, $d_{xz}$, and $d_{yz}$ with $\Delta E=-4D_q$) and
two $e_g$ orbitals ($d_{x^2-y^2}$ and $d_{3z^2-r^2}$ with $\Delta
E=+6D_q$). When such cobalt ion acquires the LS state, all its six
valence electrons occupy the degenerate $t_{2g}$ orbitals (three
$d^\alpha_{xy, xz, yz}$ spin-up and three $d^\beta_{xy, xz, yz}$
spin-down). Every electron therefore has the same energy, and the total
electron energy per Co$^{3+}$ ion, which takes into account the on-site
interactions and the crystal field splitting, can be expressed as
$E^{LS}_{oct}=E_0 -24D_q -6J_{\text{H}} +3U +12U^\prime$, where $E_0$ is
a reference energy.

In order to bring the Co$^{3+}$ ion into the IS state, one electron
should go from a $t_{2g}$ orbital to a higher-located $e_g$ one, for
example, from $d^\beta_{xy}$ to $d^\alpha_{3z^2-r^2}$. This leads to an
additional splitting of the one-electron energy levels that removes the
initial degeneracy of $t_{2g}$ orbitals: the intra-atomic exchange is
modified, causing every spin-up electron to lower its energy by
$J_{\text{H}}$, while every spin-down electron increases the energy by
$J_{\text{H}}$. Besides, the Coulomb matrix element $U$ between the
$d^\alpha_{xy}$ and $d^\beta_{xy}$ orbitals is replaced with $U^\prime$
between the $d^\alpha_{xy}$ and $d^\alpha_{3z^2-r^2}$ orbitals.
Consequently, the total electron energy for the IS state becomes
$E^{IS}_{oct}=E_0 -14D_q
-7J_{\text{H}} +2U+13U^\prime$. And finally, in the case of the
Co$^{3+}$ HS state in the octahedral environment, one more electron is
transferred from the $t_{2g}$ to $e_g$ shell (from $d^\beta_{xz}$ to
$d^\alpha_{x^2-y^2}$ in Fig. 30(a)), causing a further splitting of the
energy levels. The total electron energy for the HS state can be written
as $E^{HS}_{oct}=E_0
-4D_q -10J_{\text{H}} +U +14U^\prime$.

The energy levels of cobalt ions in the pyramidal environment ($C_{4v}$
symmetry) differ from the previous cases because the crystal field
causes a different energy splitting: +9.14$D_q$ ($d_{x^2-y^2}$),
+0.86$D_q$ ($d_{3z^2-r^2}$), -0.86$D_q$ ($d_{xy}$) and -4.57$D_q$
($d_{xz}$ and $d_{yz}$) for a square pyramid.\cite{Huheey} Consequently,
the total electron energy for the IS state, for instance, becomes
$E^{IS}_{pyr}=E_0-18.28D_q -7J_{\text{H}} +2U+13U^\prime$.

In order to evaluate which spin states of Co$^{3+}$ ions should be
preferable, we need to compare their relative energies. Given that the
difference between the intra- and inter-orbital Coulomb matrix elements
is $U-U^\prime=2J_{\text{H}}$, only two parameters $J_{\text{H}}$ and
$D_q$ appear to determine the one-electron energy levels of localized
Co$^{3+}$ states in the framework of the ionic model. Quite reasonable
parameters for Co$^{3+}$ ions are $J_{\text{H}}\sim 0.5$ eV and $D_q\sim
0.25$ eV.\cite{Huheey} Following this approach, one can easily find out
that the IS state in octahedrons is completely unstable with respect to
either the LS or HS state, while either of the latter states may win,
depending on the parameters. However, for the pyramidal positions, the
IS state appears to be stable in a very broad range of
$J_{\text{H}}$/$D_q$ (from 0.57 to 2.74), as has been concluded by
Pouchard {\it et al.}.\cite{Pouchard}

It is worth noting also that the result of this spin-state competition,
that is, which one of the spin states closely located in energy will
win, may well depend on temperature. For example, the non-magnetic LS
state in octahedral positions experimentally appears to be the ground
state, and therefore it dominates at low temperatures. However, the
spin-orbital degeneracy of the HS state is much larger,\cite{Koshibae}
and thus with increasing temperature, the HS state should sooner or
later take over the LS state because of the entropy terms associated
with the spin-orbital degeneracy; this mechanism is likely staying
behind the spin-state transition observed at $T_{\text{MIT}}\approx 360$
K.

Another important issue is the energy-level position of occupied $e_g$
orbitals. As shown in Fig. 30(a), for IS Co$^{3+}$ ions located in the
pyramidal surrounding, the energy of $d_{3z^2-r^2}$ occupied orbital is
very close to the energy of $d_{xy}$, $d_{xz}$, and $d_{yz}$ orbitals,
mainly due to the strong splitting of $d_{3z^2-r^2}$ and $d_{x^2-y^2}$
levels by the crystal field of the C$_{4v}$ symmetry. In contrast, for
cobalt ions in the octahedral environment, the energy of $d_{3z^2-r^2}$
and $d_{x^2-y^2}$ occupied orbitals appears to be much higher even in
the HS state.

Quite naturally, the qualitative features of these simple ionic model
may be expected to be reproduced in the band structure. In the framework
of a more sophisticated ligand-field (LF) approach, the overlap of
cobalt 3$d$, 4$s$, 4$p$ and oxygen 2$d$ orbitals leads to the formation
of bonding, antibonding, and non-bonding molecular orbitals (MO), which
are broadened in a solid into energy bands. The bonding molecular
orbital, being predominantly composed of the $s$ and $p$ ``atomic
orbitals'' (AO), creates a wide valence band (indicated in Figs. 30(b)
and 30(c) as V.B.) similar to ordinary band insulators (semiconductors).
In turn, the antibonding molecular orbital, being also predominantly
composed of $s$ and $p$ AO of cobalt and oxygen ions, creates a wide
conduction band (indicated as C.B.).

The interaction of cobalt $d$ orbitals with oxygen $p$ orbitals has a
different character. In a CoO$_6$ octahedron, the cobalt $t_{2g}$
orbitals should be non-bonding, because there is no linear combination
of oxygen orbitals of the same symmetry for them to interact. As a
result, a rather narrow $\pi^*$ band occupied by almost localized
electrons appears at the top of the valence band. On the other hand, the
cobalt $e_g$ orbitals interact with a linear combination of oxygen
orbitals of the $e_g$ symmetry, increasing the energy of the antibonding
$\sigma^*$ band. A center of the $\sigma^*(e_g)$ band is located by
10D$_q$ higher than the center of the $\pi^*(t_{2g})$ band (the same
energy splitting as in the CF theory). Note that in the LS state, the
$\sigma^*$ band is empty.

The on-site electron interactions as well as distortions of the CoO$_6$
octahedron, with the CoO$_5$ pyramid being the ultimate case, shift the
energy bands in the same way as in the case of discrete levels. For IS
Co$^{3+}$ ions in the pyramidal environment, the $\sigma^*$ band
occupied by electrons with the prevailing spin projection is separated
from the conduction band and shifted down to the top of the valence band
[as shown in Fig. 30(b)]. On the other hand, in the case of the HS state
in octahedral environment, the occupied $\sigma^*$ band is situated
higher in energy and is twice as large [Fig. 30(c)].

Figure 30(b) presents a simplified electronic structure originating from
a 1:1 mixture of the electronic states of CoO$_6$ octahedrons and
CoO$_5$ pyramids with the Co$^{3+}$ ions being in the LS and IS states,
respectively. We expect this state to be realized in
GdBaCo$_{2}$O$_{5.50}$ below the metal-insulator transition. The
high-temperature electronic structure corresponding to HS Co$^{3+}$ ions
in octahedral environment and IS ions in pyramidal surrounding is
depicted in Fig. 30(c). One can see that the most prominent difference
between these two pictures is a band gap in the first case, while the
occupied $\sigma^*$ band overlaps with the conduction band in the latter
one. Whether the energy gap emerges or not is determined mainly by the
position of the $\sigma^*$ bands as well as by their widths. The
electronic structure of GdBaCo$_{2}$O$_{5.50}$ discussed here offers a
natural, though much simplified, explanation of the metal-insulator
transition that is observed to coincide with the spin-state one at
$T_{\text{MIT}}\approx 360$ K.

\subsection{Transport properties of GdBaCo$_{2}$O$_{5.50}$}

\begin{figure}[!t]
\includegraphics*[width=8.6cm]{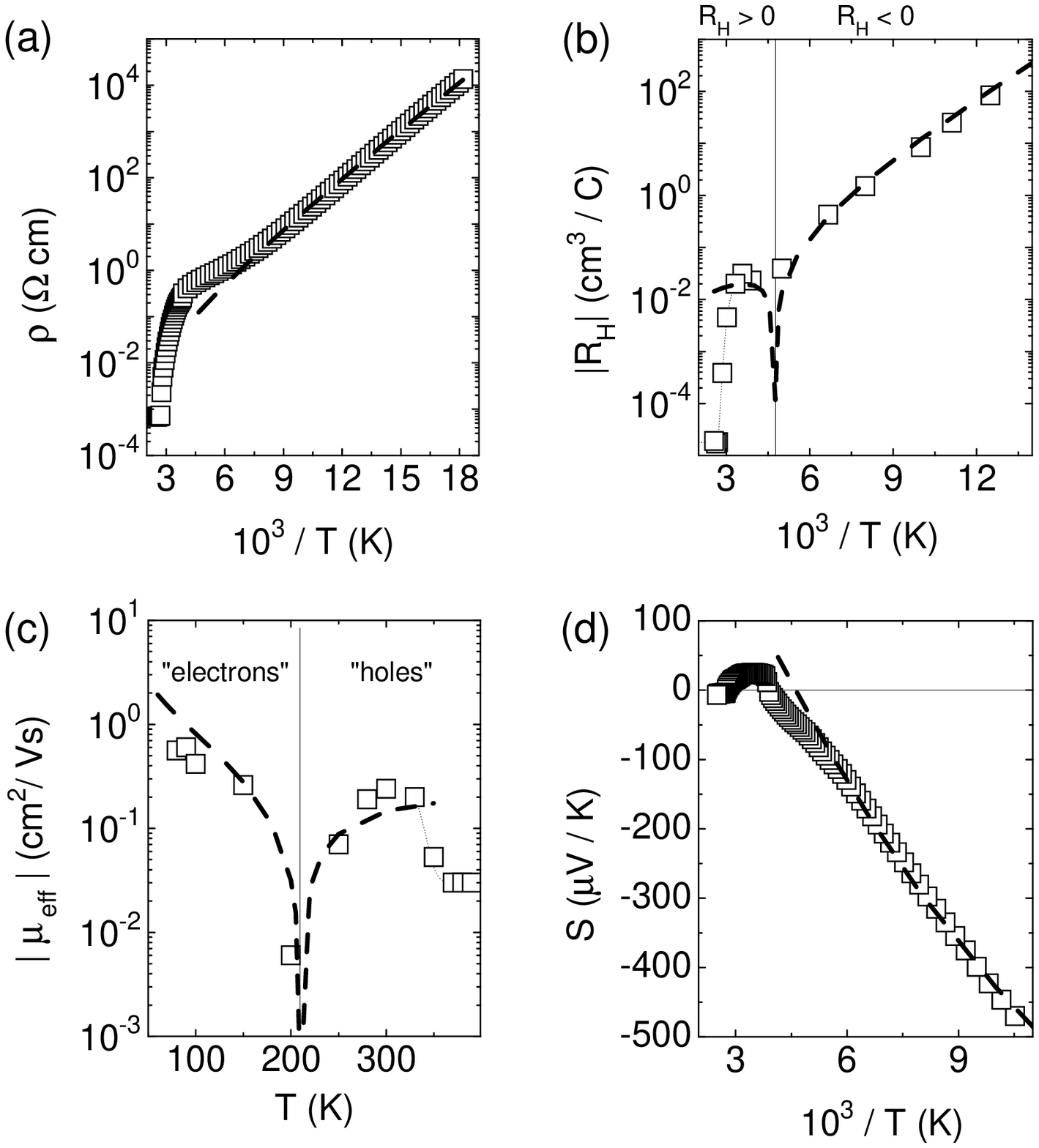}
\caption{Temperature dependences of the in-plane resistivity (a),
Hall coefficient (b), effective mobility $\mu_{\text{eff}} =
R_{\text{H}}/\rho$ (c), and thermoelectric power (d) in
GdBaCo$_{2}$O$_{5.50}$ crystals. The dashed lines are results of
simulation.}
\end{figure}

Having obtained a qualitative picture of the electronic structure for
the parent compound GdBaCo$_{2}$O$_{5.50}$, we can now review its
transport properties. The temperature dependences of the resistivity
$\rho(T)$, Hall coefficient $R_{\text{H}}(T)$, and thermoelectric power
$S(T)$ of GdBaCo$_{2}$O$_{5.50}$ are summarized in Fig. 31 together with
an ``effective'' mobility of carriers defined as $\mu_{\text{eff}} =
R_{\text{H}}/\rho$. The latter would correspond to a true Hall mobility
$\mu_{\text{H}}$, if only one type of charge carriers were present in
the system.

Apparently, all the transport properties presented in Fig. 31 behave
quite coherently. At low temperatures, both $\rho(T)$ and
$R_{\text{H}}(T)$ grow exponentially upon cooling with an activation
energy $\Delta\approx 70$ meV. An activation character of the charge
transport with the same energy gap can be also inferred from the
thermoelectric power $S(T)$ which is linear in $1/T$. On the other hand,
when the temperature increases above the metal-insulator transition at
$\approx 360$ K, all these physical quantities, $\rho$, $R_{\text{H}}$,
and $S$, acquire small and almost temperature-independent values:
$\rho_{ab} \approx \rho_c \approx 600$ $\mu\Omega$cm, $R_{\text{H}}
\approx +1.8$$\times$$10^{-5}$ cm$^3$/C, and $S \approx -4$ $\mu$V/K.
This small Hall coefficient would correspond to an unrealistically high
density of carriers of $\approx 20$ holes per Co ion, if one assumes
only one type of carriers to participate in the charge transport;
correspondingly, the apparent mobility $\mu_{\text{eff}}$ becomes
unrealistically small above $T_{\text{MIT}}$ [Fig. 31(c)]. One can
naturally conclude, therefore, that the Hall signal is strongly reduced
at high temperatures because both electrons and holes move almost
equally and their contributions to $R_{\text{H}}$, being very close in
magnitude but opposite in sign, almost cancel each other. The metallic
charge transport provided by both electrons and holes is presumably
responsible for the small value of the Seebeck coefficient as well. Note
that the sign of $S$ at high temperatures is opposite to that of the
Hall resistivity, which can easily happen when different types of charge
carriers are providing delicately balanced contributions: The holes'
contribution to $R_{\text{H}}$ is a bit larger than the electrons' one,
while in $S$ the balance appears to be opposite.

One more interesting feature in Figs. 31(b) and 31(d) is that both
$R_{\text{H}}$ and $S$ change their sign from negative to positive at $T
\approx 200$ K, long before approaching the metal-insulator transition. This
sign change indicates that the electrons and holes in
GdBaCo$_{2}$O$_{5.50}$ keep providing comparable contributions to the
conductivity down to quite low temperatures, and thus keep competing for
the dominating role in the Hall response and thermoelectric power. On one
hand, this is not surprising, given that the neutrality condition in
a non-doped semiconductor requires the number of electrons to match the
number of holes, $n=p$. On the other hand, however, the switch in sign of
$R_{\text{H}}$ and $S$ would imply that the relative mobilities of electrons
and holes should quickly change, with their ratio $\mu_n/\mu_p$ being $\gg
1$ at low temperatures but $\mu_n/\mu_p \ll 1$ at high temperatures.

One possibility is that such asymmetric mobility evolution actually
takes place, though we cannot point to any obvious reason for that.
Another possibility to be considered is that the mobility ratio is
essentially temperature independent, but instead there are more than two
types of carriers in the system. Namely, there may be two distinct types
of holes (the total number of holes $p=p_1 + p_2=n$) with very different
mobilities $\mu_{p1} \ll \mu_n \ll \mu_{p2}$ and different activation
energies. Whenever the temperature is modified, the relative shares of
the two kinds of holes should change, and their average mobility should
change accordingly, mimicking a gradual mobility evolution.

In order to understand where the different types of carries may be
coming from, we can refer to the qualitative electronic structure of
GdBaCo$_{2}$O$_{5.50}$ sketched in Fig. 30(b). The bands nearest to the
Fermi level are the conduction band and a narrow band composed of
occupied $\pi^*$ and $\sigma^*$ cobalt $d$-bands, with $E_{\text{F}}$
being located exactly in the center of the band gap at $T = 0$.
Apparently, at the lowest temperatures, only excitations from the top of
the $d$-band to the bottom of the conduction band may take place, and
thus the carriers in these bands should govern the transport properties.
It is reasonable to expect the mobility of holes in the narrow $d$-band
composed of almost localized states ($\mu_{p1}$) to be much lower than
that of electrons in the conduction band ($\mu_n$), though the latter is
not very high either, since the conduction band also includes $\sigma^*$
cobalt states (unoccupied). Under these conditions, the electrons
thermally activated into the conduction band should make a predominant
contribution to the low-temperature charge transport, providing a simple
activation behavior for the resistivity and the Hall coefficient, $\rho
\approx en\mu_n$ and $R_{\text{H}} \approx -1/en$. The thermoelectric
power for electrons in the conduction band, $S_n$, should follow a
conventional law:
\begin{equation}
S_n = -\frac{k_{\text{B}}}{e}\: \left [ \:\frac{E_{\text{C}} -
E_{\text{F}}}{k_{\text{B}}T}+ \left ( r+\frac{5}{2} \right )\: \right ],
\end{equation}
where the energy $E_{\text{C}}$ corresponds to the conduction-band
bottom, and $r$ is determined by the scattering mechanism (through the
energy dependence of the relaxation time $\tau\sim E^{r}$). Note that a
similar expression is valid for the thermoelectric power of holes,
$S_p$, as well, but their low mobility does not allow them to compete
with electrons at low temperatures, because the total thermoelectric
power for the two-carrier system is calculated, in the simplest model,
as
\begin{equation}
S_{tot} = \frac{\sigma_nS_n + \sigma_{p}S_{p}}{\sigma_n+ \sigma_{p}},
\end{equation}
where conductivity of electrons, $\sigma_n$ (as well as holes,
$\sigma_p$) is directly proportional to their mobility.

Upon increasing temperature, a wider energy range of electronic states
becomes available for thermal excitation, and at some point, a number of
electrons excited from the {\it valence} band to the conduction band
should become considerable. While the behavior of electrons in the
conduction band is the same regardless of where they are activated from,
the holes generated in the valence band clearly differ from those in the
$d$ band, because the valence band is much wider than the $d$ band and
the mobility is expected to be larger in the wider band. As soon as the
valence-band holes join the transport, the share of holes in the
conductivity, Hall coefficient, and thermoelectric power quickly
increases, and at some temperature, it exceeds the contribution of
electrons owing to the high mobility $\mu_{p_{2}}\gg\mu_n$; as a result,
$R_{\text{H}}(T)$ and $S(T)$ change their sign from negative to
positive.

To illustrate the latter picture, in Fig. 31 we show results of a
simulation (dashed lines) for $\rho(T)$, $R_{\text{H}}(T)$,
$\mu_{eff}(T)$, and $S(T)$ obtained for the three-band model, where the
mobility ratios were taken as $\mu_{p_{1}} : \mu_n : \mu_{p_{2}} = 0.1 :
1 :7$. One can see that even such very simplified model provides a
reasonable description for all the measured transport properties on the
insulator side of the metal-insulator transition. We can conclude,
therefore, that the electronic structure of GdBaCo$_{2}$O$_{5.50}$
proposed in Sec.\ \ref{sec:elstruct} gives a consistent explanation not
only for the metal-insulator transition, originating from the spin-state
transition, but also can account for the behavior of resistivity, Hall
coefficient and thermoelectric power as well.

\subsection{The origin of giant magnetoresistance}

At this point, it is interesting to consider the mechanism responsible
for the giant magnetoresistance observed in GdBaCo$_{2}$O$_{5+x}$ and
other isostructural compounds. Although the MR behavior in
GdBaCo$_{2}$O$_{5+x}$ bears some resemblance to that in manganites (in
particular, the MR in both systems is related to the AF-FM transition),
there are several dissimilar features pointing to different MR
mechanisms. Most important is that GdBaCo$_{2}$O$_{5+x}$, in contrast to
manganites, remains to be an insulator even in the FM state, indicating
that neither the double-exchange mechanism\cite{Imada, CMR_rev, Double}
nor percolation through some metallic phase is relevant
here;\cite{CMR_rev} the MR does not seem to originate from the
``spin-valve'' effects, i.e., from a tunneling between two
spin-polarized metallic regions, either. Our analysis of the resistivity
in the AF and FM states (Fig. 25) has shown that the activation energy
for carriers diminishes considerably upon the AF-FM transition, and it
is mainly a change in the density of carriers, rather than in their
mobility, that affects the resistivity.

By examining the crystal and magnetic structure of
GdBaCo$_{2}$O$_{5.50}$ (Fig. 29), one can easily find out that the
charge transport, at least along the $c$ axis, should be governed by
nominally non-magnetic CoO$_2$ layers, which go along the $ac$ plane and
are composed of cobalt ions with octahedral oxygen environment. It
appears, therefore, that the charge-carrier doping of these octahedral
CoO$_2$ planes is to a large extent determined by the relative
orientation of the magnetic moments in neighboring FM ladders. To
distinguish this intriguing MR scheme from other MR mechanisms, in a
recent paper\cite{GBC_PRL} we have coined a term ``magnetic field-effect
transistor'', based on a hypothetic structure where the charge-carrier
injection into a 2D semiconducting channel is controlled not by an
electric field, but by a magnetic state of neighboring ``ligands''.

One might wonder how the reorientation of very weakly coupled FM ladders
may affect the charge transport, bringing about a giant MR, particularly
in the nominally non-magnetic CoO$_2$ $ac$ layers. Qualitatively, the MR
mechanism appears to be quite simple: GdBaCo$_{2}$O$_{5.50}$ is a
narrow-gap insulator, where the carrier generation goes through
formation of electron-hole pairs, i.e., through formation of {\it
inevitably magnetic} Co$^{2+}$ and Co$^{4+}$ states, whose energy must
depend on the surrounding magnetic order.\cite{GBC_PRL} In other words,
the energy gap that opens at the Fermi level due to the crystal-field
splitting and on-site interactions must be further modified by the
spin-dependent exchange interactions with neighboring Co ions. As we
already know, the carriers generated by thermal excitations or doping
always favor the FM coupling between ladders (Fig. 28); thus, the FM
arrangement of ladders should, in turn, also make the carrier generation
easier. One can conceive this interrelation in a way that the energy for
generating carriers in the AF state also includes a penalty that should
be paid upon frustrating the relative order of the
antiferromagnetically-coupled ladders. When a magnetic field aligns the
FM ladders, this penalty is removed, which lowers the barrier for
carrier generation. One can expect the reduction of the insulating gap
upon the field-induced AF-FM transition to be of the same order as the
inter-atomic exchange energy $J\sim 25$ meV, which is indeed observed
experimentally (Fig. 25).

Lastly, an interesting point to be emphasized in the MR behavior is a
strong amplification effect, whereby an apparently small energy of the
magnetic field $g\mu_{\text{B}}H/k_{\text{B}}\sim 1$ K becomes capable
of changing the carriers' activation energy by several hundreds K. This
gives one more evidence for the cooperative nature of the observed MR.

\subsection{Evolution of physical properties of GdBaCo$_{2}$O$_{5+x}$
with oxygen content}

The main motif governing the behavior of the GdBaCo$_{2}$O$_{5+x}$
compound is obviously a competition of various kinds of ordering,
involving charge, spin, orbital and structural degrees of freedom. While
for several particular compositions ($x\approx 0, 0.5, 0.7$, etc.), this
ordering competition is won by certain homogeneous states, the majority
of the phase diagram appears to be occupied by mixtures of various
ordered phases (Fig. 28).

In the end-point $x=0$ compound, which corresponds to the 50\% electron
doping, the Co$^{2+}$ and Co$^{3+}$ ions order into alternating lines
running along the $a$ axis, where each cobalt spin is aligned
antiferromagnetically to six nearest neighbors to form a G-type AF
structure.\cite{Vogt, Suard} This means that in GdBaCo$_{2}$O$_{5.00}$
all the nearest-neighbor superexchange interactions, be they between the
Co$^{2+}$ ions or Co$^{3+}$ ions, are antiferromagnetic. Under such
conditions, one would expect the charge doping, which converts some
Co$^{2+}$ ions into Co$^{3+}$ ones, to cause no considerable effect on
the magnetic behavior; this should certainly be the case if just the
size of AF-ordered spins is modified. In reality, however, the oxygen
doping immediately brings in some sort of ferromagnetism. Since
GdBaCo$_{2}$O$_{5+x}$ retains its insulating behavior all the way up to
$x\sim 0.5$, one has to invoke a qualitative change in the orbital order
to account for the emerging FM behavior. Apparently, the orbital order
that underlies the AF spin ordering in GdBaCo$_{2}$O$_{5.00}$ is stable
only for the particular unidirectional charge order which is realized at
the 50\% electron doping. When the holes are introduced upon oxygen
doping, they immediately frustrate the pattern of the charge and orbital
order in CoO$_2$ planes. In fact, this doping-induced frustration offers
a clue to the origin of nanoscopic phase separation, whose spectacular
manifestations appear in the magnetic behavior: To avoid frustration,
the charge-ordered AF matrix expel the doped holes, which are then
segregated in small droplets of a new FM phase, as sketched in Fig. 32;
the evidence for this phase segregation is found in the magnetization
data, as we discussed in Section III D.

\begin{figure}[!t]
\includegraphics*[width=7.8cm]{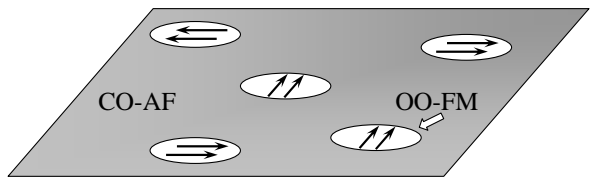}
\caption{A sketch of the nanoscopic-phase-separated state in
GdBaCo$_{2}$O$_{5+x}$ at oxygen concentrations $x\sim 0.1-0.2$. Upon
oxygen doping, nanoscopic droplets of the orbital-ordered FM (OO-FM)
phase emerge in the matrix of the charge-ordered antiferromagnet
(CO-AF). As $x$ increases, these droplets grow and should reach
percolation at $x\approx 0.25$, when a half of the AF phase is switched
into the FM one. Note that even in small droplets, the FM moments should
be aligned along one of the in-plane tetragonal axis because of the
strong spin-orbit coupling.}
\vspace{-7pt}
\end{figure}

An electronic phase separation, arising from the tendency of an ordered
state to avoid frustration, is quite common for strongly correlated
systems. Usually, the frustration is induced by the charge carrier
motion over an AF insulating background; the system, therefore, tends to
segregate the mobile carriers into metallic regions with a modified
magnetic order.\cite{CMR_rev, Kivelson} What is peculiar in
GdBaCo$_{2}$O$_{5.00}$ is that the AF and FM phases here are both
insulating; in this case, they differ only in the charge and orbital
order. Presumably, these phases correspond to the charge-ordered
insulator (50\% electron doping) and the undoped band insulator (zero
doping).

An interesting issue is whether the oxygen ions participate in the phase
separation or it is mostly an electronic phenomenon. Based on the very
small size of the FM-phase droplets -- they are not detected by the
X-ray scattering (Sec.\ \ref{sec:Str}), and exhibit just a
superparamagnetic behavior at moderate oxygen doping -- we can conclude
that the oxygen ions are not involved in the large-scale redistribution.
Though the negatively-charged oxygen ions may rearrange locally, being
dragged by the Coulomb interaction with the doped carriers, this motion
is apparently insufficient to cause the domains to grow. Consequently,
the observed nanoscopic phase separation at $x<0.5$ is most likely
governed by the electronic interactions in the CoO$_2$ planes and thus
may be a generic feature of layered cobalt oxides.

The phase-separated state depicted in Fig. 32 gives a consistent
explanation for the magnetic and transport properties of
GdBaCo$_{2}$O$_{5+x}$ in the electron-doped region ($0<x<0.5$). As long
as the FM droplets are small and located far apart, they exhibit a
superparamagnetic behavior with thermomagnetic irreversibility features
originating from the strong spin-orbit coupling. With increasing oxygen
content, the FM droplets grow until they achieve a percolation within
the CoO$_2$ planes at $x\approx 0.25$; from that point, the initial AF
phase is kept in inclusions embedded in the FM matrix. Eventually, the
system becomes homogeneous again as the composition approaches that of
the undoped parent insulator, $x=0.50$. Given that both the FM and AF
phases are insulators, it is not surprising that the conductivity of
crystals is virtually independent on doping in the entire electron-doped
region, and that it goes through variable-range hoping of localized
carriers.

While in the electron-doped region, GdBaCo$_{2}$O$_{5+x}$ tends to phase
separate into two insulating components, namely, the undoped and
50\%-electron doped phases, the hole doping ($x>0.5$) also quickly
destabilizes the homogeneous parent state. However, the phase separation
in the hole-doped region bears several important differences from the
behavior of electron-doped crystals. First, the scale of resulting
domains is much larger, implying that the oxygen ions are involved in
the macroscopic (or mesoscopic) rearrangement process and redistribute
over the crystal together with the doped carriers; otherwise, the
long-range Coulomb interactions would prevent the formation of such a
macroscopically inhomogeneous state. Second, the nearest stable phase is
located at a considerably lower doping level of $\sim 18-20$\% of holes
($x\approx 0.68-0.70$). And finally, the most important is that the
stable hole-doped phase appears to be {\it metallic}, in contrast to
insulating electron-doped compositions. The latter explains the
remarkable asymmetry in the doping dependence of the conductivity in
Fig. 7.

Now, we can only speculate on the nature of the metallic hole-doped
phase, since a detailed information on the crystal structure of
GdBaCo$_{2}$O$_{5+x}$ at $x\approx 0.68-0.70$ is still lacking.
Apparently, that phase should be well structurally ordered, since one
can hardly imagine so sharp a FM-AF transition, as shown in Fig. 18(c),
to takes place in a system with a significant disorder. Therefore, some
sort of oxygen ordering is expected to occur in GdO$_x$ layers, similar
to the formation of perfect oxygen chains in the parent compound with
$x=0.50$; one can immediately find out that the composition $x\approx
0.68-070$ best fits the ``Ortho-III'' superstructure where two filled
oxygen chains in GdO$_x$ layers alternate with one empty chain.

If the Ortho-III structure is actually realized in GdBaCo$_{2}$O$_{5+x}$
with $x\sim 0.7$, and the FM-AF transition has the same purely magnetic
origin as in the parent compound with $x=0.50$, one inevitably arrives
at the conclusion that some kind of weakly coupled FM ladders or layers
are formed in the $x\sim 0.7$ phase as well. One can further speculate
that the double octahedral CoO$_2$ $ac$ layers are still non-magnetic in
this Ortho-III phase, and the FM ladders are just located farther apart.
The main difference with the parent $x=0.50$ composition is that for
$x\sim 0.7$, about 20\% of cobalt ions should acquire the Co$^{4+}$
state. If all the holes are gathered in the FM layers, where they can
readily move, those layers would possess a quarter-filled metallic band.
Note that on the macroscopic scale, the conductivity would exhibit a
true metallic behavior only if the crystal has a perfectly ordered
oxygen subsystem; otherwise, the metallic ladders/layers would be always
interrupted by non-conducting regions. In the latter case, a
variable-range hopping between metallic fragments should be
expected.\cite{Fogler} The measured resistivity of $x=0.70$ crystals
indeed exhibits an unusual behavior, tending to saturate at low
temperatures at finite but surprisingly large values (Fig. 6).

While the suggested very speculative picture of the metallic $x\sim 0.7$
phase can easily account for the observed magnetic and transport
properties, more work is obviously necessary to clarify whether it has
anything to do with the actual microscopic state. In any case, this
intriguing phase with self-organized metallic paths bears a clear
resemblance to the stripe structures in high-$T_c$
cuprates,\cite{Kivelson} and the phase-separated conducting phases in
manganites,\cite{CMR_rev} and thus its study may help in understanding
the unusual transport in other oxides as well.

\section{SUMMARY}

In order to gain insight into the general behavior of square-lattice
layered cobalt oxides, we have performed a systematic study of the
transport, magnetic, and thermoelectric properties of
GdBaCo$_{2}$O$_{5+x}$ single crystals over a wide range of oxygen
content, $0\leq x \leq 0.77$. The high-quality crystals were grown by
the floating-zone method and their oxygen content was precisely tuned to
required values by successive annealing treatments so that the entire
phase diagram could be spanned; particularly, in the critical regions of
the diagram, the composition was modified with steps as small as $\Delta
x=0.001-0.01$.

The ``parent'' compound GdBaCo$_{2}$O$_{5.50}$ -- the composition with
all cobalt ions in the Co$^{3+}$ state -- is shown to be a metal at high
temperatures, but it switches into a band insulator (a narrow-gap
semiconductor) below the metal-insulator transition at
$T_{\text{MIT}}\approx360$ K, which is manifested in step-like anomalies
in all the measured properties, $\rho(T)$, $S(T)$, $R_{\text{H}}(T)$,
and $M(T)$. The gap opening at the Fermi level is associated with a
cooperative spin-state transition, which seems to involve a half of the
cobalt ions, while the other half keeps its spin state unchanged. Two
non-equivalent cobalt positions are created by the oxygen ions which
order in GdO$_{0.5}$ layers into alternating filled and empty rows
running along the $a$ axis. Consequently, the CoO$_2$ planes also
develop a spin-state order consisting of alternating rows of Co$^{3+}$
ions in the $S=1$ and $S=0$ states. The overall magnetic structure of
GdBaCo$_{2}$O$_{5.50}$ below $T_{\text{MIT}}$ may be conceived of as a
set of magnetic ``2-leg ladders'', which are composed of $S=1$ Co$^{3+}$
ions and are separated from each other by non-magnetic layers. Upon
cooling below $T_{\text{C}}\approx 300$ K, these ladders develop a
ferromagnetic order; however, whether a macroscopic magnetic moment will
emerge or not after the FM ladders are formed depends on inter-ladder
coupling. According to the magnetization measurements performed on
detwinned single crystals, the cobalt spins exhibit a remarkably strong
Ising-like anisotropy, being aligned along the oxygen-chain direction;
thus the FM ladders in GdBaCo$_{2}$O$_{5.50}$ can {\it only} form a
relative ferromagnetic or antiferromagnetic arrangement. Owing to the
very weak strength of this coupling between ladder, their relative
magnetic order can be easily altered by temperature, doping, or magnetic
fields, bringing about FM$\leftrightarrow$AF transitions and a giant
magnetoresistance.

A study of GdBaCo$_{2}$O$_{5+x}$ crystals with the oxygen content
deviating from $x=0.50$ has revealed that this layered compound is a
unique filling-control system that allows a {\it continuous ambipolar}
doping of the CoO$_2$ planes, that is, the doping level can be
continuously driven across the parent insulating state. This continuous
doping is manifested in spectacular singularities of the transport
properties as the system approaches the undoped state: the
thermoelectric power, for instance, tends to diverge at $x=0.50$, and
abruptly changes its sign upon crossing this peculiar point.

As soon as the CoO$_2$ planes are doped with more than a few percent of
charge carriers, be they electrons or holes, their homogeneous state is
found to become unstable, and the system exhibits a very strong tendency
to separation into several peculiar ordered phases. The resulting nano-
or mesoscopic phase mixture is clearly manifested in the transport and
magnetic properties of the crystals. For the entire electron-doped
region ($x<0.5$), GdBaCo$_{2}$O$_{5+x}$ tends to phase separate into two
insulating components, the undoped phase and the charge-ordered AF
insulator with 50\% electron doping. As a result, the conduction in
GdBaCo$_{2}$O$_{5+x}$ crystals is almost doping independent in the
entire range $0\leq x \leq 0.45$, and goes through a variable range
hopping of electrons. In the hole-doped region, the CoO$_2$ planes also
expel the doped carriers, protecting the undoped domains, but the second
phase generated upon the phase separation turns out to be {\it
metallic}, in contrast to the electron-doped crystals. Consequently, the
GdBaCo$_{2}$O$_{5+x}$ crystals gradually evolve towards a metallic state
with hole doping, and thus the doping dependence of the conductivity
exhibits a remarkable asymmetry with respect to whether the CoO$_2$
planes are doped with electrons or holes.

Although a conclusive picture of the metallic hole-doped phase has not
been established yet, such metallic state self-organized within the
electronically inhomogeneous system bears a clear resemblance to the
behavior of CMR manganites and high-$T_c$ cuprates, and thus its study
may offer a clue to understanding the unusual charge transport in
transition-metal oxides.

To conclude, the GdBaCo$_{2}$O$_{5+x}$ compound has proved to be an
interesting system with a very rich phase diagram originating from the
competition of various spin-charge-orbital ordered phases. What turns it
into a model system and makes it particularly interesting for a detailed
study is the possibility of continuous ambipolar doping of CoO$_2$
planes and the exceptionally strong cobalt-spin anisotropy, which
dramatically narrows down the range of possible spin arrangements.

\begin{acknowledgments}
We thank S. Komiya and K. Segawa for invaluable technical assistance.
\end{acknowledgments}


\begin{thebibliography}{99}

\bibitem{Imada} For review, see M. Imada, A. Fujimori, and Y.Tokura,
Rev. Mod. Phys. {\bf 70}, 1039 (1998).

\bibitem{CMR_rev} For review, see E. Dagotto, T. Hotta, and A. Moreo,
Phys. Rep. {\bf 344}, 1 (2001); E. L. Nagaev, Phys. Usp. {\bf 39}, 781
(1996).

\bibitem{Kivelson} For review, see E. W. Carlson, V. J. Emery,
S. A. Kivelson, and D. Orgad, cond-mat/0206217.

\bibitem{Sr2RuO4} Y. Maeno, H. Hashimoto, K. Yoshida, S. Nishizaki,
T. Fujita, J. G. Bednorz, and F. Lichtenberg, Nature {\bf 372}, 532
(1994).

\bibitem{SC} K. Takada, H. Sakurai, E. Takayama-Muromachi, F. Izumi,
R. A. Dilanian, and T. Sasaki, Nature {\bf 422}, 53 (2003).

\bibitem{LaNiO} J. M. Tranquada, D. J. Buttrey, V. Sachan,
and J. E. Lorenzo, Phys. Rev. Lett. {\bf 73}, 1003 (1994).

\bibitem{CO_Co214} I. A. Zaliznyak, J. P. Hill, J. M. Tranquada,
R. Erwin, and Y. Moritomo, Phys. Rev. Lett. {\bf 85}, 4353 (2000); I. A.
Zaliznyak, J. M. Tranquada, R. Erwin, and Y. Moritomo, Phys. Rev. B {\bf
64}, 195117 (2001).

\bibitem{Vogt} T. Vogt, P. M. Woodward, P. Karen, B. A. Hunter,
P. Henning, and A. R. Moodenbaugh, Phys. Rev. Lett. {\bf 84}, 2969
(2000).

\bibitem{Suard} E. Suard, F. Fauth, V. Caignaert, and I. Mirebeau,
G. Baldinozzi, Phys. Rev. B {\bf 61}, R11871 (2000).

\bibitem{cubic_Co} M. A. Se\~{n}ar\'{i}s-Rodr\'{i}guez and
J. B. Goodenough, J. Solid State Chem. {\bf 118}, 323 (1995); S.
Yamaguchi, Y. Okimoto, H. Taniguchi, and Y. Tokura, Phys. Rev. B {\bf
53}, R2926 (1996); J. Wu and C. Leighton, Phys. Rev. B {\bf 67}, 174408
(2003); and references therein.

\bibitem{BiCo2201} K. J. Thomas, Y. S. Lee, F. C. Chou, B. Khaykovich,
P. A. Lee, M. A. Kastner, R. J. Cava, and J. W. Lynn, Phys. Rev. B {\bf
66}, 054415 (2002).

\bibitem{Martin} C. Martin, A. Maignan, D. Pelloquin, N. Nguyen,
and B. Raveau, Appl. Phys. Lett. {\bf 71}, 1421 (1997).

\bibitem{Troy1} I. O. Troyanchuk, N. V. Kasper, D. D. Khalyavin,
H. Szymczak, R. Szymczak, and M. Baran, Phys. Rev. Lett. {\bf 80}, 3380
(1998).

\bibitem{Troy2} I. O. Troyanchuk, N. V. Kasper, D. D. Khalyavin,
H. Szymczak, R. Szymczak, and M. Baran, Phys. Rev. B {\bf 58}, 2418
(1998).

\bibitem{Maignan} A. Maignan, C. Martin, D. Pelloquin, N. Nguyen,
and B. Raveau, J. Solid State Chem. {\bf 142}, 247 (1999).

\bibitem{Respaud} M. Respaud, C. Frontera, J. L. Garc\'{i}a-Mu\~{n}oz,
M. A. G. Aranda, B. Raquet, J. M. Broto, H. Rakoto, M. Goiran, A.
Llobet, and J. Rodr\'{i}guez-Carvajal, Phys. Rev. B {\bf 64}, 214401
(2001).

\bibitem{GBC_PRL} A. A. Taskin, A. N. Lavrov, and Y. Ando,
Phys. Rev. Lett. {\bf 90}, 227201 (2003).

\bibitem{NaCoO} I. Terasaki, Y. Sasago and K. Uchinokura,
Phys. Rev. B {\bf 56}, R12685 (1997).

\bibitem{S_book} {\it Oxide thermoelectrics}, edited by K. Koumoto,
I. Terasaki, and N. Murayama (Research Signpost, Kerala, 2002).

\bibitem{Koshibae} W. Koshibae, K. Tsutsui, and S. Maekawa,
Phys. Rev. B {\bf 62}, 6869 (2000); W. Koshibae and S. Maekawa, Phys.
Rev. Lett. {\bf 87}, 236603 (2001).

\bibitem{Ando_specific_heat} Y. Ando, N. Miyamoto, K. Segawa, T. Kawata
and I. Terasaki, Phys. Rev. B {\bf 60}, 10 580 (1999).

\bibitem{OpticalSm} T. Saito, T. Arima, Y. Okimoto, and Y. Tokura,
J. Phys. Soc. Jpn. {\bf 69}, 3525 (2000).

\bibitem{Moritomo} Y. Moritomo, T. Akimoto, M. Takeo, A. Machida ,
E. Nishibori, M. Takata, M. Sakata, K. Ohoyama, and A. Nakamura, Phys.
Rev. B {\bf 61}, R13325 (2000).

\bibitem{Moritomo2} Y. Moritomo, M. Takeo, X. J. Liu, T. Akimoto,
and A. Nakamura, Phys. Rev. B {\bf 58}, R13334 (1998).

\bibitem{Co_214} T. Matsuura, J. Tabuchi, J. Mizusaki, S. Yamauchi, and
K. Fueki, J. Phys. Chem. Solids {\bf 49}, 1403 (1988); Y. Moritomo, K.
Higashi, K. Matsuda, and A. Nakamura, Phys. Rev. B {\bf 55}, R14725
(1997).

\bibitem{misfitRc} I. Tsukada, T. Yamamoto, M. Takagi, T. Tsubone,
S. Konno, and K. Uchinokura, J. Phys. Soc. Jpn. {\bf 70}, 834 (2001); T.
Valla, P. D. Johnson, Z. Yusof, B. Wells, Q. Li, S. M. Loureiro, R. J.
Cava, M. Mikami, Y. Mori, M. Yoshimura, and T. Sasaki, Nature {\bf 417}
627 (2002 ).

\bibitem{misfit} S. H\'{e}bert, S. Lambert, D. Pelloquin, and A. Maignan,
Phys. Rev. B {\bf 64}, 172101 (2001); T. Yamamoto, K. Uchinokura, and I.
Tsukada, Phys. Rev. B {\bf 65}, 184434 (2002).

\bibitem{Akahoshi} D. Akahoshi and Y. Ueda,
J. Solid State Chem. {\bf 156}, 355 (2001).

\bibitem{Kusuya} H. Kusuya, A. Machida, Y. Moritomo, K. Kato,
E. Nishibori, M. Takata, M. Sakata, and A. Nakamura, J. Phys. Soc. Jpn.
{\bf 70}, 3577 (2001)

\bibitem{Frontera} C. Frontera, J. L. Garc\'{i}a-Mu\~{n}oz, A. Llobet,
and M. A. G. Aranda, Phys. Rev. B {\bf 65}, 180405(R) (2002).

\bibitem{YLaLu} A. N. Lavrov, A. A. Taskin, and Y. Ando (unpublished).
Our study has shown that it is virtually impossible to grow
RBaCo$_{2}$O$_{5+x}$ crystals with R = Y, La, or Lu. For R = La, in
contrast to what was claimed in Ref. \onlinecite{Moritomo2}, the
compound crystallizes into a cubic La$_{0.5}$Ba$_{0.5}$CoO$_{3-\delta}$
structure instead of forming the ordered LaBaCo$_{2}$O$_{5+x}$ phase
when sintered in air or O$_2$. Reducing oxygen partial pressure would
favor the ordered 112 phase, but at some point the compound decomposes
into the perovskite phase La$_{2-y}$Ba$_y$CoO$_4$ and an eutectic with
approximate composition of BaCo$_2$O$_z$. According to our data, the
stability range of LaBaCo$_{2}$O$_{5+x}$ in the P - T diagram should be
very small, if any. For R = Y, one can easily prepare single-phase
YBaCo$_{2}$O$_{5+x}$ ceramic, but upon heating it melts incongruently,
decomposing into pure Y$_2$O$_3$ and the BaCo$_2$O$_z$ eutectic. Again,
the stability range of YBaCo$_{2}$O$_{5+x}$ appears to be too narrow for
the crystal growth, since its decomposition temperature almost coincides
with the melting point of the BaCo$_2$O$_z$ eutectic. And finally for R
= Lu, the 112 phase does not seem to be formed.

\bibitem{Yasuda} I. Yasuda and M. Hishinuma,
J. Solid State Chem. {\bf 123}, 382 (1996).

\bibitem{Lane} J.A. Lane and J.A. Kilner,
Solid State Ionics {\bf 136-137}, 997 (2000).

\bibitem{cond-mat} A. A. Taskin, A. N. Lavrov, and Y. Ando, cond-mat/0501127.

\bibitem{YBCO} S. J. Rothman, J. L. Routbort, U. Welp, and J. E. Baker,
Phys. Rev. B {\bf 44}, 2326 (1991).

\bibitem{twins} Y. Ando, K. Segawa, S. Komiya, and A. N. Lavrov,
Phys. Rev. Lett. {\bf 88}, 137005 (2002); A. N. Lavrov, S. Komiya, and
Y. Ando, Nature {\bf 418}, 385 (2002).

\bibitem{MR_accur} Y. Ando and K. Segawa, Phys. Rev. Lett. {\bf 88},
167005 (2002).

\bibitem{oxygen} The highest oxygen concentration $x=0.77$ was achieved
only for ceramic samples, since it required annealing at low
temperatures where oxygen diffusion was already inhibited. It has turned
out, however, that the resistivity of ceramics with $x=0.77$ was still
of insulating type at $T < 100$ K, being quite similar to that of a
single crystal with $x=0.70$ (Fig. 6).

\bibitem{Mott} N. F. Mott and E. A. Davis, {\it Electronic Processes in
Non-Crystalline Materials}, 2nd ed. (Clarendon Press, Oxford, 1979).

\bibitem{Efros}  B. I. Shklovskii,  and  A. L. Efros, {\it Electronic
Properties of Doped Semiconductors} (Springer-Verlag, Berlin, 1984)

\bibitem{Mesosc} Y. Ando, A. N. Lavrov, S. Komiya, K. Segawa,
and X. F. Sun , Phys. Rev. Lett. {\bf 87}, 017001 (2001).

\bibitem{cupr_Rc} Y. Ando, G. S. Boebinger, A. Passner, N. L. Wang,
C. Geibel, and F. Steglich, Phys. Rev. Lett. {\bf 77}, 2065 (1996); A.
N. Lavrov, M. Yu. Kameneva, and L. P. Kozeeva, Phys. Rev. Lett. {\bf
81}, 5636 (1998).

\bibitem{mang_Rc} T. Kimura,  Y. Tomioka,  H. Kuwahara,  A. Asamitsu,
M. Tamura, Y. Tokura, Science {\bf 274}, 1698 (1996).

\bibitem{LaSrTiO} T. Okuda, K. Nakanishi, S. Miyasaka, and Y. Tokura,
Phys. Rev. B {\bf 63}, 113104 (2001).

\bibitem{Proceed} A. A. Taskin, A. N. Lavrov, and Y. Ando, in
{\it Proceedings of 22-nd International Conference on Thermoelectrics,
H\'{e}rault, France, 2003}, p.196.

\bibitem{S_cer} In the close vicinity of $x=0.500$, even a very subtle
modification of the oxygen concentration (by merely $\delta x \sim
0.001$ or less) results in a significant change of the thermoelectric
power $S(T)$. However, the accuracy of measuring the oxygen content in
fairly small single crystals is lower than required for such precise
positioning of the crystals on the $x$ scale. We had therefore to
measure $S(T)$ in both single-crystal samples and samples cut from large
ceramic blocks, where the oxygen content could be determined with the
required precision. Fortunately, the thermoelectric power, in contrast
to resistivity, does not depend on the sample's geometry; also, we found
no noticeable anisotropy in the thermoelectric power of
RBaCo$_{2}$O$_{5+x}$. We observed no considerable difference in the
thermoelectric behavior of our single crystals and ceramics, and thus
used the data for both kinds of samples to establish the detailed doping
dependence $S(x)$.

\bibitem{Jonker_P} M. H. Kane, N. Mansourian-Hadavi, T. O. Mason,
W. Sinkler, L. D. Marks, K. D. Otzschi, D. Ko, and K. R. Poeppelmeier,
J. Solid State Chem. {\bf 148} 3 (1999).

\bibitem{FeCrSe} G. J. Snyder, T. Caillat, and J.-P. Fleurial,
Phys. Rev. B {\bf 62}, 10185 (2000).

\bibitem{Jonk_cup} M.-Y. Su, C. E. Elsbernd, and T. O. Mason,
J. Am. Ceram. Soc. {\bf 73}, 415 (1990).

\bibitem{SG} K. Binder and A. P. Young,
Rev. Mod. Phys. {\bf 58}, 801 (1986).

\bibitem{Soda} M. Soda, Y. Yasui, M. Ito, S. Iikubo, M. Sato,
and K. Kakurai, J. Phys. Soc. Jpn. {\bf 73}, 464 (2004).

\bibitem{LCO} T. Thio, T. R. Thurston, N. W. Preyer, P. J. Picone,
M. A. Kastner, H. P. Jenssen, D. R. Gabbe, C. Y. Chen, R. J. Birgeneau,
A. Aharony, Phys. Rev. B {\bf 38}, R905 (1988); A. N. Lavrov, Y. Ando,
S. Komiya, and I. Tsukada, Phys. Rev. Lett. {\bf 87}, 017007 (2001); S.
Ono, S. Komiya, A. N. Lavrov, Y. Ando, F. F. Balakirev, J. B. Betts, and
G. S. Boebinger, Phys. Rev. B {\bf 70}, 184527 (2004).

\bibitem{homogen} It should be emphasized that the observed two-phase
state can by no means originate from macroscopic composition gradients
in the annealed crystals; it must be an intrinsic instability of the
homogeneous state in GdBaCo$_2$O$_{5+x}$ for some range of doping that
brings about the mesoscopic phase separation. Our experiments clearly
support this conclusion. Namely, one and the same crystal annealed
successively to provide the oxygen concentrations of $x\approx 0.50$,
$x\approx 0.60$, and then $x\approx 0.70$, exhibited a single-phase
behavior for the initial and final states, but behaved as a two-phase
mixture for the intermediate composition of $x\approx 0.60$. Such
non-monotonic evolution cannot be account for by a composition gradient
that would appear in the crystal because of the slow oxygen diffusion;
for the described experiment, the composition gradient would develop
irreversibly, given that the annealing temperature was monotonically
decreased in order to get higher $x$ values, and that the oxygen
diffusion dramatically slows down upon cooling.

\bibitem{Fauth} F. Fauth, E. Suard, and V. Caignaert, Phys.
Rev. B {\bf 65}, 060401(R) (2001).

\bibitem{Double} C. Zener, Phys. Rev. {\bf 82}, 403 (1951);
P.-G. de Gennes, {\it ibid.} 118, 141 (1960).

\bibitem{largest_MA} P. Gambardella, S. Rusponi, M. Veronese,
S. S. Dhesi, C. Grazioli, A. Dallmeyer, I. Cabria, R. Zeller, P. H.
Dederichs, K. Kern, C. Carbone, and H. Brune, Science {\bf 300}, 1130
(2003).

\bibitem{typical_MA} D. Weller and A. Moser,
IEEE Trans. Mag. {\bf 35}, 4423 (1999).

\bibitem{Fauth2} F. Fauth, E. Suard, V. Caignaert, and I. Mirebeau,
Phys. Rev. B {\bf 66}, 184421 (2002).

\bibitem{Neutron2} M. Soda, Y. Yasui, T. Fujita, T. Miyashita, M. Sato,
and K. Kakurai, J. Phys. Soc. Jpn. {\bf 72}, 1729 (2003).

\bibitem{Neutron3} V. P. Plakhty, Yu. P. Chernenkov, S. N. Barilo,
A. Podlesnyak, E. Pomjakushina, D. D. Khalyavin, D. Sheptyakov, E. V.
Moskvin, S. V. Gavrilov, and A. Furrer, cond-mat/0407010.

\bibitem{Neutron4} J. C. Burley, J. F. Mitchell, S. Short, D. Miller,
and Y. Tang, J. Solid State Chem. {\bf 170}, 339 (2003).

\bibitem{J_Tc} $T_{\text{C}} = S(S+1)J_0/3k_{\text{B}}$,
where $S=1$ for IS-Co$^{3+}$ ion and $J_0=3J$, counting only nearest
neighbors.

\bibitem{Fisher} M. E. Fisher, Rep. Progr. Phys. {\bf 30}, 615 (1967).

\bibitem{Anderson} P. W. Anderson, Phys. Rev. {\bf 115}, 2 ( 1959).

\bibitem{Goodenough} J. B. Goodenough, Phys. Rev. {\bf 100}, 564 (1955).

\bibitem{Kanamori} J. Kanamori, J. Phys. Chem. Solids {\bf 10},
87 (1959).

\bibitem{Weihe} H. Weihe and H. U. G\"{u}del,
Inorg. Chem. {\bf 36}, 3632 (1997).

\bibitem{Pouchard} M. Pouchard, A. Villesuzanne, and J.-P. Doumerc,
J. Solid State Chem. {\bf 162}, 282 (2001).

\bibitem{Huheey} J. E. Huheey, E. A. Keiter, and R. L. Keiter,
{\it Inorganic Chemistry: Principles of structure and reactivity, 4th
edn.} (Harper Collins, New York, 1993).

\bibitem{Fogler} M. M. Fogler, S. Teber, and B. I. Shklovskii,
Phys. Rev. B {\bf 69}, 035413 (2004).

\end{thebibliography}
\end{document}